\RequirePackage[T1]{fontenc}
\documentclass[12pt]{article}
\pdfoutput=1 
\usepackage[height=8.85in,width=6.45in]{geometry}

\usepackage[utf8]{inputenc}
\usepackage{amsmath}
\usepackage{amssymb}
\usepackage{mathtools}
\numberwithin{equation}{section}
\usepackage{slashed}
\usepackage{braket}
\usepackage[svgnames]{xcolor}
\usepackage[colorlinks,linktocpage=true,citecolor=DarkGreen,linkcolor=FireBrick]{hyperref}
\usepackage{cite}
\usepackage{graphicx}
\usepackage{tikz}
\usepackage{tikz-cd}
\usepackage{times}
\usepackage[scaled]{couriers}
\usepackage{bm}
\usepackage{subfig}
\usepackage{mathrsfs}
\usepackage{amsthm}

\usepackage{xcolor}
\usepackage{mdframed}

\newcommand{\lsim}{ \mathop{}_{\textstyle \sim}^{\textstyle <} }
\newcommand{\vev}[1]{ \left\langle {#1} \right\rangle }

\DeclareMathOperator{\tr}{tr}

\usepackage{slashed}

\def\cN{{\cal N}}
\def\cO{{\cal O}}

\def\bR{{\mathbb R}}

\def\bZ{{\mathbb Z}}

\def\U{\mathrm{U}}
\def\SU{\mathrm{SU}}

\def\SO{\mathrm{SO}}
\def\Sp{\mathrm{Sp}}

\def\Spin{\mathrm{Spin}}

\def\beq#1\eeq{\begin{align}#1\end{align}}

\newcommand{\GeV}{\  {\rm GeV} }

\newcommand{\lmk}{\left(}  
\newcommand{\rmk}{\right)}

\newcommand{\la}{\left\langle} 
\newcommand{\ra}{\right\rangle}

\newcommand{\abs}[1]{\left\vert {#1} \right\vert}

\newcommand{\eq}[1]{Eq.~(\ref{#1})}

\begin{document}

\begin{titlepage}

\begin{flushright}
TU-1152
\end{flushright}

\vskip 3cm

\begin{center}

{\Large \bfseries Cosmic strings from pure Yang--Mills theory }

\vskip 1cm

Masaki~Yamada$^{1,2}$
and 
Kazuya~Yonekura$^{1}$

\vskip 1cm

\begin{tabular}{ll}
$^1$ Department of Physics, Tohoku University, Sendai 980-8578, Japan
\\
$^2$ FRIS, Tohoku University, Sendai, Miyagi 980-8578, Japan
\end{tabular}

\vskip 1cm

\end{center}

\noindent
We discuss the formation of cosmic strings or macroscopic color flux tubes at the phase transition from the deconfinement to confinement phase in pure Yang--Mills (YM) theory, such as $\SU(N)$, $\Sp(N)$, $\SO(N)$, and $\Spin(N)$, based on the current understanding of theoretical physics. 
According to the holographic dual descriptions, the cosmic strings are dual to fundamental strings or wrapped D-branes in the gravity side depending on the structure of the gauge group, and the reconnection probability is suppressed by $\mathcal{O}(N^{-2})$ and $e^{-\mathcal{O}(N)}$, respectively. 
The pure YM theory thus provides a simple realization of cosmic F- and D-strings without the need for a brane-inflationary scenario or extra dimension. 
We also review the stability of cosmic strings based on the concept of 1-form symmetry, which further implies the existence of a baryon vertex in some YM theory. 
We calculate the gravitational wave spectrum that is emitted from the cosmic strings based on an extended velocity-dependent one-scale model and discuss its detectability based on ongoing and planned gravitational-wave experiments.
In particular, the SKA and LISA can observe gravitational signals if the confinement scale is higher than $\mathcal{O}(10^{12})\,\mathrm{GeV}$ and $\mathcal{O}(10^{10})\,\mathrm{GeV}$ for $\SU(N)$ with $N = \mathcal{O}(1)$, respectively.

\end{titlepage}

\setcounter{tocdepth}{2}

\newpage

\tableofcontents


\section{Introduction}

The strong dynamics of gauge theory is an important topic in theoretical physics; it helps understand the nature of mesons and baryons in QCD. 
The Universe is characterized as a unique application domain or system because it is cooled down from a very high temperature wherein the gauge theory is weakly coupled. 
The deconfinement/confinement phase transition should have occurred at a certain period of time in the cosmological history, which provides a rich phenomenology in cosmology. 
For instance, 
even an $\SU(N)$ Yang--Mills (YM) theory with or without a quark that is heavier than the confinement scale is extensively considered in the literature~\cite{Okun:1979tgr,Okun:1980mu,Strassler:2006im,Kang:2008ea}. 
Further, glueballs are formed at the deconfinement/confinement phase transition~\cite{Morningstar:1999rf,Lucini:2010nv,Curtin:2022tou}. 
The glueballs can be dark matter (DM) if their lifetime is sufficiently long ~\cite{Faraggi:2000pv,Feng:2011ik,Boddy:2014yra,Boddy:2014qxa,Soni:2016gzf,Kribs:2016cew,Forestell:2016qhc,Soni:2017nlm,Forestell:2017wov,Jo:2020ggs}. 
They can also be self-interacting DM that may address the problems of small-scale structure in cosmology~\cite{Spergel:1999mh,Weinberg:2013aya}. 
The overproduction problem and decay of glueballs are considered in Ref.~\cite{Juknevich:2009ji,Juknevich:2009gg,Halverson:2016nfq,Asadi:2022vkc}. 
The gravitational wave (GW) signals from the first-order phase transition has also been investigated~\cite{Reichert:2021cvs}.

In this paper, we demonstrate that cosmic strings or macroscopic color flux tubes form at the phase transition of pure YM theory from the deconfinement to the confinement phase. 
We explain the generalized symmetry or higher-form symmetry~\cite{Gaiotto:2014kfa} that ensures the stability of topological objects, including the cosmic strings in the pure YM theory. 
This concept is a generalization of ordinal (0-form) symmetry that ensures the stability of a (0-dimensional) particle. One-form symmetry ensures the stability of cosmic strings that are one-dimensional objects. 
For example, the pure $\SU(N)$ gauge theory has discrete one-form symmetry, which is referred to as $\bZ_N^{[1]}$. 
This implies the existence of one-dimensional objects charged under $\bZ_N^{[1]}$. It is further identified as the color flux tube in the confinement phase. Such color flux tubes should form at the deconfinement/confinement phase transition with a cosmological scale, and they can be regarded as cosmic strings.%
\footnote{
Stability of color flux tubes owing to a center symmetry are discussed by 't~Hooft ~\cite{tHooft:1977nqb,tHooft:1979rtg},
and its implications for cosmic strings are also considered in Ref.~\cite{Witten:1985fp} before generalized symmetry was discovered. 
}
$\bZ_N^{[1]}$ symmetry implies that $N$ strings can joined at a vertex, referred to as a baryon vertex. This is similar to the $\bZ_N$-string, which is considered in field-theory models~\cite{Vachaspati:1986cc}. 
However, our cosmic strings are qualitatively different from those produced in weakly coupled field-theory models, and we will provide a detailed explanation in the forthcoming sections. 
In some gauge groups, such as $\Sp(N)$, 
the one-form symmetry is $\bZ_2^{[1]}$ for any $N$ and 
the baryon vertex, even if it exists, does not play a significant role in cosmology.

We also discuss some other properties of cosmic strings in pure YM theory according to electric--magnetic duality, large $N$ limit, and holographic dual descriptions. 
The electric--magnetic duality implies that the confinement is dual to Higgsing~\cite{Seiberg:1994rs}, such that the cosmic strings should form at the phase transition similar to the case where a $\U(1)$ symmetry is spontaneously broken in weakly coupled field-theory models. 
The reconnection probability of cosmic strings can be estimated in the large $N$ limit such as $1/N^2$ for $\SU(N)$~\cite{tHooft:1973alw} (see Ref.~\cite{Coleman:1985rnk} for a review). 
According to the holographic dual descriptions, 
the cosmic strings are dual to a fundamental string or a wrapped D-brane in the gravity side depending on the structure of the gauge group (see, e.g., Refs.~\cite{Witten:1998zw,Polchinski:2000uf,Klebanov:2000hb,Maldacena:2000yy,Vafa:2000wi}). We refer to these as F-string and D-string, respectively. 
The reconnection probability scales as $1/N^2$ for F-strings and $e^{-\mathcal{O}(1)}$ for D-strings, which is consistent with the argument of the large $N$ limit. 
The suppressed reconnection probability is similar to that of the cosmic superstrings that form after the brane inflationary scenario~\cite{Dvali:2003zj,Copeland:2003bj}. In our case, however, the brane inflation is not required. The pure YM theory provides a cosmic string with a small reconnection probability using a simple setup.

Cosmic strings can be indirectly detected by observing GWs emitted from stochastic dynamics~\cite{Vilenkin:1981bx,Vachaspati:1984gt}. 
We calculate the GW spectrum that is emitted from string loops with a small reconnection (or intercommutation) probability. The cosmic string network can be described by 
extending the velocity-dependent one-scale (VOS) model~\cite{Kibble:1984hp,Martins:1995tg,Martins:1996jp,Martins:2000cs} to consider the small intercommutation probability. 
The extended VOS model was originally proposed in Ref.~\cite{Avgoustidis:2005nv}, where they also confirm its consistency with numerical simulations. 
The resulting GW signal can be within the reach of the future sensitivity of 
the Square Kilometer Array (SKA)~\cite{Janssen:2014dka} and LISA~\cite{LISA:2017pwj} 
if the confinement scale $\Lambda$ is higher than $\mathcal{O}(10^{12})$ and $\mathcal{O}(10^{10})$, respectively. 
We can determine the $N$ and the confinement scale $\Lambda$ by observing the GW spectrum.

The rest of this paper is organized as follows. 
In Sec.~\ref{sec:theory}, we review the idea of 1-form symmetries. This topic has attracted significant attention in theoretical physics and is also important for particle cosmology that treats topological defects and (pseudo-) NG modes (see, e.g., Refs.~\cite{Hidaka:2019mfm,Hidaka:2020ucc,Hidaka:2020iaz,Hidaka:2020izy,Hidaka:2021mml,Hidaka:2021kkf,Nitta:2022ahj,Cordova:2022rer,Yamamoto:2022vrh}). 
We particularly show that the pure $\SU(N)$ gauge theory has a 1-form $\bZ_N^{[1]}$ symmetry under which the color flux tube or cosmic string is charged. 
In Sec.~\ref{sec:heuristic}, we use other theoretical tools to demonstrate the properties of cosmic strings in the pure YM theory. 
Based on the electric--magnetic duality, we expect that the cosmic strings form at the deconfinement/confinement phase transition, similar to the formation of cosmic strings in ordinal field-theory models with a spontaneously broken $\U(1)$ symmetry. 
In the large $N$ limit, we explain that the intercommutation probability should scale as $1/N^2$ for $\SU(N)$. 
Such a small intercommutation probability is consistent with the holographic dual descriptions because cosmic strings can be identified as fundamental strings in the gravity side. 
We also discuss the differences for other gauge groups, such as $\Sp(N)$, $\SO(N)$, and $\Spin(N)$. 
In Sec.~\ref{sec:VOS}, 
we first summarize the properties of strings and qualitatively discuss the phenomenological consequence. We then solve the dynamics of long strings based on the extended VOS model. 
In Sec.~\ref{sec:GW}, we calculate the GW spectrum that is emitted from cosmic string loops. The resulting spectrum is shown with the future sensitivity curves, including SKA and LISA. 
Sec.~\ref{sec:conclusion} comprises the discussion and conclusions of the paper.

\section{Cosmic strings and 1-form symmetries}
\label{sec:theory}

The stability of a particle can be guaranteed by considering the specific symmetry under which the particle is charged.
For example, the stability of dark matter candidates can be ensured by introducing (global or gauge) symmetries such as $\bZ_N$ or $\U(1)$.
In the same way, the stability of a (cosmic) string can be ensured by 1-form symmetries~\cite{Gaiotto:2014kfa}. 
We review this concept in this section.
In the next section, we discuss some qualitative properties of color flux tubes in gauge theories. 
If the reader is interested only in the phenomenology of cosmic strings in the pure YM theory, she/he may skip the next two sections and directly refer to Section~\ref{sec:VOS}.

\subsection{1-form symmetries and operators}
In this section, we review the concept of 1-form symmetries in the operator formalism with a fixed time direction.
We particularly consider operators that act on the Hilbert space. 
The study in \cite{Gaiotto:2014kfa} provides the general spacetime descriptions with time-ordered operators or in the Euclidean signature spacetime
wherein the time direction need not be assumed. In the following discussions, we do not consider time ordering unless otherwise stated.

First, we recall how an ordinary $\U(1)$ symmetry can be described. 
We have the corresponding conserved Noether current $J^\mu$.
Through the integration of this term, we obtain a conserved charge operator
\beq
Q(\Sigma) = \int_{\Sigma} d\Sigma_\mu J^\mu
\eeq
where $\Sigma$ is a 3-dimensional surface inside the 4-dimensional spacetime
such as the constant time slice $\Sigma=\{t={\rm const.}\}$,
and $d\Sigma_\mu$ is set such that its length is the volume element of $\Sigma$ and its direction is orthogonal to the surface $\Sigma$.
The conservation equation $\nabla_\mu J^\mu=0$ ensures that the charge operator $Q(\Sigma)$ depends only on the topology of $\Sigma$,
and thus, it is invariant under a continuous change in $\Sigma$. This topological invariance is the abstract description of the usual charge conservation.
For example, the topological invariance states that when we take two time slices $\Sigma_1= \{ t = t_1 \}$ and $\Sigma_2= \{ t = t_2 \}$,
we obtain $Q(\Sigma_1) = Q(\Sigma_2)$ with the assumption that there is no flux of charge at spatial infinity.

Using the charge operator $Q(\Sigma)$, we can construct a unitary operator $U(\Sigma, \alpha)$ as
\beq
U(\Sigma, \alpha) = \exp[ i \alpha Q(\Sigma)],
\eeq
where $\alpha$ is an arbitrary parameter.
This operator induces the symmetry transformation as follows.
If ${\cal O}(x)$ is an operator whose charge is $q$ under the $\U(1)$, and if $\Sigma$ is topologicaly equivalent to a time slice $\Sigma=\{ t={\rm const.} \}$,
then we get
\beq
U(\Sigma, \alpha) {\cal O}(x) U(\Sigma, \alpha)^\dagger  = e^{iq\alpha} {\cal O}(x). \label{eq:symmtransf1}
\eeq
When the symmetry is not $\U(1)$ but a discrete group such as $\bZ_N$, the current $J^\mu$ and the charge $Q(\Sigma)$ are not conserved. 
However, the unitary operator $U(\Sigma, \alpha)$ still persists, and it possesses the desired topological invariance related to $\Sigma$ provided that the parameter $\alpha$ is considered as $\alpha = 2\pi k /N $, where $k \in \bZ_N$.
Thus, we can characterize the symmetry using such operators $U(\Sigma, \alpha)$. (See \cite{Gaiotto:2014kfa} for a complete description.) 

Now, we can extend the aforementioned discussions to 1-form symmetries. First, let us consider a 1-form $\U(1)$ symmetry, which may be denoted as $\U(1)^{[1]}$.
This symmetry is characterized by a current operator $J^{\mu\nu}$, which is antisymmetric $J^{\mu\nu} = -J^{\nu\mu} $; it is conserved in the sense that $\nabla_\mu J^{\mu \nu} =0$. 
We will discuss an explicit example later. The charge $Q(\Sigma)$ is defined as
\beq
Q(\Sigma) = \int_{\Sigma}\frac12 d\Sigma_{\mu\nu} J^{\mu\nu} \label{eq:charge2},
\eeq
where $\Sigma$ is now a 2-dimensional surface, and $d\Sigma_{\mu\nu}$ is such that it is antisymmetric $d\Sigma_{\mu\nu} = -d\Sigma_{\nu\mu}$,
its length $\sqrt{\frac12 d\Sigma_{\mu\nu} d\Sigma^{\mu\nu}}$
is the volume element of $\Sigma$, and its directions are orthogonal to $\Sigma$.
(We conveniently define a 2-form $J = \frac{1}{2^2} \epsilon_{\mu\nu\rho\sigma} J^{\mu\nu} dx^\rho \wedge dx^\sigma$ using differential forms, and then
$Q(\Sigma) = \int_{\Sigma} J$.) Based on the conservation equation $\nabla_\mu J^{\mu \nu} =0$ (or $d J =0$ in the differential form notation),
this operator depends only on the topology of $\Sigma$. 
More precisely, if we have two surfaces $\Sigma_1$ and $\Sigma_2$ such that there exists a 3-dimensional surface $\Gamma$ whose
boundary is $\partial \Gamma = \Sigma_1 - \Sigma_2$, we obtain $Q(\Sigma_1) = Q(\Sigma_2) $
(since $Q(\Sigma_1) - Q(\Sigma_2) = \int_\Gamma dJ=0$ using the Stokes theorem and the conservation equation $dJ=0$).
In particular, $Q(\Sigma)$ is invariant under the continuous change of $\Sigma$. 

We can also define $U(\Sigma, \alpha) = \exp[ i \alpha Q(\Sigma)]$ similar to the case of ordinary symmetries.
Now, in the case of ordinary symmetries, point operators such as ${\cal O}(x)$ are charged under $U(\Sigma,\alpha)$ as expressed in \eqref{eq:symmtransf1}.
In the case of 1-form symmetries, loop operators are charged under it. Here, loop operators ${\cal O}(C)$ are defined on
1-dimensional loops $C$. 
An example which is relevant for our later purposes is the Wilson loop operator in gauge theories, 
\beq
W_R(C) = \tr_R P\exp \left( i \int_C A_\mu dx^\mu \right)  \label{eq:wilson1}
\eeq
where $A_\mu$ is the gauge field, $P$ is the path ordering, and the trace $\tr_R$ is taken in a representation $R$.
These kinds of operators are referred to as loop operators because they are supported on a loop $C$ rather than a point $x$ as $\cO(x)$.

Now, what corresponds to \eqref{eq:symmtransf1} is the following. For simplicity, we suppose that
both $C$ and $\Sigma$ are contained in a fixed time slice, say, $t = 0$. 
Let $\vev{C,\Sigma} $ be the number of intersections between $C$ and $\Sigma$ (including the sign) within that time slice.
For example, let us take $C$ to be a straight line in the $x^1$ direction as $C=\{ (x^1, x^2,x^3)\, |\, x^2=x^3=0 \}$ with some orientation,
and take $\Sigma$ to be a surface at $x^1=0$ as $\Sigma =\{ (x^1, x^2, x^3)\, |\, x^1=0 \}$ with some orientation. 
In this case, they intersect at the single point $(x^1,x^2,x^3)=(0,0,0)$, and thus, we obtain $\vev{C,\Sigma} = \pm 1$, where the sign depends on the orientation of $C$ and $\Sigma$. 
Using the intersection number $\vev{C,\Sigma} $, we obtain
\beq
U(\Sigma, \alpha) {\cal O}(C) U(\Sigma, \alpha)^\dagger  = e^{iq\alpha \vev{C,\Sigma} } {\cal O}(C). \label{eq:symmtransf2}
\eeq
This represents the generalization of \eqref{eq:symmtransf1} to 1-form symmetries.\footnote{ 
For a clear understanding of the same in the spacetime formulation, we may place the operators 
$U(\Sigma,\alpha) ,\,{\cal O}(C),\, U(\Sigma,\alpha)^\dagger $ at $t = t_2, t_1, t_0$ with $t_2> t_1 >t_0$, respectively. 
Because of the topological invariance regarding $\Sigma$, we can arrange operators in such a time-ordered manner. 
In this case, the intersection number $ \vev{C,\Sigma} $ in 3-dimensional space is equal to the linking number
between $\{t_2\} \times \Sigma - \{t_0\} \times \Sigma$ and $C$
in 4-dimensional spacetime. This is similar to the formulation discussed in \cite{Gaiotto:2014kfa}.}

Similar to the case of ordinary symmetries, we can consider not only $\U(1)^{[1]}$ 
but also discrete 1-form symmetries $\bZ_N^{[1]}$. This represents the 1-form symmetry corresponding to the group $\bZ_N$.
For this group, we restrict $\alpha$ to be of the form $\alpha = 2\pi k /N$.

\subsection{Dynamical charged objects}
\begin{figure}
\centering
\includegraphics[width=0.45\textwidth]{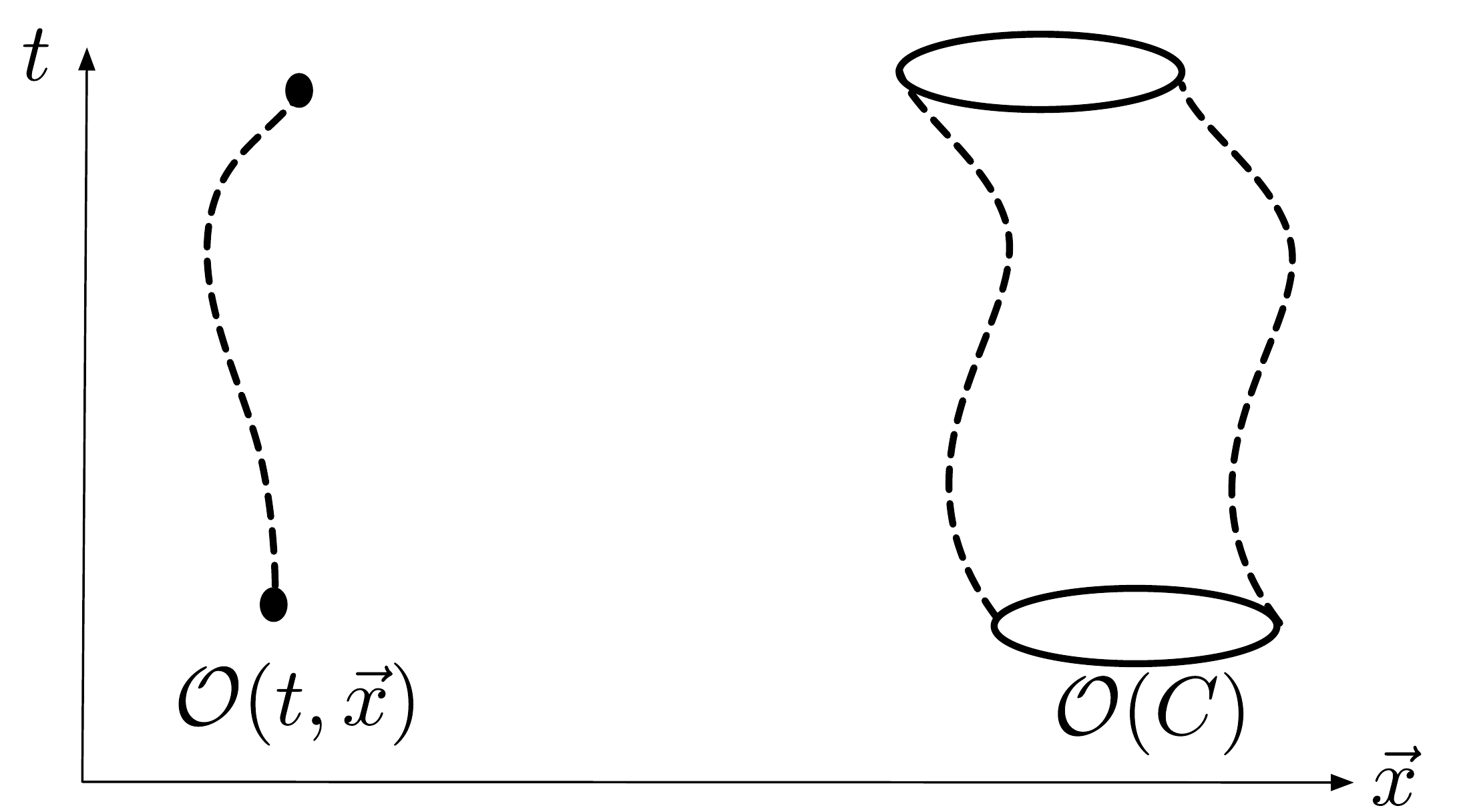}
\caption{ Left: creation of a particle from the vacuum $\ket{\Omega}$ by a point operator $\cO(t,\vec x)$. Right: creation of a string from the vacuum $\ket{\Omega}$ 
by a loop operator $\cO(C)$. \label{fig:creation}}
\end{figure}

An ordinary symmetry is either unbroken or spontaneously broken.
When it is unbroken, the vacuum expectation values $\bra{\Omega} {\cal O}(x) \ket{\Omega}$ of
all charged operators are zero. The action of the operator ${\cal O}(x)$ to the vacuum state $\ket{\Omega}$
creates a particle (or particles) whose (total) charge is the same as that of ${\cal O}(x)$.
Intuitively, when the particle has a mass $m$ and ${\cal O}(x)$ creates a single particle,
${\cal O}(t,\vec x) ) \ket{\Omega}$ describes a single particle state that is localized 
near the point $\vec x$ at the time $t$, with the uncertainty of the position that is given by the Compton wavelength $m^{-1}$.
By considering a sufficiently heavy (i.e. nonrelativistic) particle, we can regard it to be localized at $\vec x$ when the time is $t$. 
See the left side of Fig.~\ref{fig:creation}.

Similarly, a 1-form symmetry is either unbroken or spontaneously broken. 
The question of spontaneous breaking is subtle, and for concreteness, we consider the case where the space is not $\bR^3$ but $S^1 \times \bR^2$,
where $S^1$ is the $x^1$ direction with the periodic boundary condition $x^1 \sim x^1 + L$.
We expect to recover the results on $\bR^3$ in the limit $L \to \infty$. 
Let us consider a loop
\beq
C = \{ (x^0, x^1, x^2, x^3)\, |\, x^0=0, x^2=0, x^3=0 \} \label{eq:Cloop1}
\eeq
It is wrapped on the $S^1$. 
We take $\Sigma$ to be transverse to $C$ within $t=0$ as 
\beq
\Sigma =\{ (x^0, x^1, x^2, x^3)\, |\, x^0=0, x^1=0 \}.
\eeq
However, it should be noted
that $U(\Sigma,\alpha)$ only depends on the topology of $\Sigma$.
When the 1-form symmetry is unbroken (or more exactly, when the symmetry described by $U(\Sigma,\alpha)$ for our choice of $\Sigma$ is unbroken), 
all vacuum expectation values of loop operators $\bra{\Omega} {\cal O}(C) \ket{\Omega}$
are zero. The action of the operator ${\cal O}(C)$ on the vacuum state $\ket{\Omega}$
now creates a string-like object, which is wrapped on $S^1$. This string (or strings) has the same (total) charge
under $U(\Sigma,\alpha)$ as that of the operator ${\cal O}(C)$, as seen from \eqref{eq:symmtransf2} and the condition of unbroken symmetry 
$U(\Sigma, \alpha)\ket{\Omega}=\ket{\Omega}$.
Assuming that ${\cal O}(C)$ creates a single string, and that the tension of the string is sufficiently large,
the state ${\cal O}(C)\ket{\Omega}$ descries a string that is localized near $C$. See the right side of Fig.~\ref{fig:creation}.

When we consider more general $C$, the string created by ${\cal O}(C)$ can disappear if and only if $C$
is topologically trivial. For example, we may consider
\beq
C' = \{ (x^0, \vec x )\, |\, x^0=0, |\vec x|^2=R^2 \}
\eeq
where $R$ is a constant parameter. In this case, a string is created by the action of ${\cal O}(C')$ to the vacuum state,
and it is initially localized near $C'$. After some time, 
the string can shrink to zero size and eventually disappear. From the point of view of the 1-form symmetry,
this process is possible because such a topologically trivial $C$ has the trivial intersection number $ \vev{C,\Sigma}=0$ irrespective of $\Sigma$. 
However, the 1-form symmetry guarantees that there is no local mechanism that makes the string unstable.
When the symmetry is explicitly broken, the string can decay even if it is wrapped on a topologically nontrivial cycle, such as $S^1$ mentioned earlier. We will discuss this decay process later.

\subsection{Examples}
Here, we discuss two examples of 1-form symmetries. 
For concreteness, we impose the periodic boundary condition $x^1 \sim x^1 + L$ with a sufficiently large $L$ as mentioned earlier,
which ensures that the space is $S^1 \times \bR^2$ rather than $\bR^3$. 

For the first example, we consider a $\U(1)$ gauge theory, which is coupled to 
a scalar field $\phi(x)$ with charge 1 under the gauge $\U(1)$ symmetry. 
This theory has a 1-form $\U(1)^{[1]}$ symmetry with the conserved current 
\beq
J^{\mu\nu} = \frac{1}{4\pi}F_{\rho\sigma} \epsilon^{\rho\sigma\mu\nu}.
\eeq
This is conserved $\nabla_\mu J^{\mu\nu} =0$ because of the Bianchi identity $\epsilon^{\rho\sigma\mu\nu} \nabla_{\mu}F_{\rho\sigma} =0$.

The operator which is charged under this 1-form symmetry is known as the 't~Hooft operator $H(C)$~\cite{tHooft:1977nqb}, and because it is defined in an abstract way, we do not try to explain it. However, the string that is created by the 't~Hooft operator is well-known.
Suppose that the $\U(1)$ gauge symmetry is spontaneously broken by a vacuum expectation value of $\phi$.
For example, we can assume the potential to be $V(\phi) = - \mu^2 |\phi|^2 + \lambda |\phi|^4$ although its details are not important for
our discussions of the 1-form symmetry $\U(1)^{[1]}$. Then, we obtain the usual vortex string.
For concreteness, we assume that the string is stretched in the direction $x^1$, and it is localized at $x^2=x^3=0$.
Then, using the polar coordinates $x^2 + ix^3 = r e^{i\theta}$, we can consider a vortex configuration
$\phi(x) \to v e^{i k \theta} $ at $r \to \infty$,
where $v$ is the vacuum expectation value of $\phi$, and $k \in \bZ$ is some integer. 
To minimize the energy coming from the kinetic term $D_\theta \phi = (\partial_\theta - i A_\theta) \phi$,
we set the $\theta$-component of the gauge field to be
$A_\theta \to k $ at $r \to \infty$.
Now, let $\tilde C$ be a large loop $\tilde C = \{ (x^1,x^2,x^3) \, | \, x^1 =0,~ |x^2|^2 + |x^3|^2 = a^2  \}$ for some large constant $a$
which we take to be infinity $a \to \infty$. 
Moreover, let $\Sigma$ be the disk filling $\tilde C$, i.e. $\Sigma = \{ (x^1,x^2,x^3) \, | \, x^1 =0,~ |x^2|^2 + |x^3|^2 \leq a^2 \}$
with $a \to \infty$.
From the definition of the charge $Q(\Sigma)$ given in \eqref{eq:charge2}, we obtain
\beq
Q(\Sigma)  
= \int_{\Sigma} dx^2dx^3 J^{01} 
= \frac{1}{2\pi} \int_{\Sigma}dx^2 dx^3 F_{23} 
= \frac{1}{2\pi} \int_{\tilde C} A_\mu dx^\mu =  \frac{1}{2\pi} \int_{\tilde C} A_\theta d\theta = k.
\eeq
Thus, the aforementioned vortex configuration has a charge $k$ under $Q(\Sigma)$. 

We conclude that in the Higgs phase, the dynamical object charged under the 1-form $\U(1)^{[1]}$ symmetry is the usual vortex string.
Conversely, when the $\U(1)$ gauge symmetry is not spontaneously broken, 
it is known that 't~Hooft operators $H(C)$ have non-zero expectation values $\bra{\Omega} H(C) \ket{\Omega} \neq 0$.
Then, the 1-form symmetry $\U(1)^{[1]}$ is spontaneously broken, and no clear dynamical objects are charged under it. 

The $\U(1)^{[1]}$ symmetry can be explicitly broken if a magnetic monopole is introduced in the theory.
In this case, the Bianchi identity $\epsilon^{\rho\sigma\mu\nu} \nabla_{\mu}F_{\rho\sigma} =0$ is no longer true,
and thus, $J^{\mu\nu}$ is not conserved. In fact, if the monopole has the unit magnetic charge,
the vortex string can end on a monopole or antimonopole. Thus, 
the string can decay through the pair creation of a monopole and an antimonopole, and thus, it is not conserved.
However, if the monopole charge is $N>1$, then the $\bZ_N^{[1]}$ subgroup of $\U(1)^{[1]}$ remains unbroken, and
the string is still stable.

The next example, which is the main subject of interest in this paper, is the YM theory with gauge group $G$. 
We can also add matter fields provided they transform trivially under the center of the gauge group, such as the adjoint representation. 
For concreteness, we mainly discuss the case $G=\SU(N)$. However, the generalizations to other groups are also straightforward. 

The theory with the group $\SU(N)$ has a 1-form $\bZ_N^{[1]}$ symmetry, called the center symmetry. 
The operator $U(\Sigma, \alpha)$ should be defined in an abstract manner.
We do not explain the details (see \cite{Gaiotto:2014kfa} for modern descriptions of center symmetries),
but we briefly discuss it in the operator formalism within a fixed time, or in the ``Schr\"odinger picture'' at (say) $t=0$.
We take a 2-dimensional surface $\Sigma$ as in the previous section.
Whenever we cross this surface within the 3-dimensional space (at $t=0$),
we perform a transformation by a center $e^{\pm 2\pi i k /N} I_N \in \SU(N)$, where the sign in the exponent depends on the 
orientation of $\Sigma$. This gauge transformation at $\Sigma$
is the definition of $U(\Sigma, \alpha)$ with $\alpha = 2\pi k/N$.

For example, suppose that the surface is given by $\Sigma =\{ (x^1, x^2, x^3)\, |\, x^1=0 \}$.
For a field in the fundamental representation $\phi$, we impose $\phi(x^1=+\epsilon, x^2, x^3) =e^{ 2\pi i k /N}  \phi(x^1=-\epsilon, x^2, x^3)$
for an infinitesimally small $\epsilon$. This jump by the gauge transformation $e^{ 2\pi i k /N} I_N$
is the definition of $U(\Sigma,\alpha)$ with $\alpha = 2\pi k/N$.\footnote{
In the spacetime formulation, we impose this discontinuity in the time region $t < t_0$, where $t_0$ 
is the time at which the operator is inserted. In general, we obtain the phase $e^{ 2\pi i k /N} $ whenever we go along a small loop 
around the codimension-2 surface $\Sigma$.}

We assumed that the theory has no dynamical field that transforms nontrivially under the center
$e^{ 2\pi i k /N} I_N \in \SU(N)$.
Then, the presence of the jump by $e^{ 2\pi i k /N} I_N$
does not have any effect on the local dynamics of the fields that transform trivially under $e^{ 2\pi i k /N} I_N \in \SU(N)$.
However, its existence has a global effect on the gauge field configuration~\cite{tHooft:1977nqb,tHooft:1979rtg}
(see also \cite{Witten:1999ds,Witten:2000nv} for additional explanations), 
and thus, this operator is not completely trivial. For example, with the aforementioned choice of $\Sigma$, the holonomies (Wilson lines) of the gauge field around the $S^1$ direction of $x^1$
receive an additional contribution given by the center $e^{ 2\pi i k /N}$ after the action of $U(\Sigma,\alpha)$.
The fact that the jump has no effect on local dynamics indicates that it has the topological invariance
under continuous deformation of $\Sigma$. This topological invariance is the desired property for $U(\Sigma, \alpha)$ as mentioned previously.

Now, we consider a 1-dimensional loop $C$ 
and we take the Wilson loop operator $W_R(C)$ as in \eqref{eq:wilson1}. 
At the intersection of $C$ and $\Sigma$, the Wilson loop yields an additional phase factor because of the cut by $e^{ 2\pi i k /N} I_N \in \SU(N)$, and hence,
we obtain
\beq
U(\Sigma,\alpha) W_R(C) U(\Sigma,\alpha)^\dagger  = e^{i q_R \alpha \vev{C,\Sigma} } W _R(C)
\eeq
where the value of the integer $q_R$ is such that the center $e^{ 2\pi i k /N} I_N \in \SU(N)$ acts in the representation $R$ as $e^{ 2\pi i k q_R /N} $.\footnote{
In the spacetime formulation, this equation can be easily interpreted if the operators 
$U(\Sigma,\alpha) ,\,W_R(C),\, U(\Sigma,\alpha)^\dagger $ are placed at $t = t_2, t_1, t_0$ with $t_2> t_1 >t_0$, respectively. 
Because of the topological invariance about $\Sigma$, we can arrange operators in this time-ordered manner. Then, the branch cut that is mentioned 
in the previous footnote produces the phase $ e^{i q_R \alpha \vev{C,\Sigma} } $.}
For the fundamental representation $F$, we have $q_F=1$.

We have discussed the presence of the 1-form symmetry $\bZ_N^{[1]}$, which is referred to as the center symmetry,
and found loop operators $W_R$ that are charged under it.
We should check if they are spontaneously broken. When the theory is weakly coupled, 
it is usually spontaneously broken, mainly because of the following reason.\footnote{The following argument is not always true when spacetime is compactified. 
It is possible to consider a setup wherein the theory is weakly coupled but the center symmetry is unbroken.
For a recent review, see \cite{Poppitz:2021cxe}. 
} 
At the leading order of perturbation theory, the gauge field
is approximately zero if $A_\mu=0$ is (one of) the leading saddle point of the path integral. 
Then, the Wilson loop operator is $W_R(C) \sim \tr_R 1 + \cdots$ where the ellipses are higher order terms in perturbation theory.
Therefore, its vacuum expectation value $\bra{\Omega} W_R(C) \ket{\Omega}$ is nonzero and the center symmetry is spontaneously broken. 

Conversely, the vanishing expectation values of $W_R(C)$ (for topologically nontrivial $C$ such as the loop \eqref{eq:Cloop1})
serves as the condition for confinement. The definition of confinement in the language of 1-form symmetry is that
the theory is confined if the 1-form center symmetry is not spontaneously broken. 
In this case, we obtain a string-like dynamical object when we act $W_R(C)$ on the vacuum state $\ket{\Omega}$ as discussed before. 
What is it?

The string that is charged under the 1-form center symmetry is the color flux tube of the confining $\SU(N)$ gauge theory. 
The color flux tube is often described as a tube between a fundamental quark and an antiquark. 
In the current theory, we assumed that the center of $\SU(N)$ acts trivially on dynamical fields, and thus, there is no dynamical fundamental quark.
However, a Wilson loop operator is interpreted as the worldline of an ``external'' quark. We recall that our string is produced by Wilson loops, as shown in the right side of Fig.~\ref{fig:creation}.
This means that if we act $W_F(C)$ (where we have taken the fundamental representation $F$) on $\ket{\Omega}$,
we create a string localized at $C$ at a given time. In the spacetime picture, 
this means that the string worldsheet ends on the loop $C$, and this loop was interpreted 
as the worldline of an external quark. Therefore, our string ends on a quark (or antiquark) and can thus be interpreted as the color flux tube.
This color flux tube is stable because of the 1-form symmetry $\bZ_N^{[1]}$.
The stable color flux tube was discussed by 't~Hooft~\cite{tHooft:1977nqb,tHooft:1979rtg} 
even before the formulation of the concept of 1-form symmetries, but the modern formulation in terms of the 1-form symmetry 
may give a clearer perspective of its stability.

If we introduce matter fields in the fundamental representation,
the 1-form center symmetry is explicitly broken. In the presence of such matter fields, the aforementioned operator $U(\Sigma,\alpha)$ has a considerable impact on the local dynamics of fields, and thus, it is no longer topologically invariant under deformations of $\Sigma$. 
Color flux tubes can end on dynamical quarks, and thus, they can decay via a pair creation of
a quark and an antiquark. This is the usual case in QCD. However,
if the fundamental quark mass is significantly larger than the dynamical scale of the $\SU(N)$ gauge field,
we obtain an approximate 1-form symmetry in low energy, and the color flux tubes are metastable.

We recall a rough estimate of its decay rate~\cite{Vilenkin:1982hm}.
The 1-form symmetries are explicitly broken when there exists a particle on which dynamical strings can end. 
In the case of a $\U(1)$ gauge theory with a Higgs field $\phi$, the relevant particle is a magnetic monopole.
In the case of a $\SU(N)$ gauge theory, the relevant particle is a fundamental quark. 
We denote the string tension as $\mu$ and the particle mass as $m$, and assume that $m \gg \sqrt{\mu}$. 
We then consider a Euclidean spacetime configuration that represents the decay of a string.

Without decay, we assume that the string is located at $x^2=x^3=0$.
It is extended in the directions $x^0, x^1$, where $x^0$ means the Wick rotated Euclidean time direction.

Now, the decay is realized through a bubble that is created by a particle. 
We consider the case that this bubble is created at zero temperature. 
Similar to the thin wall approximation of the vacuum decay rate,
we consider the following configuration.
\begin{enumerate}
\item In the region $|x^0|^2 + |x^1|^2 < r^2$, where $r$ is a constant parameter, there is no string. This corresponds to a true vacuum.
\item On the circle $|x^0|^2 + |x^1|^2 = r^2$, we have a virtual particle loop. This corresponds to a thin wall.
\item In the region $|x^0|^2 + |x^1|^2 > r^2$, we have the string. This corresponds to a metastable vacuum.
\end{enumerate}
By translation symmetry, we also obtain configurations whose center is at an arbitrary point on the string. 
The Euclidean action of this configuration relative to the action without the bubble is given as a function of $r$ by
$
S(r) = -\pi \mu r^2 + 2\pi m r.
$
It has an unstable saddle point at $r = r_*:= m/\mu$, and the value of the action is
$
S(r_*) = - \pi \frac{m^2}{\mu}.
$
Therefore, the decay rate of the string per volume at zero temperature is estimated as
\beq
\label{eq:decayrate}
\Gamma \sim \tilde{\mu} \exp\left(  - \pi \frac{m^2}{\mu} \right) ,
\eeq
where $\tilde{\mu}$ is a typical mass squared scale related to the string and the particle. 

The thin wall approximation is valid provided that $r_*$ is significantly larger than the typical size 
of the particle $1/m$ and the typical size of the string $1/\sqrt{\mu}$. This condition is satisfied
in the parameter region $m \gg \sqrt{\mu}$. The decay rate is well-suppressed in this parameter region.

Next, we analyze the case of a finite temperature. We denote the inverse temperature as $\beta$.
The Euclidean time direction $x^0$ has the periodicity $x^0 \sim x^0 + \beta$, and hence, it is a circle $S^1_\beta$. 
When $r_* \ll \beta$, the aforementioned estimate is still valid. Conversely, when $r_* \gg \beta$,
we expect that a particle and an antiparticle move around the circle $S^1_\beta$ in almost straight lines.
In this case, we cannot find a desired unstable saddle point using the aforementioned semiclassical method.
The action increases monotonically as the distance between the particle and the antiparticle becomes smaller.
However, we expect that a contribution $2m\beta = m\beta + m\beta$ can be obtained in the action, and this would come from the particle loop and the antiparticle loop around $S^1_\beta$.
If the order of the action is not significantly different, we may have
\beq
\Gamma \sim \tilde{\mu}(\beta) \exp\left( -cm\beta \right),
\eeq
where $c$ is an order 1 constant, and $\tilde{\mu}(\beta)$ is a mass-squared scale, which is now also a function of $\beta$. 
Assuming that the $\beta$ dependence of $\tilde{\mu}(\beta)$ is not significant, the decay rate is the largest at the highest possible temperature. The highest temperature is the critical temperature for the phase transition between the Higgs and Coulomb phases
in the case of the $\U(1)$ gauge theory, and the confinement phase and the deconfinement phase in the case of the $\SU(N)$ theory.
Let $\beta_*$ be the critical value. The decay rate at this temperature is given by $\tilde{\mu}(\beta_*) \exp\left( -cm\beta_* \right)$.
It is suppressed if $m \gg (\beta_*)^{-1}$. 

In the case of the $\SU(N)$ gauge theory, both $\sqrt{\mu}$ and $(\beta_*)^{-1}$ are of the order of the dynamical scale $\Lambda$ of the gauge theory.
Thus, the decay rate can be sufficiently suppressed if we have $m \gg \Lambda$.

\section{Other pictures and some properties of strings}
\label{sec:heuristic}

Even though direct computations are difficult in strongly coupled gauge theories, some properties are understood well 
for YM theories.
In this section, we would like to briefly recall them.

\subsection{Electric--magnetic duality}

The YM theory in the confining phase is a strongly-coupled theory, and 
it is not immediately obvious what intuition we should have for the creation of cosmic strings in the phase transition from the deconfinement phase at high temperatures
to the confinement phase at low temperatures. Thus, it is helpful to consider other theories
that are qualitatively the same as YM theory but are more calculable.

As mentioned previously, we can add matter fields to the theory provided they transform trivially under the center of $\SU(N)$.
(If matter fields that transform nontrivially under the center exist, we assume that their mass is much larger than the dynamical scale such that they are decoupled at the critical temperature of phase transition.)   
Assuming that the theory still confines, the qualitative behavior is similar to the pure YM as far as the color flux tube is concerned. 
If we add some appropriate matter fields, there is electric--magnetic dual description wherein the theory is weakly coupled and Higgsed.

One such example is given by mass-deformed $\cN=2$ supersymmetric YM theory. 
For concreteness, we consider the case that the gauge group is $\SU(2)$.
The theory contains some scalar fields and fermions in the adjoint representation of the gauge
group to ensure that their dimensionless interactions preserve $\cN=2$ supersymmetry. We can add supersymmetry-breaking mass terms.
This theory has a dual description in low energy by a $\U(1)$ gauge theory with a Higgs field $\phi$~\cite{Seiberg:1994rs}.
This $\U(1)$ is the electric--magnetic dual of the Cartan subgroup $\U(1) \subset \SU(2)$ of the original gauge group.
The Higgs field in the dual theory is a magnetic monopole of the original $\SU(2)$ theory, and the confinement is dual to the Higgsing.
The color flux tube of the $\SU(2)$ gauge theory is dual to the vortex string of the dual $\U(1)$ theory.
The $\U(1)$ theory with only the Higgs field has a 1-form $\U(1)^{[1]}$ symmetry, as discussed before.
However, it also contains a heavy magnetic monopole with a magnetic charge of 2. 
This monopole is a $W$-boson of the original $\SU(2)$ theory and it explicitly breaks the 1-form
symmetry from $\U(1)^{[1]}$ to $\bZ_2^{[1]}$. This duality suggests that we can qualitatively intuit similarly 
for the color flux tube string as the usual vortex string in the $\U(1)$ Higgs phase. 
Some differences at the quantitative level are discussed later in the paper. 

Another (but related) example of a duality is as follows. We consider the mass-deformed $\cN=4$ supersymmetric $\SU(N) $ YM theory, which contains some scalars and fermions in the adjoint representation.
This theory is almost self-dual. The dual theory has exactly the same matter content as that in the original one; the Lagrangians are also qualitatively the same
with some different parameters.
However, the gauge group topology is $\SU(N)/\bZ_N$ rather than $\SU(N)$.
The confining vacuum of the original theory is dual to the Higgs vacuum of the dual theory~\cite{Donagi:1995cf,Polchinski:2000uf}. These vacua are described as follows.
There are three complex scalar fields $\Phi_i~(i=1,2,3)$ in the adjoint representation, and we regard them as $N \times N$ 
traceless matrices. 
The potential energy is given by
\beq
\label{eq:EMdual}
V(\Phi) = \left| m \Phi_i + \frac{i}{2} g  \epsilon_{ijk} [\Phi_j, \Phi_k] \right|^2
\eeq
where $\epsilon_{ijk}~(i,j,k=1,2,3)$ is the totally antisymmetric tensor, $m$ is the mass parameter that breaks some of supersymmetry, 
and $g$ is a dimensionless coupling (which is actually the same as the gauge coupling in $\cN=4$ Super-YM). 
One of the minima of the potential is $\Phi_i=0$. At this point, the matter fields are massive, the gauge group is unbroken, and the theory is confined. 
This is the confining vacuum. Another vacuum is described as follows.
We take $\Phi_i$ to be $\Phi_i = (m/g) t_i$,
where $t_i~ (i=1,2,3)$ are the generators of the $\SU(2)$ Lie algebra in the irreducible $N$-dimensional (i.e. ``spin'' $\frac12 (N-1)$) representation.
In particular, they satisfy the matrix commutation relation $[t_i, t_j]= i \epsilon_{ijk} t_k$.
It can be inferred that the aforementioned potential is minimized. At this vacuum, the gauge group $\SU(N)/\bZ_N$ is completely Higgsed.\footnote{
For $N>2$, there are various other vacua that are a mixture of confinement and Higgsing, by taking $t_i$ to be in some reducible representation of $\SU(2)$ algebra. 
We do not discuss them because we are interested in the confinement phase of the original theory ($\Phi_i=0$) and the Higgs phase of the dual theory ($\Phi_i = (m/g) t_i$). } 
In the Higgs phase, a string is associated to the homotopy group $\pi_1(\SU(N)/\bZ_N) = \bZ_N$ in a similar way in which the stability
of the vortex string in the $\U(1)$ Abelian-Higgs model is related to $\pi_1(\U(1))=\bZ$.
This string is dual to the color flux tube in the confining vacuum. 
It should be noted that the stability of the string in the Higgs phase is determined using $\pi_1(\SU(N)/\bZ_N) = \bZ_N$.
This can also be understood as a magnetic 1-form symmetry $\bZ_N^{[1]}$ for $\SU(N)/\bZ_N$,
which is similar to the $\U(1)$ case with $J^{\mu\nu} = \frac{1}{4\pi}F_{\rho\sigma} \epsilon^{\rho\sigma\mu\nu}$; however, we do not review it here.

From these dualities, we expect that cosmic strings are produced in 
the thermal phase transition from the deconfinement phase at high temperatures to the confinement phase
at low temperatures even though direct computations in strongly coupled theories are difficult. 
We believe that this qualitative conclusion is valid for theories that do not have explicit dual descriptions, such as the pure non-supersymmetric YM theory.
The reason is as follows. For theories which have the explicit electric-magnetic dual, 
the dual theory at low temperatures is in the Higgs phase for which the usual Kibble-Zurek argument may apply. 
Now in the original theory, we change the mass terms of the additional matter fields
to be very large. The theory reduces to the pure YM. If we change the mass parameters of the original theory,
some parameters of the dual theory are also changed. However, the Kibble-Zurek argument does not
rely on the details of parameters as far as infinitely long strings are concerned. 
Thus we expect that infinitely long cosmic strings are produced in the phase transition.
It would be interesting if this expectation from the electric-magnetic duality could be confirmed more directly by simulations of pure YM.

\subsection{Large \textit{N} limit}

When two strings collide transversely, there is a possibility of reconnection with some probability. 
This is determined using the string coupling $g_s$, which is estimated as follows.

\begin{figure}
\centering
\includegraphics[width=0.9\textwidth]{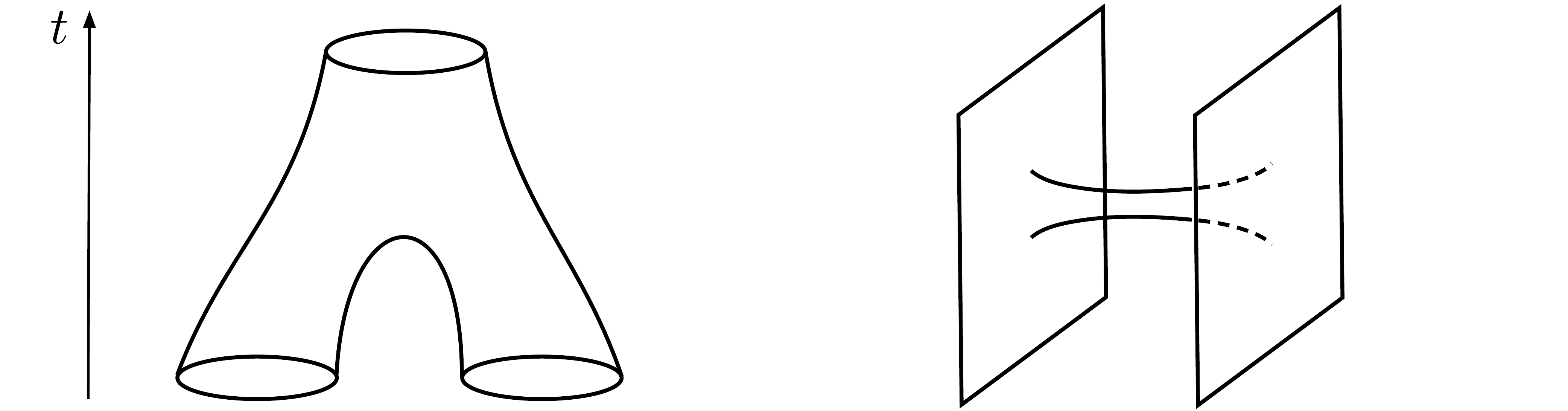}
\caption{ Left: Two strings colliding at a single point to form a single string. The worldsheet of this process in spacetime
is topologically a sphere with three holes. Right: Interactions of two string worldsheets via the tube-like region (handle), which 
smoothly connects two worldsheets. \label{fig:pairofpants}}
\end{figure}

We consider a situation where each of the two strings in the initial state is in the form of a circle $S^1$.
Suppose that these strings collide at a single point, and they reconnect thereafter. 
In the final state, we have one string in the form of a circle $S^1$. See the left side of Fig.~\ref{fig:pairofpants}.
This is a process whose initial state consists of two circles and the final state is one circle. 
The $1+1$-dimensional worldsheet of the string in spacetime is topologically a sphere with three holes.

In the theory of large $N$ counting, the behavior of the amplitude in the large $N$ limit is well known ~\cite{tHooft:1973alw}. (See \cite{Coleman:1985rnk} for a detailed review.)
For this purpose, we may regard each of the three holes to be an external quark loop. Indeed,
we have argued that a string is created by the action of a Wilson loop operator $W_F(C)$ on the vacuum,
and this Wilson loop can be regarded as an external quark loop. 
The bulk of the string consists of various gluon propagators. In the large $N$ limit,
the amplitude of a general process, whose two dimensional surface has the Euler number $\chi$, is known to behave as $N^\chi$.
The Euler number is given by $\chi = 2 -g - h$, where $g$ is the genus of the surface without considering the holes, 
and $h$ is the number of holes.
In the process of the left of Fig.~\ref{fig:pairofpants}, 
we obtain $g=0$ (for the sphere) and $h=3$ (for the three holes). Thus, this value is proportional to $N^{-1}$.

From the locality, we expect that only the region near the colliding point is relevant for the amplitude if the strings are sufficiently large compared to the
inverse of the dynamical scale $\Lambda^{-1}$
of the theory.
Except $N$ and the relative velocity and angle of the initial strings, there are no dimensionless parameters in the pure $\SU(N)$ YM theory.
Thus, we conclude that the amplitude for the reconnection is proportional to $g_s \sim N^{-1}$.
The detailed computations of the amplitude in the case of cosmic superstrings can be found in \cite{Jackson:2004zg}.
If $N$ is large, the probability $P$ for reconnection is roughly $P \sim N^{-2}$. 
This is different from the usual weakly coupled vortex string in the $\U(1)$ Abelian Higgs model.

We can also find the large $N$ behavior of the binding energy of two or more strings.\footnote{
This paragraph discusses only the interactions caused by the strong dynamics. 
Other interactions such as gravity are negligible. 
}
The interactions of two string worldsheets proceed through the diagram of the form shown in the right of Fig.~\ref{fig:pairofpants},
wherein two worldsheets are connected by a tube-like region, which is referred to as a handle in mathematics. 
Compared to the diagram without interactions (i.e. two independent worldsheets), the handle increases the genus $g$
by one unit, and thus, the process is suppressed by $N^{-2}$. This implies that the bound state of $k$ strings (if it exists as a stable object)
has a tension $\mu_k$, which is given by
\beq
\label{eq:kstrings}
\mu_k=k\mu  + \cO( N^{-2}) 
\eeq
where the first term on the right-hand side is obtained from the sum of the tensions of $k$ strings, 
and the second term is the effect of interactions. 
This is the tension of the string with $\bZ_N^{[1]}$ charge $k$ (if the binding energy is negative and the bound state exists). 
For example, in some gauge theory models~\cite{Douglas:1995nw,Hanany:1997hr,Herzog:2001fq,Firouzjahi:2006vp}, 
the tension is approximately given by 
\beq
\mu_k = \mu \cdot \frac{ \sin \frac{k}{N}}{\sin \frac{1}{N}}  = k\mu-\frac{(k^3-1)}{6N^2} \mu+\cdots   \qquad  ( 1 \leq k \leq N-1),\label{eq:boundtension}
\eeq 
which is consistent with the above large $N$ argument.

The implication of the binding energy is as follows. If we consider a configuration for which
the charge of the center symmetry $\bZ^{[1]}_N$ is given by $k$ where $k=1,2,\cdots, N-1$, 
there are several possibilities for the lowest energy configuration.
One possibility is to have $k$ separated strings each of which has $\bZ^{[1]}_N$ charge $1$. Another possibility is 
to have a single bound state with $\bZ^{[1]}_N$ charge $k$. If \eqref{eq:boundtension} is valid,
the second one gives the lowest energy configuration. However,
the binding energy is suppressed by $N^{-2}$ if $k=\cO(1)$.
Thus, if we consider the large $N$ limit, we expect (and assume in this paper) that the 
effect of the binding is not so significant for the cosmological evolution of string networks.

\subsection{Holographic dual descriptions}

We discuss the results of large $N$ from another related perspective.
For some $\SU(N)$ gauge theories that are qualitatively similar to the pure $\SU(N)$ YM
such as the mass-deformed $\cN=4$ Super-YM discussed previously, holographic dual descriptions are available.
See, e.g., Refs.~\cite{Witten:1998zw,Polchinski:2000uf,Klebanov:2000hb,Maldacena:2000yy,Vafa:2000wi}.
The color flux tube in gauge theories is dual to the fundamental string (or some object that 
has the charge of the fundamental string
such as a wrapped D3-brane with a flux) in the gravity side. 
The coupling $g_s$ is then literally the string coupling in the sense of the superstring theory. 
In the holographic dual, we fix the 't~Hooft coupling $Ng_s$ in the large $N$ limit. 
Then, $g_s$ scales as $1/N$ in the large $N$ limit. 

The diagrams such as the ones shown in Fig.~\ref{fig:pairofpants} are more intuitively understood in the holographic dual  
because they represent the basic string interactions. Thus, many qualitative properties, which hold true in cosmic superstrings
are also expected to hold true for color flux tubes.

There are also some differences from cosmic superstrings.
One difference is the effect of finite (rather than infinite) $N$. 
For instance, there can be bound states as argued around \eqref{eq:boundtension},
although the binding energy goes to zero when $N \to \infty$. 
The regime $N \to \infty$ is more close to the case of cosmic superstrings in a weakly coupled, large volume limit.
Another difference is about what types of stable strings exist. In cosmic superstring scenarios,
we usually have F-strings, D-strings, and their composite $(p,q)$ strings. However, in YM theories, the types of stable strings depend on the gauge group,
and also the details of matter contents if we consider non-pure YM. For instance,
Klebanov and Strassler constructed a very explicit dual between a certain gauge theory and a gravity theory~\cite{Klebanov:2000hb}.  
Their gauge theory describes some warped throat region of string theory~\cite{Giddings:2001yu}. 
Both F- and D-strings are present in their model. But the D-string may be associated in the gauge theory side to 
the spontaneous breaking of a $\U(1)$ global symmetry~\cite{Aharony:2000pp,Gubser:2004qj}
which is not present in the pure $\SU(N)$ YM. Thus we do not expect D-strings in the pure $\SU(N)$ YM.\footnote{
However, a string with a large $\bZ^{[1]}_N$ charge $k \sim \cO(N)$ may be regarded as a D-string, or more precisely
a D-brane wrapped on the internal manifold~\cite{Herzog:2001fq}. We do not consider such large $k$ in this paper.
}
However, we will discuss in the next subsection that there are strings which can be qualitatively regarded as D-strings
if we consider the gauge group $\SO(N)$ (or its universal cover $\Spin(N)$). The classification of stable strings is different from cosmic superstrings and
is discussed in the next subsection.

\subsection{Other gauge groups}
Until now, we have mainly discussed $\SU(N)$ gauge group. We now discuss some other gauge groups.

For other gauge groups such as
$\SO(N)$ or $\Sp(N)$, the center symmetry is different.
In general, the center $C_G$ of a group $G$ is the subgroup 
$C_G \subset G$ whose elements commute with any element of $G$.
For example, the center of $\SU(N)$ is given by 
\beq
C_{\SU(N)} = \{ e^{ 2\pi i k /N} I_N ~|~k \in \bZ_N \}, 
\eeq
which we have used in the previous section.
The centers of $\SU(N)$, $\Spin(N)$ (which is the simply connected double cover of $\SO(N)=\Spin(N)/\bZ_2$) and $\Sp(N)$ are given by
\beq
\SU(N) \supset \bZ_N, \qquad 
\Spin(N) \supset \left\{
\begin{array}{cl} \bZ_2 \times \bZ_2 & (N=4K) \\ \bZ_4 & (N=4K+2) \\ \bZ_2 & (N=2K+1) 
\end{array} \right. , 
\qquad 
\Sp(N) \supset \bZ_2.
\label{eq:one-formsyms}
\eeq
The 1-form center symmetry of the pure $G$ theory for the simply-connected $G$ (i.e. $\pi_1(G)=0$) 
is determined in terms of the center $C_G$, which we may denote as $C^{[1]}_G$.

More explicitly, these centers are described as follows.
\begin{itemize}

\item $\Sp(N)$ : 
The $\Sp(N)$ consists of $2N \times 2N$ unitary matrices $A$ satisfying $A^T J A = J$, where $J$ is the invariant tensor of $\Sp(N)$
given by 
\beq
J= \begin{pmatrix} 0 & I_N \\ -I_N &0 \end{pmatrix} .
\eeq
This group has the center given by $C_{\Sp(N)} = \{\pm I_{2N}\} \cong \bZ_2$.
Thus, the 1-form center symmetry is $\bZ_2^{[1]}$, and
there is only one type of string, which is charged nontrivially under $\bZ_2^{[1]}$.
This is created by the Wilson loop operator $W_F$ in the fundamental $2N$-dimensional representation $F$ of $\Sp(N)$.

\item $\Spin(N)$ with $N=4K$ : The $\Spin(N)$ for $N =\text{even}$ has two spinor representations, 
which we denote as $S_1$ and
$S_2$. The center $C_{\Spin(4K)} = \bZ_2 \times \bZ_2$ is given as follows. 
We denote the two $\bZ_2$'s as $\bZ_2(1)$ and $\bZ_2(2)$. The nontrivial element of $\bZ_2(1)$ acts
on $S_1$ as $(-1)$, whereas it acts trivially on $S_2$ as $(+1)$. 
Similarly, the nontrivial element of $\bZ_2(2)$ acts
on $S_2$ as $(-1)$, whereas it acts trivially on $S_1$ as $(+1)$. 
Both the $\bZ_2(1)$ and $\bZ_2(2)$ act as $(-1)$ on the fundamental $N$-dimensional representation $F$ of $\Spin(N)$
because $F$ is contained in the tensor product $S_1 \otimes S_2$.
The stable strings are classified by the charges under $ \bZ_2^{[1]} \times \bZ_2^{[1]}$,
and we may denote the charge as $(k_1,k_2)$, where $k_1 =0,1$ and $k_2=0,1$ are integers modulo 2. 
The string with a charge $(1,0)$ is created by the Wilson loop operator $W_{S_1}$,
whereas the string with charge $(0,1)$ is created by $W_{S_2}$.
The string with charge $(1,1)$ is created by $W_F$.
The strings $(1,0)$ and $(0,1)$ are related by a symmetry (which is the outer automorphism of $\Spin(N)$
for $N=\textrm{even}$),
and they have the same tension. The tension is of order $N\Lambda^2$, as we discuss below. 
The string $(1,1)$ may be regarded as a bound state of $(1,0)$ and $(0,1)$, 
and its tension is of order $\Lambda^2$. If $N \gg 1$, the binding energy is negative, which indicates the presence of the bound state.\footnote{
\label{footnote:Spin4K}
Here we comment on the case with a small values of $N$.
For $N=4$, which is the smallest value of the form $N=4K$,
we obtain $\Spin(4) = \SU(2) \times \SU(2)$, and $(1,0)$, and $(0,1)$ comes from independent gauge groups; 
their bound state does not exist. 
For $\Spin(8)$, the outer automorphism group is enhanced to the symmetric group $\mathrm{S}_3$, which permutes the three representations $S_1, S_2$ and $F$. Thus, the tensions of $(1,0)$, $(0,1)$ and $(1,1)$ are the same.
}

\item $\Spin(N)$ with $N=4K+2$ : The two spinor representations $S_1$ and $S_2$
are complex conjugate representations of each other for $N=4K+2$. 
The generator of the center $C_{\Spin(4K+2)} = \bZ_4$ acts as $(+i) =\sqrt{-1}$ on $S_1$,
and $(-i)$ on $S_2$. It acts on the $N$-dimensional fundamental representation $F$ as $(-1)$
because $F$ is contained in $S_1 \otimes S_1$ or $S_2 \otimes S_2$.
The stable string is classified by the $ \bZ_4^{[1]}$ charge $k= 0,1,2,3$, which is an integer modulo 4.
The string with charge $k=1$ is created by $W_{S_1}$, whereas the string with charge $k=3$ is created by $W_{S_2}$. 
The string with charge $k=2$ is created by $W_F$.
The tensions of strings with $k=1$ and $k=3$ are the same because of the symmetry (outer automorphism), and are of order $N\Lambda^2$,
while the tension of the string with $k=2$ is of order $\Lambda^2$. The string with $k=2$ may be a bound state of two $k=1$ strings,
and the binding energy is negative if $N \gg 1$.\footnote{
It should be noted that $\Spin(6)=\SU(4)$. If the formula \eqref{eq:boundtension} holds true, the bound state exists. 
}

\item $\Spin(N)$ with $N=2K+1$ : There is only one spinor representation $S$ for $N=\textrm{odd}$.
The generator of the center $C_{\Spin(2K+1)} = \bZ_2$ acts on $S$ as $(-1)$, but it acts trivially on $F$ as $(+1)$.
Only one kind of stable string charged is present under $ \bZ_2^{[1]}$, and it is created by $W_S$;
it has the tension of order $N\Lambda^2$.\footnote{It should be noted that $\Spin(3) = \SU(2)=\Sp(1)$ and $\Spin(5) = \Sp(2)$.}
However, we will later argue that there may be a metastable string created by $W_F$.
\end{itemize}

For the strings created by $W_F$, the large $N$ behavior is similar to that of $\SU(N)$. 
Their tensions are of order $\mu \sim \Lambda^2$, and string couplings are of order $g_s \sim 1/N$. 
We refer to them as fundamental strings or F-strings. 

However, the strings created by the spinor representations of $\Spin(N)$ show significantly different $N$ behavior,
and we refer to them as D-strings. 
For concreteness, we discuss the case of $\Spin(2K+1)$ although the cases of $\Spin(N)$ for $N=\textrm{even}$ are similar. 

For a clearer understanding, we focus on the subgroup 
$\SU(K) \subset \SO(2K) \subset \SO(2K+1) = \Spin(2K+1)/\bZ_2$. 
We can capture the qualitative behavior of the large $N$ (or large $K$) limit
by considering only this subgroup $\SU(K)$. 
The spinor representation $S$ of $\Spin(2K+1)$ decomposes under $\SU(K)$ as
\beq
S \to \bigoplus_{n=0}^{K} \wedge^n F,
\eeq
where $\wedge^n F$ is the $n$th antisymmetric representation of $\SU(K)$.
It contains representations with large $\SU(K)$ charges $n \sim \mathcal{O}(K)$.
Thus, it possesses qualitatively similar properties as ``baryons'' of $\SU(K)$ in the sense that it consists of the antisymmetrization of large 
numbers of fundamental representations.
The tension of the string associated with $S$ behaves as 
\beq
\mu \sim N \Lambda^2 
\eeq
because it contains order $N \sim K$ fundamental strings
of $\SU(K)$. This phenomenon is confirmed in a holographic dual description~\cite{Witten:1998xy}. 
In the dual side, the color flux associated with $S$
is given by a (wrapped) D-brane, whose tension is proportional to $g_s^{-1} \sim N$.

Such ``baryonic'' objects might have an exponentially suppressed reconnection probability 
when the relative velocity of the string $v$ is not significantly small,
\beq
P \sim e^{-cN}  \label{eq:expsuppression},
\eeq
where $c$ is expected to be $\mathcal{O}(1)$ if $v$ is not small.
Such an exponential suppression was discussed in the case of baryon particles for $\SU(N)$ gauge theories in Sec.~8.3 of Ref.~\cite{Witten:1979kh}.
We have further mentioned that the color flux tube is given by a (wrapped) D-brane in the holographic dual.\footnote{
Baryons (if present) are also some wrapped D-branes. If D-branes possess some universal features
in the large $N$ limit irrespective of their dimension, shape, etc., that universal feature
is expected to hold true both for the baryons and strings created by the spinor representation.
}
An exponential suppression $e^{-c/g_s}$ was found for the case of D1-brane~\cite{Jackson:2004zg}.
The coefficient $c$ was also determined as a function of the velocity $v$ and the angle $\theta$
for the case of the D1-brane; however, we are unsure of the universality of this explicit functional form. 
We believe that this would be an interesting research topic.

There may be another reason to expect the behavior \eqref{eq:expsuppression}.
If we consider the effective action of the string, we expect that there is an overall factor of $N$ owing to the aforementioned reason.
Thus, the effective action (with only minimal degrees of freedom on the string worldsheet) may be
\beq
S_\textrm{eff} = - \frac{N}{4\pi}\left( \int \sqrt{-\gamma} \left(\hat{\mu} + \frac{c}{2} R +\cdots \right)d^2\sigma   \right)
\eeq
where $\hat{\mu}$ is a constant of order $\mathcal{O}(\Lambda^2)$, $c$ is a constant of order $\mathcal{O}(1)$,
and $R$ is the Ricci scalar of the string worldsheet. The first term proportional to $\hat{\mu}$ is the Nambu-Goto action which is a ``cosmological constant term'' on the worldsheet,
and the second term proportional to $R$ is the ``Einstein-Hilbert term'' on the worldsheet.\footnote{The factor of $1/4\pi$ in front of the action
is put so that it is consistent with the naive dimensional analysis of strong dynamics.} 
We also expect higher order terms, but we
assume that they do not dramatically change the following argument.
If the aforementioned action is qualitatively valid, the Einstein-Hilbert term gives $\exp( cN \chi/2)$, where $\chi = \frac{1}{4\pi} \int \sqrt{-\gamma} R$ is the Euler number of the worldsheet.
Thus, the string coupling may be $e^{- cN/2}$, and the reconnection probability may behave as $P \sim e^{- c N}$.
This result is obtained based on the assumption that there are only minimal degrees of freedom on the worldsheet. 
This, however, may be changed if we also include other degrees of freedom. For example, the D-branes in superstring theory contain several degrees of freedom
other than the motion of the string. Some degrees of freedom are caused by the supersymmetry, which is absent in color flux tubes of YM theories.
Provided that the effects of those possible degrees freedom and the higher order terms are not too significant, this result is expected to be qualitatively valid.

\subsection{Baryon vertex and other dynamical objects}
\begin{figure}
\centering
\includegraphics[width=0.7\textwidth]{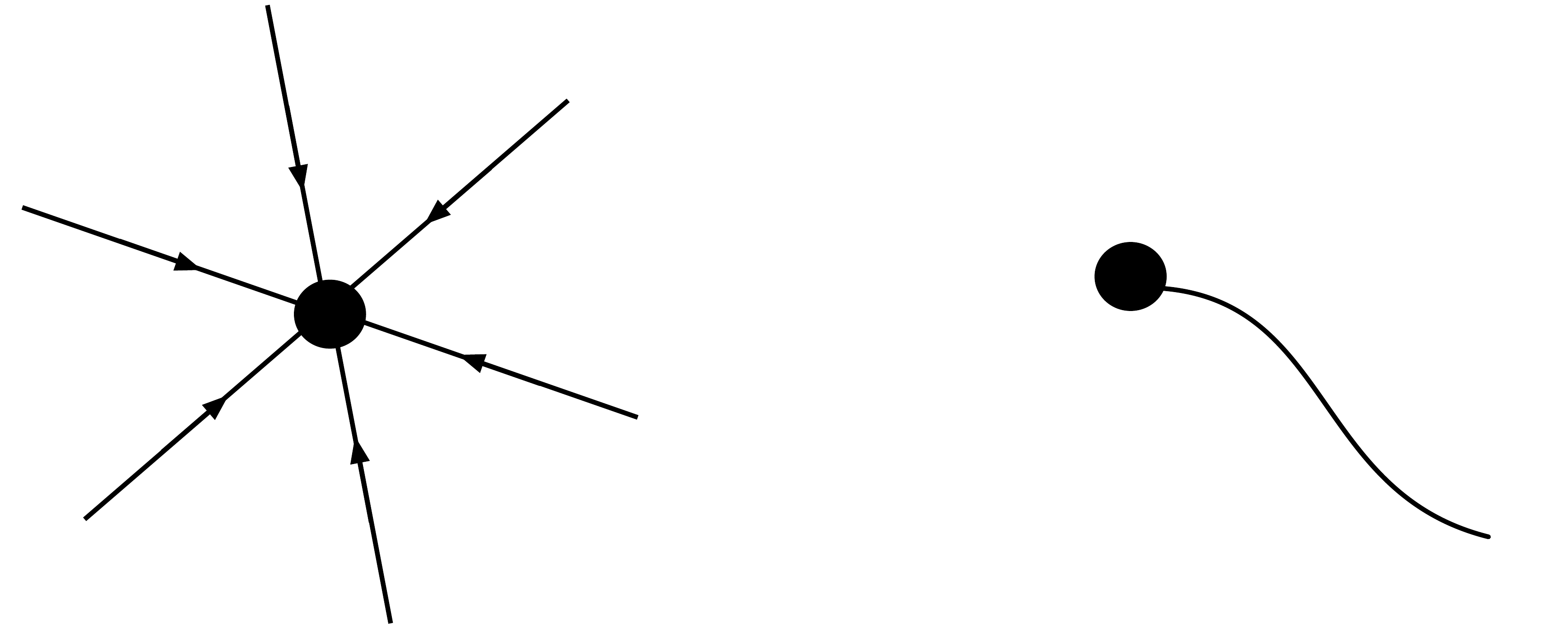}
\caption{Left: $N$ strings in $\SU(N)$ YM can end on a single vertex. In this figure, we have taken $N=6$. This vertex is called the baryon vertex. 
Right: Metastable color flux tube associated to the fundamental representation of $\SO(2K+1)$ can end on a colored baryon, which is a kind of baryon vertex
with only a single string attached to it.
\label{fig:baryonvertex}}
\end{figure}

We discuss another property of YM theories.
The 1-form center symmetry of $\SU(N)$ is $\bZ^{[1]}_N$, and this suggests that if there are $N$ strings, they can end on a single vertex,
as seen in the left side of Fig.~\ref{fig:baryonvertex}. Such a vertex is referred to as a baryon vertex, and it should not be confused with a baryon particle.
This is a vertex where $N$ strings can end. Its existence is explicitly seen in some holographic dual descriptions of gauge theories~\cite{Witten:1998xy}. 

This property of baryon vertex is special to the gauge group $\SU(N)$. 
For other gauge groups such as
$\SO(N)$ or $\Sp(N)$, the center symmetry is quite different as we have seen earlier,
and we do not have a baryon vertex that connects order $N$ strings.

It should also be noted that there are baryon particles in $\Spin(2K)$ or $\SO(2K)$ theories, and these are constructed using the totally antisymmetric tensor $\epsilon_{i_1 \cdots i_{2K}} $ of $\SO(2K)$.
For instance, we have gauge invariant operators 
\beq
\epsilon_{i_1 \cdots i_{2K}} F^{i_1i_2}_{\mu_1\mu_2} \cdots F^{i_{2K-1}i_{2K}}_{\mu_{2K-1} \mu_{2K}}
\eeq
where $i_K$'s are gauge indices and $\mu_k$'s are spacetime indices. 
The corresponding particle is the baryon particle, which is constructed purely from gluons. 

In the $\Spin(2K+1)$ theory, only one stable string is associated to the spinor representation $S$.
However, in addition to the stable string,
a metastable string may also be associated to the fundamental representation $F$. 
It should be noted that there is a colored baryon in the fundamental representation of the gauge group, which may be created by an operator
\beq
\label{eq:coloredbaryon}
B_{i, \mu_1 \cdots \mu_{2K}}  = \epsilon_{i i_1 \cdots i_{2K}} F^{i_1i_2}_{\mu_1\mu_2} \cdots F^{i_{2K-1}i_{2K}}_{\mu_{2K-1} \mu_{2K}}.
\eeq
It has the color index $i$, and thus, the color flux tube associated to the fundamental representation can end on it and decay via the pair productions.
However, the mass of baryons are of order $N\Lambda$~\cite{Witten:1979kh}, and thus, the decay is suppressed; this color flux tube is metastable.

\section{Dynamics of cosmic string network}
\label{sec:VOS}

\subsection{Properties of cosmic strings}
\label{sec:properties}

From the discussion in Sec.~\ref{sec:heuristic}, 
cosmic strings should be formed at the phase transition from the deconfinement phase to the confinment phase in the pure 
$\SU(N)$ (as well as other) YM theory. 
Depending on the structure of gauge group, 
there are two types of cosmic strings: the one being dual to a fundamental string and the one being dual to a wrapped D-brane in the gravity side by the holographic dual descriptions. We refer to them as an F-string and a D-string, respectively. There are only F-strings in $\SU(N)$ and $\Sp(N)$, whereas there are both F-strings and D-strings in $\Spin(N)$ and $\SO(N)$ ($= \Spin(N)/\bZ_2$).\footnote{
For small $N$, the distinction between F-strings and D-strings is ambiguous. 
For example, the stable string of the $\SU(2)$ YM is regarded as an F-string if the gauge group is regarded as $\SU(N)$
with $N=2$, but it is also regarded as a D-string from the point of view of $\Spin(4) = \SU(2) \times \SU(2)$.
}
For the case of $\Spin(2K+1)$ and $\SO(2K+1)$, the F-string is metastable.
Here, we summarize the properties of those cosmic strings and phenomenological implications to their dynamics. 

\begin{itemize}

\item Naturally small tension 

The string tension $\mu$ is given by the confinement scale of YM theory $\Lambda$ such as 
\beq
\label{eq:mu}
 \mu \sim 
 \left\{
 \begin{aligned}
 &\Lambda^2 &\quad \text{for F-string}
\\
 &N \Lambda^2 &\quad \text{for D-string}
 \end{aligned}
 \right. \,. 
\eeq
In particular, for the case of $\SU(N)$, 
the numerical factor is determined using the lattice simulations as~\cite{Athenodorou:2021qvs}
\beq
\label{eq:tension}
 \frac{\Lambda}{\sqrt{\mu}} = 0.5055(7)[250] + \frac{0.306(12)}{N^2}, 
\eeq
where the round (square) brackets represent statistical (systematic) errors. 
Here, the dynamical scale is determined using the renormalization group such as 
\beq
 \Lambda = \mu_0 e^{-8 \pi^2 / (b g_0^2)}, 
\eeq
where $b$ ($= 11N/3$ for $\SU(N)$) is the one-loop coefficient of the beta function and $\mu_0$ is a renormalization scale. The scale of $\Lambda$ is exponentially sensitive to the fundamental gauge coupling $g_0$, and thus, it can be naturally small and does not require fine-tuning mainly because of the dimensional transmutation. This is in contrast to cosmic strings in weakly coupled field-theory models, where the string tension is determined using a Higgs scale that cannot be naturally small at least in non-supersymmetric models. 

If two F-strings connect and form a bound state with $k=2$ winding number, 
its tension should scale as $2 \mu(1 + \mathcal{O}(1/N^2))$ (see \eq{eq:kstrings}). 
This is consistent with the lattice result~\cite{Athenodorou:2021qvs}. 
We can thus neglect the formation of bound state with a winding number larger than unity for a sufficiently large $N$.

\item Small intercommutation probability 

From the discussion of large $N$ limit and holographic dual descriptions, 
the intercommutation (or reconnection) probability of strings is not $\mathcal{O}(1)$ but scales as 
\beq
\label{eq:P}
 P \sim 
 \left\{
 \begin{aligned}
 &N^{-2} &\quad \text{for F-string}
\\
 &\exp(-c N) &\quad \text{for D-string}
 \end{aligned}
 \right. \,, 
\eeq
for a large $N$. 
Here, $c = \mathcal{O}(1)$ when the relative velocity of the string is not small. 
This is a unique property of our cosmic string contrary to weakly-coupled field-theory cosmic strings, wherein the strings almost always reconnect after the intersection. 
A small intercommutation probability is also realized for F- and D-strings that are formed after D-brane inflation~\cite{Polchinski:1988cn,Jackson:2004zg,Hanany:2005bc}. 
Our cosmic string, namely the color flux tube in gauge theories, is dual to the F- and D-strings in the gravity side according to the holographic dual descriptions.

The impact of exponentially-suppressed intercommutation probability for D-strings is quite non-trivial because the exponential factor $c$ depends on not only $N$ but also the relative velocity and the relative angle of strings. 
Because we are interested in the statistical properties of string network, 
we consider an average over the relative velocity and relative angle. 
As a result, the intercommutation probability may not be significantly suppressed like the above exponential form. 
Because the factor of $c$ cannot be determined in the current understanding of strong dynamics, we do not discuss the form of intercommutation probability further. In the subsequent analysis, we consider $P$ as a free parameter to calculate the GW signals from cosmic strings.

Even if we take $P$ as a free parameter, 
the consequence of small intercommutation probability is non-trivial because of a small wiggly structure of cosmic strings. 
Let us consider an intersection event of two long strings. 
The small wiggles on the strings move as fast as $1/\sqrt{2}$, whereas 
the relative velocity between the long strings is not that fast according to numerical simulations~\cite{Martins:2005es,Avgoustidis:2005nv}. 
This results in many intersections of small wiggles within the time scale of collision of long strings. 
As a result, the intercommutation probability between long strings is effectively enhanced by a factor of $N_{\rm scat} \sim 10$~\cite{Avgoustidis:2005nv}. 
Denoting the intercommutation probability of (ideal straight) strings as $P$, 
we obtain the effective intercommutation probability of (realistic) wiggly strings such as 
\beq
\label{eq:Peff}
 P_{\rm eff} = 1 - (1-P)^{N_{\rm scat}}. 
\eeq
This gives $P_{\rm eff} \sim 10 /N^2$ for a sufficiently large $N$ for an F-string.

Both F-strings and D-strings exist in $\SO(N)$ and $\Spin(N)$, and in this case, 
the bound state may form after the collision of two different types of strings. 
This process can be interpreted as the transition from a closed F-string to open F-strings connected by D-strings. 
The probability of this process is expected to be $N^{-1}$. 
In those models, the number of bound state is finite because of the discrete 1-form symmetry. 
This is in contrast to the cosmic superstrings that form after brane inflation~\cite{Dvali:2003zj,Copeland:2003bj}. 
In the brane-inflationary scenarios, D-strings (1-dimensional Dirichlet branes)~\cite{Jones:2002cv,Sarangi:2002yt,Dvali:2002fi,Jones:2003da,Pogosian:2003mz,Dvali:2003zj,Copeland:2003bj} as well as F-strings~\cite{Dvali:2003zj,Copeland:2003bj} form, 
which results in a complicated string network with infinite number of bound states with different tensions~\cite{Copeland:2006eh,Copeland:2006if,Copeland:2007nv,Avgoustidis:2007aa,Rajantie:2007hp,Pourtsidou:2010gu,Sousa:2016ggw}. 
In this paper, we focus on the case with a single type of cosmic strings.

The intercommutation probabilities of D- and F-strings are additionally suppressed using a volume factor if they move in a higher-dimensional space~\cite{Jones:2003da,Jackson:2004zg,Pourtsidou:2010gu}. 
This factor does not appear in our model because we consider $3+1$ dimensional spacetime.

\item Baryon vertex

Depending on the gauge theories, baryon vertex may or may not be present. 
For instance, SU($N$) gauge theory has the 1-form $\bZ^{[1]}_N$ symmetry, which the cosmic string is charged under. 
In this case, a cosmic string with $k + N$ winding number has the same tension as that with $k$ winding number. 
This specifically indicates that $N$ cosmic strings can end on a single baryon vertex as seen in the left side of Fig.~\ref{fig:baryonvertex}. 
This is similar to the case with so-called $\bZ_N$ string, where a field theory with the symmetry breaking pattern of $\SU(N)$ $\to$ $\bZ_N$ by $N$ adjoint Higgs fields is considered~\cite{Vachaspati:1986cc,Ng:2008mp}. 
It should be noted that this property is different from that of non-Abelian strings considered in Refs.~\cite{Spergel:1996ai,McGraw:1997nx,Avgoustidis:2007aa}, where 
there are many different strings that cannot pass through or reconnect with each other. In this case, the string network is frustrated and its energy density dominates the Universe.

The dynamics of string network should be qualitatively different for the cases of $\bZ^{[1]}_2$, $\bZ^{[1]}_3$ and $\bZ^{[1]}_N$ with $N>3$.

i) Case with $\bZ^{[1]}_2$: 
there may be a baryon vertex that connects two cosmic strings with opposite fluxes. A similar network is considered in weakly-coupled field-theory models with monopoles and cosmic strings, which is referred to as a necklace~\cite{Hindmarsh:1985xc,Berezinsky:1997td,Hindmarsh:2016dha}. 
Based on the numerical simulations in the weakly coupled models, the energy density of baryon vertex (which is called beads in the literature) becomes negligible compared to that of cosmic strings~\cite{Hindmarsh:2016dha}. This implies that the effect of baryon vertex to the network evolution is negligible at a later time even if the baryon vertex exists. 
This is the case for $\SU(2)$, $\Sp(N)$, $\SO(2K+1)$, and $\Spin(2K+1)$. 

Here, we discuss the case of $\SO(4K)$ and $\Spin(4K)$, where the one-form center symmetry is $\bZ^{[1]}_2 \times \bZ^{[1]}_2$. 
For $K=1$, we have two independent $\bZ^{[1]}_2$ charged strings (see footnote~\ref{footnote:Spin4K}). 
For $K \ge 2$, there are two D-strings and one bound state (that corresponds to an F-string). 
In particular, for $K=2$, all three strings have the same tensions because of the outer automorphism. We leave these cases for a future work~\cite{futurework}.

ii) Case with $\bZ^{[1]}_3$: 
a baryon vertex connects three cosmic strings with the same tensions. 
A similar network was considered in the context of hadronic string of QCD theory~\cite{Artru:1974zn} though they consider unstable strings with light quarks. 
The dynamics of cosmic strings with 
the same tensions~\cite{Copeland:2005cy,Hindmarsh:2006qn}
or with different tensions~\cite{Urrestilla:2007yw} 
are considered in the literature. 
An extended version of VOS model was also proposed~\cite{Copeland:2006if,Avgoustidis:2007aa}, and it explains the results of numerical simulations. 
In this case, a baryon vertex can be formed by the intersection of two strings even at a later time. 
According to numerical simulations, the network reaches the scaling solution. We expect that there are $\mathcal{O}(1)$ baryon vertices within a Hubble horizon.

iii) Case with $\bZ^{[1]}_{N>3}$: 
a baryon vertex connects $N$ cosmic strings with $k=1$ winding number.
In this case, it is difficult to produce the baryon vertex after the phase transition. 
The formation of baryon vertex requires that $N$ strings meet within a distance of order the string width. Such an event is negligible during the dynamics of the string network for the case of $N >3$. 
Even if such an event occurs, 
the probability of vertex formation is additionally suppressed exponentially through a tunneling factor for a large $N$ like $e^{-\mathcal{O}(N)}$~\cite{Witten:1979kh}. 
We can thus neglect the late-time formation of baryon vertex for the case $N >3$.

However, 
the baryon vertices can form at the confinement/deconfinement phase transition even for $N > 3$. 
This can be interpreted in a similar manner to the monopole production 
in the electric--magnetic dual description. 
For example, one can consider the symmetry breaking pattern of $\SU(N)/\bZ_N \to \U(1) \to 1$ in a weakly couple field theory, which leads to the formation of monopoles followed by that of cosmic strings. In this case, $\mathcal{O}(1)$ monopoles form within a correlation volume at the first phase transition. Those monopoles are attached by $N$ cosmic strings at the second phase transition. 
This case should hold true even if those phase transitions occur simultaneously at the same energy scale. 
We thus expect that at least $\mathcal{O}(1)$ baryon vertices form within a Hubble horizon. 
Then, the number of baryon vertices within a Hubble horizon increases as the Universe expands, provided that they do not annihilate. 
In other words, their number should be reduced via the annihilation to reach the scaling regime. This is possible because the baryon vertices are connected by strings and are pulled toward each other by their tensions. 

Based on the aforementioned explanation, the network reaches the scaling regime for the case of $N=3$. The main difference between $N=3$ and $N>3$ is the absence of formation of baryon vertex at a late time in the latter case. This difference is not significant for attaining the scaling regime, which requires a mechanism to {\it reduce} the number of baryon vertices. 
We thus conclude that the network for the case with $N>3$ also reaches the scaling regime at a later time, where the number of baryon vertex within a Hubble horizon is constant in time. 
This is also expected in analogy to the domain-wall and cosmic-string system that is extensively studied in the context of QCD axion models. 
According to field-theory simulations~\cite{Spergel:1990ee,Hiramatsu:2010yn,Hiramatsu:2012sc,Kawasaki:2014sqa}, 
even if a single cosmic string is attached by multiple domain walls, the system is not frustrated but it follows the scaling solution wherein the numbers of domain walls as well as the cosmic strings are of order unity within the horizon. 
The similarity of our system to this system can be highlighted by reducing the dimension of the topological defects, that is, by replacing the cosmic strings by baryon vertices and the domain walls by cosmic strings. 

In the scaling regime, 
the number of baryon vertex should be of order unity within the horizon 
if the intercommutation probability is of the order of unity. 
If the intercommutation probability is significantly less than unity, the numbers of baryon vertex and cosmic strings within the Hubble horizon may be larger, but they should still be constant in time. 
We can only expect that at least $\mathcal{O}(1)$ baryon vertices and $\mathcal{O}(1)\times N$ cosmic strings survive within a Hubble horizon. This sets the lower bound on the number density of long cosmic strings. 
We will discuss the consequence of this effect in detail in Sec.~\ref{sec:baryonvertex}.

In the numerical calculations, we neglect the impact of baryon vertex, 
which is justified for a large $N$ or in a theory without baryon vertex such as $\Sp(N)$.

\item Exponentially suppressed decay rate of cosmic string

In our model, the 1-form $\bZ^{[1]}_N$ symmetry is a global symmetry. In a consistent theory of quantum gravity, any global symmetries must be explicitly broken~\cite{Banks:2010zn} (see also e.g., Refs.~\cite{Montero:2017yja,Harlow:2018tng,Rudelius:2020orz,Yonekura:2020ino,Heidenreich:2020pkc,Heidenreich:2021xpr,Cordova:2022rer} for generalized global symmetry). 
The 1-form symmetry $\bZ^{[1]}_N$ is explicitly broken if one introduces quarks with fundamental representation. 
The model still reduces to our pure $\SU(N)$ YM theory if the mass of quarks are much larger than the dynamical scale. 
The cosmic string can decay if the mass of quarks is of the order of dynamical scale. 
The decay rate of the string per volume at zero temperature is estimated as \eq{eq:decayrate}: 
\beq
\Gamma \propto \exp\left(  - \pi \frac{m^2}{\mu} \right) ,
\eeq
where $m$ is the quark mass. 
Recently, the decaying cosmic strings have attracted significant attention~\cite{Buchmuller:2021mbb,Dunsky:2021tih,Lazarides:2022jgr}. 
A similar scenario can be realized in our model if there is a vector quark with mass of order the confinement scale. 
It is possible to naturally realize the parameter region like $\mathcal{O}(1) \cdot \Lambda \lsim m \lsim \mathcal{O}(10)\cdot \Lambda$,
if the number of quarks in the fundamental representation is such that the UV gauge theory is in the conformal window.
Then, the confinement occurs soon after the decoupling of the massive quarks. See e.g.
Refs.~\cite{Luty:2004ye,Ibe:2007wp,Yanagida:2010zz} for a detailed discussion of such scenarios in different contexts.

However, it is expected that the length of cosmic strings cannot be sufficiently long but suppressed by some power of number density of quarks if quarks and anti-quarks present in the thermal plasma at the deconfinement/confinement phase transition.
If quarks are abundant in the thermal plasma at the time of phase transition, 
cosmic strings tend to form such that they connect the quarks and anti-quarks. 
This results in the formation of relatively short cosmic strings 
and we expect that the number of long (superhorizon) cosmic strings is exponentially suppressed. 
For long cosmic strings to form, the number density of fermions must be significantly suppressed in the thermal plasma.%
\footnote{
See Ref.~\cite{Hamaguchi:2011kt} for the evolution of short cosmic strings that connect quarks and anti-quarks.
}
Therefore they should be diluted by inflation for this scenario to work. In other words, one has to consider the case wherein e,g., the maximal temperature of the Universe is higher than $\Lambda$ but is lower than $m$. 

For pure $\SO(2K+1)$ YM theory, 
there is a colored baryon in the fundamental representation of the gauge group with mass of order $N\Lambda$ (see discussion around  \eq{eq:coloredbaryon}). Those colored baryons are expected to form at the deconefienmen/confinement phase transition, in which case sufficiently long F-strings cannot form. We therefore expect that only D-strings should survive at a later epoch in pure $\SO(2K+1)$ YM theory.

In this paper, we consider stable cosmic strings in the pure YM theory.

\item No new composite state of F- and D-strings

For cosmic superstrings in brane inflationary scenarios,
one usually has F-strings, D-strings, and their composite $(p,q)$ strings. However, in YM theories, the types of stable strings depend on the gauge group. 
They also depend on the details of matter contents if we consider non-pure YM. 
For instance,
we do not expect D-strings in the pure $\SU(N)$ YM. 
In $\SO(N)$ and $\Spin(N)$ gauge theories, 
there are both stable F- and D-strings for $N = 4K$ and $4K+2$. 
The bound state of those strings are not similar to the cosmic superstrings. 
For the case of $N = 4K$, the center symmetry is $\bZ^{[1]}_2 \times \bZ^{[1]}_2$ (see Eq.~(\ref{eq:one-formsyms}))), which means that 
only three different types of cosmic strings exist. 
The two of them, whose charges are $(1,0)$ and $(0,1)$ under $\bZ^{[1]}_2 \times \bZ^{[1]}_2$, are identified as D-strings whereas the remaining one, whose charge is $(1,1)$, is an F-string. 
The latter one can be created from the collision of $(1,0)$ and $(0,1)$ D-strings. 
However, there is no new composite state of F- and D-strings because 
the $\bZ^{[1]}_2 \times \bZ^{[1]}_2$ symmetry implies for instance that the composite of strings with charges $(1,0)$
and $(1,1)$ is just the string with charge $(0,1)$. 
One can consider a system of both F- and D-strings in this gauge theory, but the whole network is not as complicated as the cosmic superstrings.
A similar conclusion holds 
for the case of $N = 4K +2$, in which case the center symmetry is 
$\bZ^{[1]}_4$ and there again exist only one F-string and two D-strings strings without a new composite state. 
The system with both F- and D-strings can be described by multiple VOS equations used in Refs.~\cite{Avgoustidis:2007aa,Rajantie:2007hp,Pourtsidou:2010gu}. 
The interaction term between F- and D-strings may be introduced, but we do not need the other composite states.

\end{itemize}

\subsection{Extended VOS model}
\label{sec:VOS2}

It is known that the network of cosmic strings reaches a scaling regime in a finite time scale. 
The statistical properties can be described by the VOS model~\cite{Kibble:1984hp,Martins:1995tg,Martins:1996jp,Martins:2000cs}, which is supported by numerical simulations~\cite{Ringeval:2005kr,Blanco-Pillado:2011egf,Blanco-Pillado:2013qja,Blanco-Pillado:2017oxo,Blanco-Pillado:2017rnf}, 
and it is used to calculate GW spectrum~\cite{Caldwell:1991jj,DePies:2007bm,Sanidas:2012ee,Sousa:2013aaa,Sousa:2016ggw}. 
In this paper, we use an extended VOS model to incorporate the effect of the small intercommutation probability~\cite{Avgoustidis:2005nv}. This extended model is also supported by numerical simulations.

We focus on the dynamics of cosmic strings and omit the existence of baryon vertex, which will be discussed below. 
At a late time, the width of cosmic string is significantly shorter than the curvature radius of strings. 
In addition, our strings have no long-range interactions. 
We thus expect that its dynamics can be described by the Nambu-Goto 
action. This is also consistent with the fact that our string corresponds to the fundamental string by the holographic dual descriptions. 
The action is given by 
\beq
 S = - \mu \int \sqrt{-\gamma} d \sigma^2, 
\eeq
where $\sigma^a$ is worldsheet coordinates and $\gamma_{ab}$ is the two-dimensional string worldsheet metric. 

We consider the dynamics of cosmic strings in the Friedmann-Robertson-Walker Universe: 
\beq
 ds^2 = a^2(\tau) \lmk d \tau^2 - d {\bf x}^2 \rmk,
\eeq
where $\tau$ is the conformal time and $a$ is the scale factor. 
We choose the gauge conditions of $\sigma^0 = \tau$ and $\dot{\bm x} \cdot {\bm x}' = 0$, where the dot and prime denote the derivatives with respect to $\tau$ and $\sigma^1 (\equiv \sigma)$, respectively. The equation of motion for a string is then given by 
\beq
 &\ddot{\bm x} + 2 a H \lmk 1 - \dot{\bm x}^2 \rmk \dot{\bm x}  
= \frac{1}{\epsilon} \lmk \frac{{\bm x}'}{\epsilon} \rmk', 
\\
\label{eq:epsilon}
&\dot{\epsilon} + 2 a H \dot{\bm x}^2 \epsilon = 0, 
\eeq
where 
\beq
\epsilon \equiv \sqrt{\frac{{\bm x}'^2}{1 - \dot{\bm x}^2} } 
\eeq
represents the coordinate energy per unit length. 
Because the theory is gapped and there is no massless particle in the plasma, 
the strings do not interact with the ambient plasma. Thus, we need not include the friction force in \eq{eq:epsilon} (that is, we can take the friction lengthscale infinity, $l_f \to \infty$).

Because we are interested in the dynamics of string network rather than that of individual strings, we take a spatial average over a whole observable Universe and use some statistical quantities to describe the network of cosmic strings. 
We denote the correlation length of cosmic strings as $\xi$, which is expected to be of order the Hubble horizon length. This represents the distance beyond which string directions are not correlated. 
The network of cosmic strings consists of long and short strings. 
We define a long string such that its length is longer than the correlation length $\xi$. 
We further denote the typical inter string distance of long strings as $L$, which is related to the energy density of long strings as $\rho_\infty = \mu / L^2$.

For a standard cosmic string with $\mathcal{O}(1)$ intercommutation probability, the inter string distance $L$ and the correlation length $\xi$ are of the same order with each other. 
However, this is not the case for a small effective intercommutation probability $P_{\rm eff}$ ($\ll 1$). 
The number of collisions between different long strings within one Hubble time is reduced for $P_{\rm eff} \ll 1$, which results in a large $\rho_\infty$ and small $L$. 
On the contrary, the number of self-reconnection for a single long string within one Hubble time is not that reduced even for $P_{\rm eff} \ll 1$. 
This is because the left and right movers of string perturbations collide several times because of their periodic motion, and they eventually reconnect within a long time scale of order $H^{-1}$. This effect 
makes the correlation length of order the Hubble scale, $\xi = \mathcal{O}(H^{-1})$. 
We thus need to use an extended version of VOS model where $\xi$ and $L$ evolve differently.%
\footnote{
One may instead use 
the standard VOS model with a smaller loop chopping efficiency parameter $\tilde{c} \to P_{\rm eff}^\gamma \tilde{c}$ with $\gamma = 1/3$. 
The factor of $\gamma = 1/3$ is conventionally used but is originally given by $(0.6^{+0.15}_{-0.12}) /2$ from a numerical simulation~\cite{Avgoustidis:2005nv}. 
The result of numerical simulation is also reproduced using the extended VOS model adopted in this paper, as discussed in Ref.~\cite{Avgoustidis:2005nv} even if the resulting power for a small $P_{\rm eff}$ is $1/2$ rather than $1/3$. 
}

The energy density of the long strings and the average root-mean-square string velocity are given by 
\beq
\label{eq:rhoinfty}
 &\rho_\infty = \frac{1}{V} \mu a \int \epsilon d \sigma , 
 \\
 &\bar{v}^2 \equiv \frac{\int \dot{\bm x}^2 \epsilon d \sigma}{\int \epsilon d \sigma}, 
\eeq
where the integral is taken only for the long strings. 
Taking the time derivative of these quantities, 
we obtain the evolution equations such as 
\beq
 \label{eq:drhodt}
 &\frac{d\rho_\infty}{d t} = - \lmk 2 H (1 +\bar{v}^2) \rmk \rho_\infty + \lmk \frac{d\rho_\infty}{d t}  \rmk_{\rm loop}, 
 \\
 \label{eq:dvdt}
 &\frac{d\bar{v}}{d t} = (1 -\bar{v}^2) \lmk \frac{k(\bar{v})}{R} - 2H \bar{v} \rmk,  
\eeq
where we include up to second order terms in \eq{eq:dvdt}. 
The first term in the right-hand side of \eq{eq:drhodt} is obtained from the dilution and stretching of strings by the Hubble expansion with the modulation by the redshift of the string velocity. The last term represents the energy loss via the loop production discussed below. 
The term with $k(\bar{v})/R$ in \eq{eq:dvdt} is obtained from the acceleration due to the curvature of string and the term with $2H\bar{v}$ comes from the damping due to the Hubble expansion. 
The curvature radius $R$ is defined using 
\beq
 \frac{a}{R} \hat{\bm u} = \frac{d^2 {\bm x}}{d s^2}, 
\eeq
where $s$ ($ds = \abs{{\bm x}'} d \sigma$) is the physical length along the string and $\hat{\bm u}$ is the unit vector for the direction of $d^2 {\bm x}/d s^2$. 
We assume that the curvature radius $R$ is of the same order as that of the correlation length $\xi$, whereas it is considered to be equal to $L$ in the standard VOS model. 
The momentum parameter $k(\bar{v})$ is related to the small scale structure on strings and is given by 
\beq
 k(\bar{v}) &\equiv \frac{\la \lmk \dot{\bm x} \cdot \hat{\bm u} \rmk \lmk 1 - \dot{\bm x}^2 \rmk 
  \ra}{\bar{v}(1-\bar{v}^2)} 
  \\
  &= \frac{2\sqrt{2}}{\pi} 
\frac{1-8 \bar{v}^6}{1+8 \bar{v}^6}, 
\eeq
where the second line is an analytic function that fits well with the numerical simulations~\cite{Martins:2000cs}. 
It should be noted that $k(1/\sqrt{2}) = 0$~\cite{Martins:1996jp}. 
Because we do not have a friction force, $\bar{v}$ is usually in the relativistic regime. 

According to numerical simulations of cosmic strings, 
string loops are continuously generated from the reconnection of long strings. 
This results in 
the energy loss of long loops. 
The distribution of length of loops is conventionally described using a scale-invariant loop production function $f(l_i,t)$, where $f(l_i,t) d l_i$ is the number of produced loops with length $\in (l_i,l_i+dl_i)$ 
for each intercommutation event. 
The energy-loss rate for a long string is then proportional to 
\beq
\label{eq:tildec}
 \mu \int_0^{\infty} l_i f(l_i,t)  dl_i
 \equiv \tilde{c} \mu \xi, 
\eeq
where $\tilde{c}$ is called a loop chopping efficiency parameter. 
Our definition of $\tilde{c}$ is identical to that of the conventional model for the case with $\xi = L$. 
As discussed previously, string loops are produced through the self-intercommutation, and thus, their typical length should be of order $\xi$ rather than $L$ for the case of $\xi \ne L$. 
In the scaling regime, 
we expect that the loop production function $f(l_i, t)$ scales as $f(l_i, t) = t^{-1} f(x)$ with $x= l_i / t$, which implies that $\tilde{c}$ is constant in time. 
According to the numerical simulations, 
$\tilde{c} = 0.23 \pm 0.04$ fits the numerical results in both 
the radiation dominated era (RD) and the matter dominated era (MD)~\cite{Martins:2003vd} (see also Refs.~\cite{Bennett:1989yp,Allen:1990tv,Martins:1996jp,Martins:2000cs}).

We should also estimate the number of intercommutation events per unit time. 
The number density of long strings is given by $n \sim \rho_\infty / (\mu \xi) = 1/ ( L^2 \xi)$. 
Because a string sweeps over one correlation length within the time scale of $\xi / \bar{v}$, 
the number of intersections for a given string per unit time is proportional to $n \xi^3 (\bar{v}/\xi)$. 
Using \eq{eq:tildec}, and considering the effective intercommutation probability $P_{\rm eff}$, 
we thus find that the energy density of long strings decreases because of the loop production:
\beq
 \lmk \frac{d \rho_\infty}{d t} \rmk_{\rm loop} 
 &= -P_{\rm eff} (\tilde{c} \mu \xi) n \frac{n \xi^3  \bar{v}}{\xi} , 
 \\
 &= -P_{\rm eff} \tilde{c} \bar{v} \rho_\infty \lmk \frac{\xi}{L^2} \rmk, 
 \label{eq:drhodt2}
\eeq
This can be reduced to the standard VOS equations based on the assumption that $\xi = L$ and $P_{\rm eff} =1$.%
\footnote{
\label{footnote:loop}
In Ref.~\cite{Sakellariadou:2004wq}, it was highlighted that 
small loops can be produced mainly by self-intercommutation for $P_{\rm eff} \ll 1$. 
However, we consider that a wiggly structure that later results in self-intercommutation mainly comes from the intercommulation between different long strings. 
The production process of single loop therefore proceeds based on the following steps: production of wiggly structure from an intercommutation between different strings, followed by the self-intercommutation of wiggly structure. 
The rate of the first step is estimated by \eq{eq:drhodt2}, whereas the second step is expected to be sufficiently efficient, as discussed previously \eq{eq:rhoinfty}. 
We thus assume that the energy loss is proportional to the rate of the first step, which is the bottleneck process for loop production. 
Our result of $\rho_\infty \propto P_{\rm eff}^{-1}$ is consistent with the one obtained in Ref.~\cite{Sakellariadou:2004wq}. 
}

By combining Eqs.~(\ref{eq:drhodt}) and (\ref{eq:drhodt2}), and rewrite $\rho_\infty$ in temrs of $L$, we obtain the evolution equations such as 
\beq
\label{eq:VTS1}
 &2 \frac{dL}{dt} = 2 H L \lmk 1 + \bar{v}^2 \rmk 
 + P_{\rm eff} \tilde{c} \bar{v} \lmk \frac{\xi}{L} \rmk, 
 \\
\label{eq:VTS2}
 &\frac{d\bar{v}}{dt} = \lmk 1 - \bar{v}^2 \rmk \lmk \frac{k(\bar{v})}{\xi} - 2 H \bar{v} \rmk, 
 \\
 &\xi = c_\xi t, 
\eeq
with $c_\xi = \mathcal{O}(1)$ that is specified below. 
Assuming that $H = r/t$ and $a \propto t^r$, where $r = 1/2$ in RD and $r=2/3$ in MD, 
we obtain 
the scaling solution as follows:
\beq
 &L_{\rm asym} = c_L t, \\
 &c_L \equiv 
 \sqrt{P_{\rm eff} \tilde{c}  \frac{k(\bar{v}_{\rm asym})}{4r (1-r(1+\bar{v}_{\rm asym}^2))}}, 
 \\
 &\bar{v}_{\rm asym} = \frac{k(\bar{v}_{\rm asym})}{2rc_\xi}. 
\eeq
The energy density of long strings is thus proportional to $P_{\rm eff}^{-1}$ in RD, MD, as well as in the flat spacetime ($r=0$). This is consistent with the analytic argument and numerical simulations in flat spacetime of Refs.~\cite{Sakellariadou:1990nd, Sakellariadou:2004wq}. 
Further, the result of the extended VOS model well fits the numerical simulations in RD and MD of Ref.~\cite{Avgoustidis:2005nv}, which the conventional dependence $\rho_\infty \propto P_{\rm eff}^{-2/3}$ used in the most literature is based on (see, e.g., Refs.~\cite{Pourtsidou:2010gu,Auclair:2019wcv}). 
The difference in power results from the limited range of parameters in numerical simulations. 
In Ref.~\cite{Avgoustidis:2005nv}, it was shown that the obtained numerical results can be fitted better once they include only a logarithmic correction to $P_{\rm eff}$. 
In our numerical calculations, we use both the extended VOS model and the conventional one for the purpose of comparison.

Here, we specify the $\mathcal{O}(1)$ parameter $c_\xi$. 
We assume 
\beq
 c_\xi = \left. c_L \right\vert_{P_{\rm eff} = 1, \bar{v}_{\rm asym} = \bar{v}_{\rm asym}^{\rm (sVOS)}} 
 = 
 \left\{
 \begin{aligned}
 &0.27 \quad \text{in RD}
 \\
 &0.62 \quad \text{in MD}, 
 \end{aligned}
 \right.
 \label{c_xi}
\eeq
with 
\beq
 \bar{v}_{\rm asym}^{\rm (sVOS)} 
 = 
 \left\{
 \begin{aligned}
 &0.66 \quad \text{in RD}
 \\
 &0.58 \quad \text{in MD}, 
 \end{aligned}
 \right.
\eeq
to reproduce the standard results of the VOS model in the scaling regime for the case of $P_{\rm eff} =1$. The extended VOS model is thus a smooth extension of the VOS model with a smaller $P_{\rm eff}$ and $L$.

It should be noted that the survival probability of sting loops (or the probability that a string loop does not intersect with long strings) is almost unity for a small loop. 
Thus, we can neglect the backreaction from the dynamics of string loops for the above equations except for the loop chopping efficiency parameter. The evolution equation for the energy density of string loops can be solved after deriving the solutions to the above equations.

\subsection{Effect of baryon vertex}
\label{sec:baryonvertex}

We now consider the effect of baryon vertex, which may or may not exist depending on the gauge theory. 
For instance, this effect is apparent for the pure $\SU(N)$ gauge theory. 

If the baryon vertex exists, we expect that $\mathcal{O}(1)$ baryon vertices are produced within a correlation volume at the confinement/deconfinement phase transition. 
If the baryon vertices are not annihilated, 
the number of baryon vertices within a Hubble horizon increases, and the Universe tends to be dominated by cosmic strings that are attached to the baryon vertices. 
However, as discussed in Sec.~\ref{sec:properties}, the annihilation is efficient because of the string tension that connects the baryon vertices. The network should reach the scaling regime wherein the number of baryon vertices within a Hubble horizon is constant in time. 
We thus expect that the number of baryon vertex is at least of order unity within the Hubble horizon. Because each baryon vertex connects $N$ cosmic strings, 
the energy density of those cosmic strings must be larger than the order 
\beq
 \rho_{\rm min} \sim \frac{N \mu H^{-1}}{H^{-3}} \sim \frac{N \mu}{t^2}. 
\eeq
This should be regarded as the lower bound on the energy density of long cosmic strings. 
Because $\rho_\infty \sim P_{\rm eff}^{-1}\mu /t^2 \sim N^2\mu/t^2$ for $P_{\rm eff} \ll 1$ in the extended VOS model, 
we obtain $\rho_{\rm min} \ll \rho_{\infty}$ for a sufficiently large $N$. 
This justifies the aforementioned analysis of the extended VOS model. 
Conversely, 
this suggests that the number of baryon vertex may be of order $N$ within the Hubble horizon because we expect that almost all long cosmic strings are attached by some baryon vertices.

If $N$ is not sufficiently large and $P_{\rm eff}$ is of the order unity, 
$\rho_{\rm min}$ may be comparable to $\rho_{\infty}$. 
Although the extended VOS model is not justified in this case, we expect that it can still help estimate the relevant quantities with a correction of $N$ dependence. 
The dynamics of this system should be similar to the standard NG cosmic strings that can be described by the standard VOS model, except for a correction from the factor of $N$. 
In particular, for the case of $N=2$, the effect of baryon vertex can be neglected, and we can use the standard VOS model. 
The extended VOS model is reduced to the standard VOS model for $P_{\rm eff} \simeq 1$, and thus, we can use the extended VOS model in this case. 
For a larger (but not too large) $N$, 
we use the same result but with a correction to $\rho_\infty$ by a factor of $N/2$. 
This is mainly because of the fact that the total length of cosmic strings attached to a baryon vertex within a Hubble horizon scales as $N/2$. 
This procedure can be used for the case of $N=3,4$. 
In our numerical calculations, we omit this correction for simplicity because it changes the result only by a factor of the order unity.

\section{Gravitational wave signals}
\label{sec:GW}

\subsection{String loop density and GW spectrum}

The gravitational waves are mainly emitted from the string loops. 
Once we obtain the solution to $L(t)$ and $\bar{v}(t)$, we can derive the energy density of string loops by solving the evolution equation of energy density of string loops. 
The evolution equation can be read from Eqs.~(\ref{eq:drhodt}) and (\ref{eq:dvdt}) with $R \ll H^{-1}$ and $k(1/\sqrt{2}) = 0$ for small loops: 
\beq
\label{eq:VTS3}
 \dot{\rho}_{\rm loop} (l_i,t) = 
 -3 H \rho_{\rm loop} (l_i,t) 
+ P_{\rm eff} \frac{\rho_\infty(t) \bar{v}_\infty(t) l_i}{L^2(t)} f (l_i,t). 
\eeq
This equation can be clearly explained from the fact that the dynamics of small loops are decoupled from the Hubble expansion except for the dilution of loop number density. The second term in the right-hand side is the source term from chopping the long strings. 
We thus obtain 
\beq
  \rho_{\rm loop} (l_i,t) = 
P_{\rm eff} \mu l_i  \int_{t_i}^{t} dt'
  \lmk \frac{a(t')}{a(t)} \rmk^3 
   \frac{\bar{v}_\infty(t') }{L^4(t')} f (l_i,t'). 
\eeq
To demonstrate our results, we numerically calculate this equation without relying on the scaling assumption. 
For the purpose of illustration, we also derive the scaling solutions, where 
we expect that the loop production function $f(l_i, t')$ scales as $f(l_i, t') = t'^{-1} f(x)$ with $x'= l_i / t'$. 
Then using $a(t')/a(t) = (t'/t)^r$, $L(t') = c_L t'$, and $\bar{v}_\infty = \bar{v}_{\rm asym}$, we can rewrite the solution as 
\beq
\label{eq:scaling}
 &\rho_{\rm loop} (l_i,t) = \frac{P_{\rm eff}c_\xi^4}{c_L^4}
 \frac{\mu \nu }{t^{3r} l_i^{3-3r}}, 
 \\
 &\nu = \bar{v}_{\rm asym} c_\xi^{-4} \int_0^\infty x'^{3(1-r)} f(x') dx', 
 \label{eq:nur}
\eeq
in the scaling regime. 
Since $c_L \propto \sqrt{P_{\rm eff}}$, 
the energy density of string loops is $\propto P_{\rm eff}^{-1}$.

The string loops with length $l$ emit GWs in a discrete set of frequencies $f_n = 2 n / l$ that are associated with harmonic modes $n$ ($n=1,2,\dots$) on the string loop. 
The averaged energy of GWs for a mode $n$ emitted per unit time is given by 
\beq
 \frac{dE_n}{dt} = P_n G \mu^2,
 \label{eq:energyloss}
\eeq
where 
the averaged power spectra is given by 
\beq
 P_n = \frac{\Gamma}{\xi(q)} n^{-q}, 
\eeq
Here, 
$\xi(q)$ is the zeta function and 
$\Gamma \approx 50$ is a numerical factor~\cite{Vachaspati:1984gt,Burden:1985md,Garfinkle:1987yw,Blanco-Pillado:2017oxo}. 
The spectral index $q$ is given by $q = 4/3 ,5/3, 2$ for 
GWs from cusps, kinks, and kink collisions, respectively. 
Hereafter, we assume that the GWs are dominantly produced by cusps and take $q = 4/3$. 
The loop length is thus reduced by emitting GWs such as 
\beq
 \dot{l} = - \Gamma G \mu. 
\eeq
The solution to this equation is $l = l_i - \Gamma G \mu (t-t_i)$, where $l_i$ is the length of loop produced at a time $t_i$. 
By substituting this into the solution of $\rho_{\rm loop}$, we obtain the number density of string loops $n_{\rm loop}$ as follows: 
\beq
 n_{\rm loop}(l,t) dl 
 &= \frac{\rho_{\rm loop} (l_i, t)}{\mu l_i} d l_i, 
 \\
 \label{eq:scalingsolution}
 &= 
  \frac{P_{\rm eff}c_\xi^4}{c_L^4}
 \frac{\nu }{t^{3r} (l+\Gamma G \mu (t-t_i))^{4-3r}} d l_i,
\eeq
where we 
use the scaling solution of \eq{eq:scaling} in the second line. 
A similar result is obtained from the 
numerical simulations in the standard NG strings~\cite{Blanco-Pillado:2011egf,Blanco-Pillado:2013qja}.

Using the number density of cosmic string loops, we can estimate the GW spectrum emitted from those cosmic strings. 
The total energy density of GWs per unit physical frequency $f$ can be calculated using 
\beq
\label{eq:power1}
 \frac{d \rho_{\rm GW}}{df} (t) 
 = 
 \int_{t_i}^t dt' \lmk \frac{a(t')}{a(t)} \rmk^3 
 \int_0^l dl \, n_{\rm loop}(l,t')
 \, h \lmk f \frac{a(t)}{a(t')},l \rmk, 
\eeq
where $t_i$ is the formation time of cosmic strings, i.e., the time of the phase transition. 
The spectrum of GWs emitted by a loop $h(f)$ is given as 
\beq
 h(f,l) = G \mu^2 \sum_{n=1}^\infty P_n \delta (f - f_n(l)), 
\eeq
from \eq{eq:energyloss}. 
The present GW spectrum is thus given by 
\beq
 \Omega_{\rm gw} (f) 
 &\equiv 
 \frac{8 \pi G}{3 H_0^2}
  \frac{d \rho_{\rm GW}}{d \ln f}
 \\
 &= \frac{8\pi G^2}{3 H_0^2} 
 \mu^2 f \sum_{n=1}^\infty c_n (f) P_n,
 \label{eq:Omegagw}
\eeq
with 
\beq
 c_n (f) = \frac{2n}{f^2} 
 \int_0^{z_i} \frac{dz}{H(z) (1+z)^6} 
 n_{\rm loop} \lmk \frac{2n}{(1+z)f}, t'(z) \rmk,
\eeq
where $H_0$ is the Hubble parameter at present and $z$ ($=a(t_0)/a(t')-1$) is the redshift. 
The infinite summation in \eq{eq:Omegagw} is convergent but makes a logarithmic contribution up to $n \approx 10^6$~\cite{Sanidas:2012ee,Sousa:2013aaa}. 
In our numerical calculations, we directly evaluate the summation up to $n=10^2$ and approximate the part of $n >10^2$ by integral with an interpolating function from $n = 10^2$ to $10^{10}$.

Further, we need the explicit form of the loop production function $f(l_i/t)$ to calculate the number density of loops from \eq{eq:VTS3} 
(or to calculate $\nu_r$ from \eq{eq:nur} in the scaling regime). 
Based on the conventional practice in the literature, 
we assume that the loop production function is monocromatic function as 
\beq
 f(x) = \frac{\tilde{c} }{\alpha_L} \delta (x - \alpha_L c_\xi), 
 \label{eq:fx}
\eeq
where $x \equiv l_i /t$ and the overall factor follows from \eq{eq:tildec}. 
We also define $\alpha \equiv \alpha_L c_\xi$. 
Moreover, we should add a couple of phenomenological factors in the overall factor. 
The string loops are produced using a nonzero center-of-mass energy, which will decrease with the redshift. This reduces the total energy of string loops by a factor of $1/\sqrt{2}$~\cite{Vilenkin:2000jqa}. 
In addition, the monocromatic function does not represent the actual distribution of string loops. 
It is shown that the effect of its finite width can be effectively incorporated by reducing the energy density by a factor of $\mathcal{F} = \mathcal{O}(0.1)$~\cite{Sanidas:2012ee,Blanco-Pillado:2013qja}. 
We thus use 
\beq
f(x) 
= \frac{\mathcal{F}}{f_r} \frac{\tilde{c} c_\xi }{\alpha} \delta (x - \alpha),
\eeq
with $f_r = \sqrt{2}$. 
In the scaling regime, this gives 
\beq
 \nu = \frac{\mathcal{F}}{f_r} \frac{\tilde{c} }{c_\xi^3} \bar{v}_{\rm asym}  
 \alpha^{2-3r}, 
\eeq
from \eq{eq:nur}. 
We take $\mathcal{F} = 0.1$ in our numerical calculation.

It can be considered that $\alpha$ is at most $L/t$ ($= c_L \propto P_{\rm eff}^{1/2}$ in the scaling regime). 
This (or correspondingly $\alpha \sim 0.3 P_{\rm eff}^\gamma$) is often assumed in the literature~\cite{Sousa:2016ggw,Auclair:2019wcv}. 
However, we expect that $\alpha$ should scale such as $\xi/t$. 
As we previously commented in the footnote~\ref{footnote:loop}, 
the loop production process is implemented in two steps. The bottleneck process is the creation of wiggly structure based on an intercommutation between different strings. 
Because the correlation length or a typical curvature of each long string is of the order $\xi$, we expect that the loop size is proportional to $\xi$ rather than $L$ 
even if the effective intercommutation probability is significantly smaller than unity. 
The value of $\alpha$ that we should consider is under debate even for the simplest model of cosmic string. 
In this paper, we adopt the common assumption of $\alpha =0.1$, which is 
confirmed by the Nambu-Goto simulations with $P_{\rm eff} =1$~\cite{Blanco-Pillado:2013qja}. 
The result of \eq{eq:scalingsolution} is in good agreement with the numerical simulations for $P_{\rm eff} =1$ with this choice~\cite{Auclair:2019wcv}.

To calculate the GW signals, 
we numerically solve Eqs.~(\ref{eq:VTS1}), (\ref{eq:VTS2}), (\ref{eq:VTS3}), and (\ref{eq:Omegagw}) with 
\beq
 &\frac{1}{a} \frac{da}{dt} = 
 H_0 
 \lmk \Omega_\Lambda
 + \Omega_m (1+z)^3 
 + \Omega_{\rm rad} \mathcal{G}(z) (1+z)^4 
 \rmk^{1/2}, 
 \\
 &\mathcal{G}(z) \equiv 
 \frac{g_*(z) g_s^{4/3}(0)}{g_*(0) g_s^{4/3}(z)},
\eeq
where 
$\Omega_\Lambda = 0.685$, 
$\Omega_m = 0.315$~\cite{Planck:2018vyg}, 
and $g_*(z)$ ($g_s(z)$) 
is the effective number of relativistic degrees of freedom for the energy (entropy) density. 
The $\mathcal{O}(1)$ parameter $c_\xi$ is given by \eq{c_xi} for $r = 1/2$ ($2/3$) in RD (MD) with an interpolating function in the period between them. 

Before demonstrating our numerical results, we analytically calculate the amplitude of GW signals at a high frequency. 
For a very high frequency, the GWs are mainly emitted from string loops that are produced in the RD. 
The GW amplitude in this regime can be analytically calculated using the scaling solution. 
By changing the time variable from $t$ (or $z$) to $\tilde{x} = l(t) / t$ with $l(t) = 2n / (1+z(t))f$, we can rewrite $dz/H(1+z) = -dt = 2 l(x) d\tilde{x}/\tilde{x}^2$. Based on the scaling solution of \eq{eq:scalingsolution} and $(a/a_0) = (2t H_0 \sqrt{\mathcal{G}(z)\Omega_{\rm rad}})^{1/2}$  
in the RD, 
we obtain 
\beq
\label{eq:scalingresult}
 \lmk \Omega_{\rm gw} h^2 (f) \rmk^{(\rm scaling)}
 &\simeq 
 \frac{128 \pi}{9} \Omega_{\rm rad} h^2 g(f)
 \frac{P_{\rm eff}c_\xi^4 \nu}{c_L^4}
 \sqrt{\frac{G \mu}{\Gamma}} , 
 \\
 &\simeq 5.5 \times 10^{-11}
 P_{\rm eff}^{-1} \sqrt{\alpha} \lmk \frac{g(f)}{0.39} \rmk \lmk \frac{G\mu}{10^{-12}} \rmk^{1/2} 
\eeq
for $\alpha \gg \Gamma G \mu$ and a large $f$ ($\gg f_{\rm eq}$), where $h = H_0/(100\, {\rm km/sec/Mpc})$ ($\simeq 0.674$) is the reduced Hubble parameter. 
Here, $f_{\rm eq}$ is defined by the frequency at which the corresponding loop length $l$ is equal to $\Gamma G \mu t$ for the mode $n=1$ at the matter-radiation equality: 
\beq
 f_{\rm eq} 
 &\equiv \frac{1}{\Gamma G \mu t_{\rm eq} (1+z_{\rm eq})}
 \\
 &\simeq 3.7 \times 10^{-6} \, {\rm Hz} 
 \lmk \frac{G \mu}{10^{-12}} \rmk^{-1}, 
\eeq
where $t_{\rm eq}$ and $z_{\rm eq} = \Omega_{\rm m} / \Omega_{\rm rad}$ are the time and the redshift at the matter-radiation equality, respectively. 
In \eq{eq:scalingresult}, 
we include a function $g(f)$ that represents the effect of change of relativistic degrees of freedom. 
This effect further reduces the GW amplitude by a factor of  
\beq
\label{eq:g}
 g(f) = \mathcal{G}^{3/4}(z_{\rm loop}(f)) \mathcal{G}^{1/4}(z_{\rm GW}(f)), 
\eeq
where $z_{\rm loop}$ and $z_{\rm GW}$ are determined by 
\beq
 &t(z_{\rm loop}(f)) = t_{\rm eq} \frac{\Gamma G \mu}{\alpha} \lmk \frac{f}{f_{\rm eq}} \rmk^{-2}, 
 \\
 &t(z_{\rm GW}(f)) = t_{\rm eq} \lmk \frac{f}{f_{\rm eq}} \rmk^{-2}. 
\eeq
These times are related to the formation time of string loops that mainly contribute the GW signal for a frequency $f$ 
and the time at the GW emission, respectively. 
The first factor in \eq{eq:g} comes from the dilution of the number of string loops by the entropy dilution from the relativistic degrees of freedom in \eq{eq:scaling}~\cite{Blanco-Pillado:2017oxo}. 
The second factor is derived from the redshift of GWs with the effect of the change of the relativistic degrees of freedom~\cite{Battye:1997ji}.%
\footnote{
Those contributions are highlighted and considered in the literature~\cite{Battye:1997ji,Blanco-Pillado:2017oxo,Cui:2018rwi,Auclair:2019wcv} with some confusions or without any specific clarity on the frequency dependence. 
In some instances, $g(f) \simeq \mathcal{G}$ is used in the limit of large $f$. To consider the observable effect of $g(f)$ in the spectrum by new degrees of freedom, the frequency dependence of \eq{eq:g} should be taken into consideration, which does not seem to be fully discussed analytically in the literature. 
}
It can be inferred that $z_{\rm loop}(f)$ is before the BBN epoch and after the electroweak phase transition in some parameters of interest.

The result of \eq{eq:scalingresult} 
is correct only for $\alpha \gg \Gamma G \mu$, which includes the case with $\alpha = 0.1$. 
The GW amplitude during the scaling regime in RD becomes independent of $\alpha$ for a smaller $\alpha$ than $\Gamma G \mu$. 
Those behaviors are similar to the case with $P = 1$, which one can refer to, e.g., Ref.~\cite{Auclair:2019wcv}.

\subsection{Numerical results}

\begin{figure}[t]
\centering
\includegraphics[clip, width=14cm]{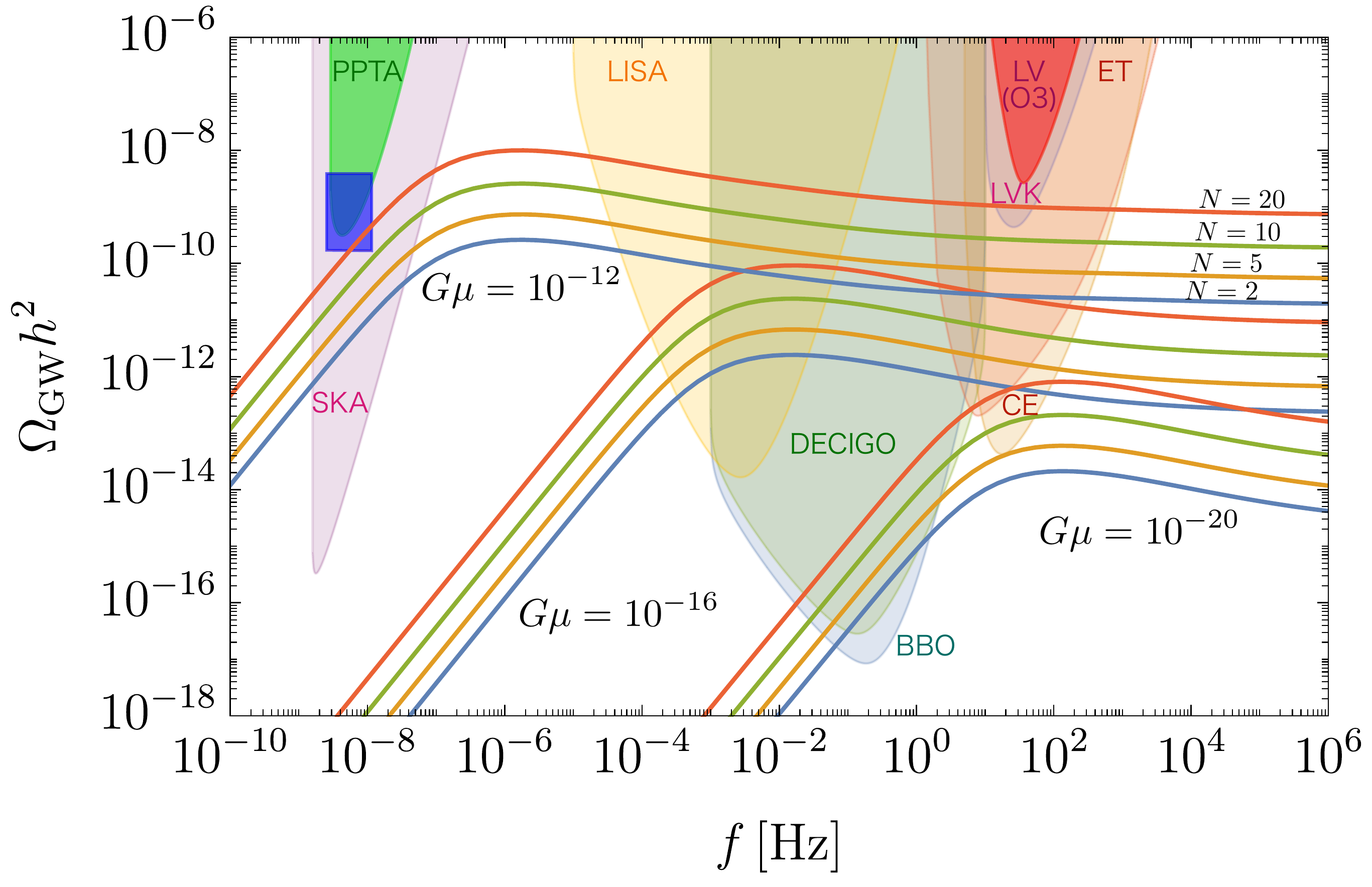}
\caption{GW spectra from cosmic string loops calculated using extended VOS model
for the case of $G\mu =  10^{-12}, 10^{-16}, 10^{-20},$ with 
$N=2,5,10,20$. We use the relation $P = 1/N^2$ for F-strings. 
Green and red shaded regions are excluded by PPTA~\cite{Shannon:2015ect} and aLIGO/aVirgo~\cite{KAGRA:2021kbb}, respectively. 
Dense blue region highlights the potential signals by NANOGrav~\cite{NANOGrav:2020bcs} and PPTA~\cite{Goncharov:2021oub}. 
The lightly shaded regions represent the future sensitivities of GW experiments. 
\label{fig:spectrum}
}
\end{figure}

We numerically solve the extended VOS equation and calculate the GW spectrum without assuming scaling solution. 
The result is shown in Fig.~\ref{fig:spectrum} 
for the case of $G\mu = 10^{-12}$, $10^{-16}$, and $10^{-20}$ with 
$N=2,5,10$, and $20$. We assume F-strings with $P = 1/N^2$. 
We consider $\alpha = 0.1$ and $N_{\rm scat} = 10$. 
In particular, the case with $N = 2$ corresponds to $P_{\rm eff} \simeq 0.94$, which is very close to unity. 
This reproduces the standard result for the case of $P_{\rm eff} = 1$ within an $\mathcal{O}(1)$ factor. The resulting spectrum is in good agreement with the one shown in, e.g., Ref.~\cite{Auclair:2019wcv}.

The power-law-integrated sensitivity curves for ongoing and planned GW experiments 
are plotted according to Ref.~\cite{Schmitz:2020syl}, 
including 
SKA~\cite{Janssen:2014dka},
LISA~\cite{LISA:2017pwj},
DECIGO~\cite{Kawamura:2011zz,Kawamura:2020pcg},
BBO~\cite{Harry:2006fi},
Einstein Telescope (ET)~\cite{Punturo:2010zz,Maggiore:2019uih},
Cosmic Explorer (CE)~\cite{Reitze:2019iox},
and aLIGO+aVirgo+KAGRA (LVK)~\cite{Somiya:2011np,KAGRA:2020cvd}.
The current constraint from Parkes Pulsar Timing Array (PPTA)~\cite{Shannon:2015ect} 
and aLIGO/aVirgo's third observing run (LV(O3))~\cite{KAGRA:2021kbb} 
are indicated by the dense green and red shaded regions, respectively. 
The blue box near the bottom of the PPTA constraint highlights the potential signals of pulsar timing array (PTA) experiments, such as NANOGrav~\cite{NANOGrav:2020bcs} and PPTA~\cite{Goncharov:2021oub}.

The GW spectrum shows a peak at a certain frequency, and below this frequency, it scales as $\Omega_{\rm GW} \propto f^{3/2}$. 
The amplitude of GW and frequency at the peak are approximately given by 
\beq
 &\lmk \Omega_{\rm GW} h^2 \rmk^{(\rm peak)} \simeq 2.5 \times 10^{-10} \times P_{\rm eff}^{-1} \lmk \frac{G\mu}{10^{-12}} \rmk^{1/2}, 
 \\
 &f^{(\rm peak)} \simeq 1.9 \times 10^{-6}  \, {\rm Hz}  \times \lmk \frac{G\mu}{10^{-12}} \rmk^{-1}. 
\eeq
It is observed that the dependence on $N$ and $\Lambda$ (or $P_{\rm eff}$ and $G \mu$) is not degenerate around the peak, and thus, we can determine both if we can observe the spectrum around the peak. Fortunately, this is within the sensitivity curve for the GW experiments, such as LISA, for the most parameter region of interest. 
We have also shown that the result of \eq{eq:scalingresult} agrees with our numerical results within a limit of large $f$.

\begin{figure}[t]
\centering
\includegraphics[clip, width=14cm]{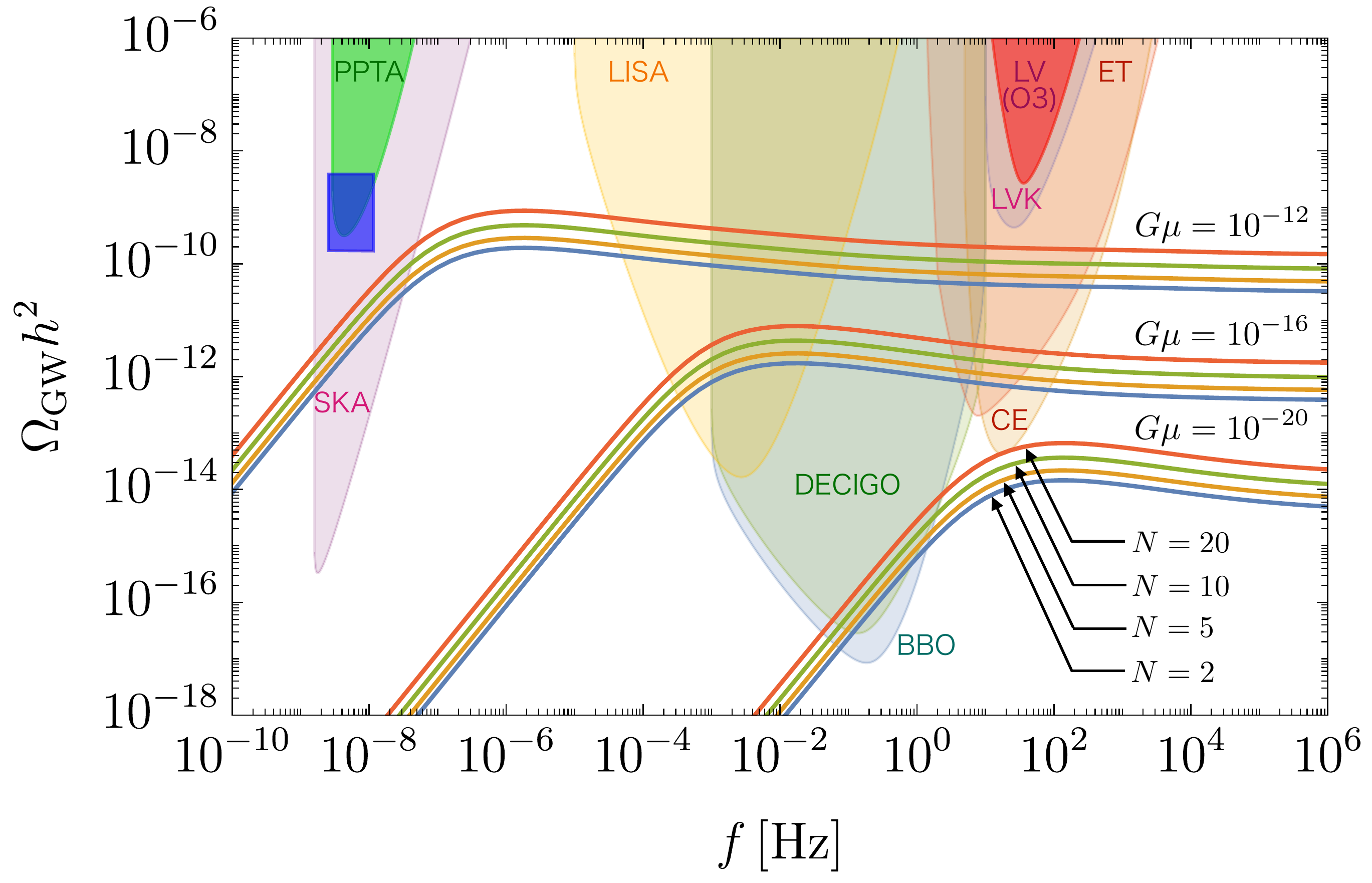}
\caption{Same as Fig.~\ref{fig:spectrum} but with the standard VOS model ($\xi = L$) and the replacement of $\tilde{c} \to \tilde{c} P_{\rm eff}^\gamma$ with $\gamma = 1/3$ and with $\alpha = 0.3 P_{\rm eff}^\gamma$. 
\label{fig:spectrum2}
}
\end{figure}

To compare the standard result for a small intercommutation probability, we also solve the standard VOS model (that is, $\xi = L$) by replacing $\tilde{c} \to \tilde{c} P_{\rm eff}^\gamma$ with $\gamma = 1/3$ and with $\alpha = 0.3 P_{\rm eff}^\gamma$~\cite{Auclair:2019wcv}. The resulting GW spectrum is shown in Fig.~\ref{fig:spectrum2} for the same parameters with Fig.~\ref{fig:spectrum}. 
The effect of small intercommutation probability is milder than the case with the extended VOS model because in this case, $(\Omega_{\rm gw} h^2)^{\rm (scaling)} \propto P_{\rm eff}^{-2\gamma} \alpha^{1/2} \propto P_{\rm eff}^{-1/2}$ in the scaling regime.

\begin{figure}[t]
\centering
\includegraphics[clip, width=14cm]{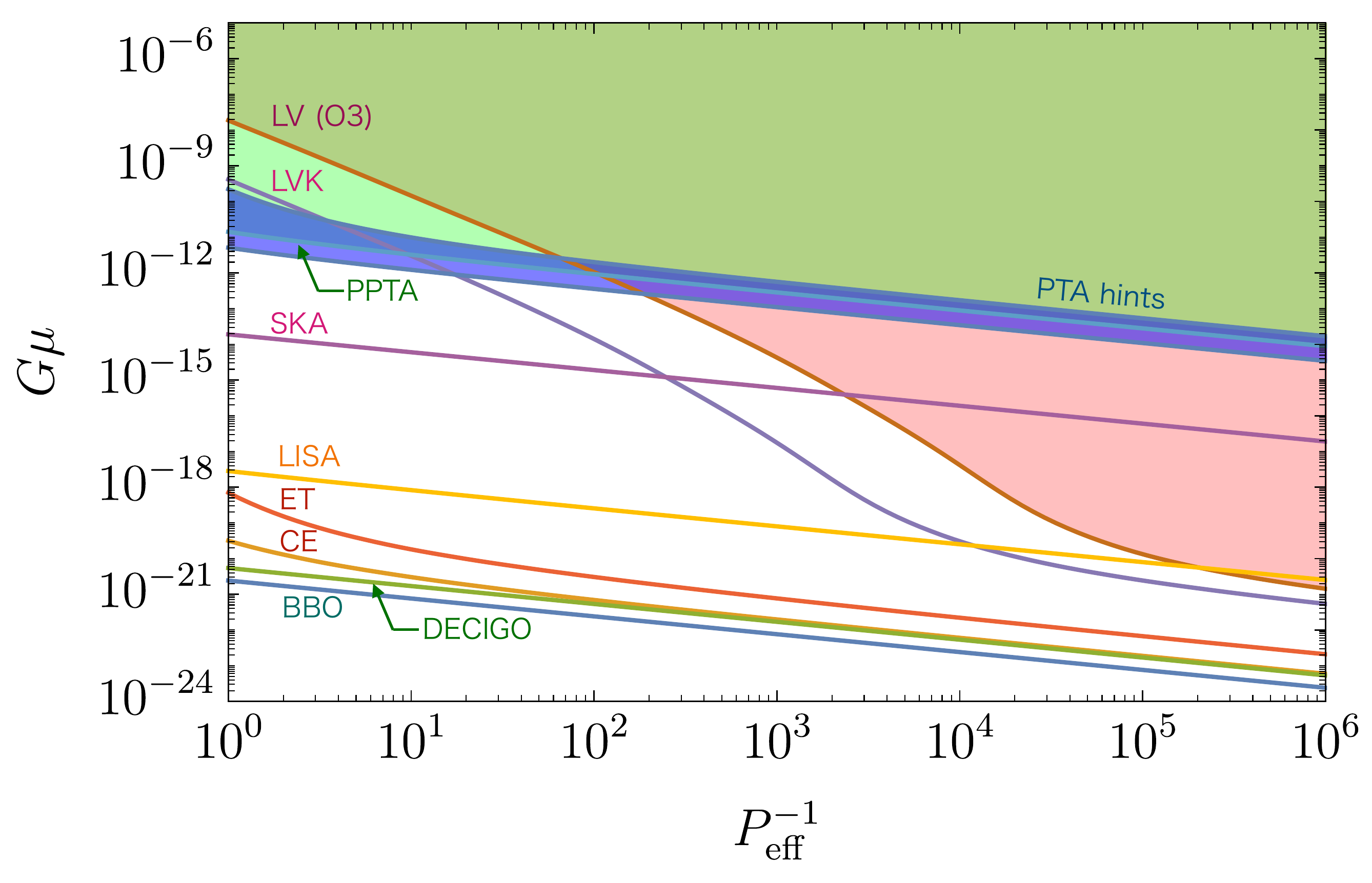}
\caption{Exclusion plot and future sensitivity curves in $G\mu$\,-\,$P_{\rm eff}^{-1}$ plane for ongoing and planned GW experiments. Green and red shaded regions are excluded by PPTA~\cite{Shannon:2015ect} and aLIGO/aVirgo~\cite{KAGRA:2021kbb}, respectively. 
Dense blue region highlights the potential signals by NANOGrav~\cite{NANOGrav:2020bcs} and PPTA~\cite{Goncharov:2021oub}.  
\label{fig:constraint1}
}
\end{figure}

\begin{figure}[t]
\centering
\includegraphics[clip, width=14cm]{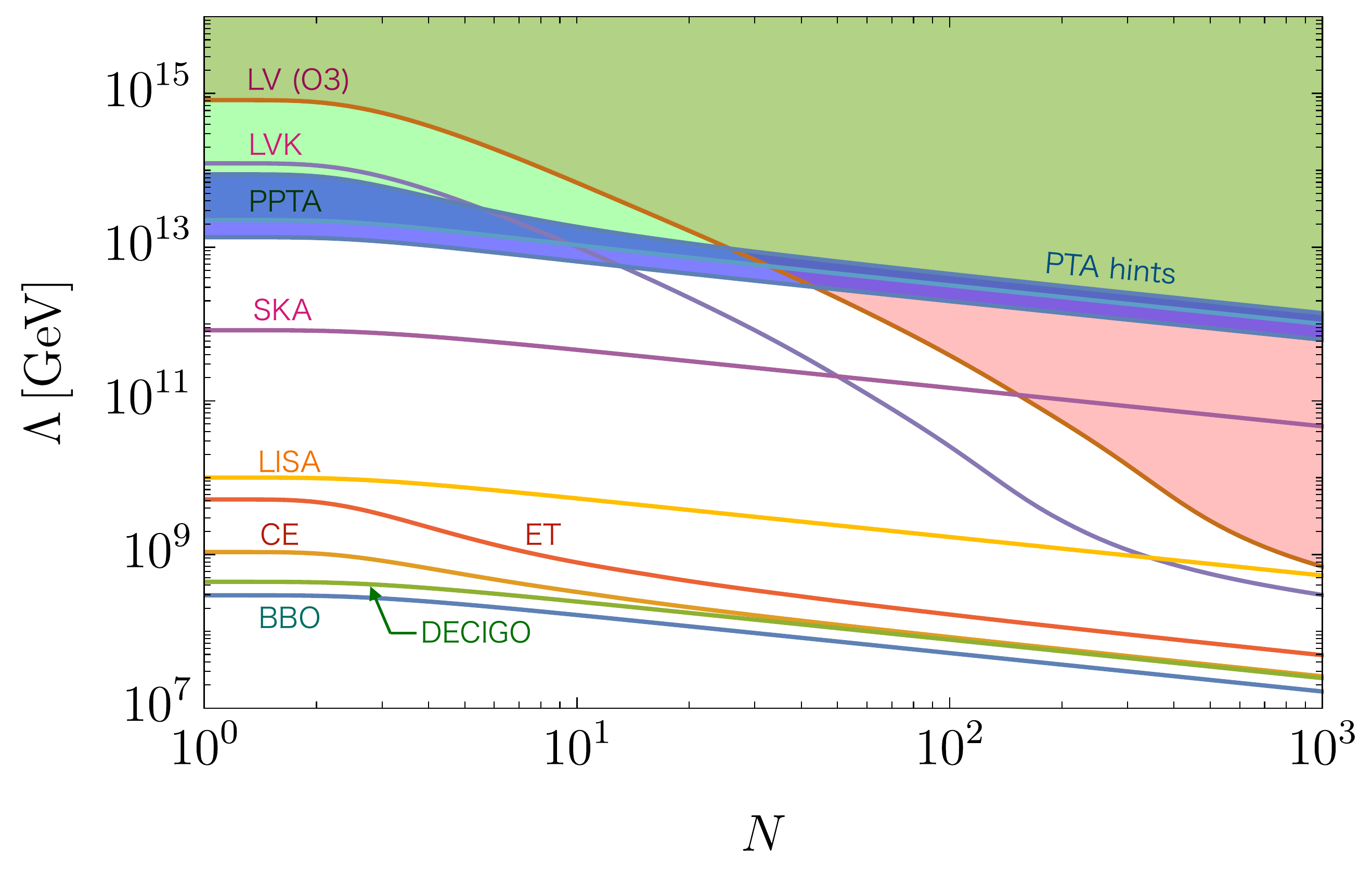}
\caption{Same as Fig.~\ref{fig:constraint1} but in $\Lambda$\,-\,$N$ plane for F-strings, wherein we 
treat $N$ as a continuous variable via the relation of $P_{\rm eff} = 1-(1-P)^{N_{\rm scat}}$ with $P = 1/N^2$ and $N_{\rm scat} = 10$, although it must be an integer in a realistic model. 
\label{fig:constraint2}
}
\end{figure}

We show the constraints and sensitivity curves in terms of $G \mu$ and $P_{\rm eff}^{-1}$ in Fig.~\ref{fig:constraint1} using the extended VOS model. Each curve represents the lower bound on the parameter space wherein the predicted $\Omega_{\rm GW}h^2$ is greater than the sensitivity curve at a specific frequency. 
The green and red shaded regions are excluded by PPTA~\cite{Shannon:2015ect} and aLIGO/aVirgo~\cite{KAGRA:2021kbb}, respectively. 
The dense blue region highlights the potential signals by NANOGrav~\cite{NANOGrav:2020bcs} and PPTA~\cite{Goncharov:2021oub}. 
The constraint on the string tension can be considerably strong for a large $P_{\rm eff}^{-1}$ by the ground-based GW experiments, such as aLIGO, aVirgo, and KAGRA. 
These experiments can search for GW signals from the scaling regime in the RD, wherein the spectrum scales as $\Omega_{\rm GW} h^2 \propto \sqrt{G\mu} / P_{\rm eff}$ (see \eq{eq:scalingresult}). 
On the contrary, 
the sensitivity curves do not depend significantly on $P_{\rm eff}^{-1}$ for other experiments. 
This is because the peak frequency of the GW spectrum is too high to be detected by the latter experiments at a small $G \mu$, and in such cases, the GW signals decrease as $\Omega_{\rm GW} h^2 \propto (G\mu)^2 / P_{\rm eff}$ for $f \ll f^{\rm (peak)}$. 
If we consider $P_{\rm eff} \lesssim 0.1$, 
we can expect signals for aLIGO, aVirgo, and KAGRA 
consistently with the constraint from the PPTA. This is a unique feature of our model; the amplitude of GW can be enhanced by $P_{\rm eff}^{-1}$ without shifting its peak frequency. 

We rewrite the sensitivities for $G \mu$ and $P_{\rm eff}^{-1}$ 
in terms of $\Lambda$ and $N$ 
by using the relations Eqs.~(\ref{eq:tension}), (\ref{eq:P}), and (\ref{eq:Peff}) for F-strings. 
Here we take $P = 1/N^2$ though it has an $\mathcal{O}(1)$ uncertainty. 
The resulting sensitivity curves are shown in Fig.~\ref{fig:constraint2}. 
Here, we treat $N$ as a continuous variable for the purpose of representation, although it must be an integer. 
We find that the present constraint puts an upper bound on the confinement scale as $\Lambda \lesssim 2 \times 10^{13} \GeV$ for $N = \mathcal{O}(1)$. 
The NANOGrav and PPTA hints favor the parameter near the threshold of this constraint. 
There have been several studies on the GW signals from cosmic strings in this context~\cite{Ellis:2020ena,Blasi:2020mfx,Blanco-Pillado:2021ygr}. 
We find that our cosmic strings could also explain the PTA hints 
when $G \mu \sim 10^{-10\,\text{-}\,12}$ 
or 
$\Lambda \sim 10^{13\,\text{-}\,14} \GeV$ for $N = \mathcal{O}(1)$. 
Here, it should be noted that 
\beq
 G \mu \simeq 2.8 \times 10^{-14} \lmk \frac{\Lambda}{10^{12} \GeV} \rmk^2, 
\eeq
where we use the lattice result of $\mu \simeq 4 \Lambda^2$ for $\SU(N)$~\cite{Athenodorou:2021qvs}.

The ongoing and planned GW experiments can search for the signals up to $\Lambda \gtrsim \mathcal{O}(10^8) \GeV$. 
A large parameter space, spanning over five orders of magnitude, can be searched through the GW experiments. 
It can be inferred that such an intermediate scale for $\Lambda$ is naturally realized mainly because of the dimensional transmutation in the pure YM theory. 

Finally, we discuss the constraint on cosmic strings from the CMB temperature and polarization data. 
The energy density of cosmic strings should be lower than the order 1\% for the total temperature anisotoropy. 
From the detailed numerical simulations, the Planck result puts an upper bound on the string tension as $G \mu < 1.1 \times 10^{-7}$ for the ordinary Nambu-Goto cosmic strings~\cite{Charnock:2016nzm}. 
The constraint should be stronger for strings with a small intercommutation probability because the energy density is $\rho_\infty \propto P_{\rm eff}^{-1}$. 
However, even if we include this enhancement factor, 
the constraint from CMB observations is much weaker than the present constraint evaluated through the GW experiments.

\section{Discussion and conclusions}
\label{sec:conclusion}

We have discussed that 
cosmic strings 
or macroscopic color flux tubes 
form at the deconfinement/confinement phase transition in pure YM theory. 
Depending on the structure of gauge group, 
these strings can be understood as fundamental (F-) strings and wrapped D-brane (which are referred to as D-strings) 
in the holographic dual descriptions, and possess a small intercommutation probability of $1/N^2$ and $e^{-\mathcal{O}(N)}$, respectively. 
We have explained that the cosmic strings have 
discrete 1-form symmetry, which further ensures
stability. 
The 1-form symmetry also implies that $N$ cosmic strings can intersect at a baryon vertex for the case of $\SU(N)$. 
The network of the cosmic strings thus possess rich properties in the pure YM theory even if we do not assume brane inflationary scenario or extra dimensions. 
We have further discussed the types of cosmic strings formed in $\SU(N)$, $\Sp(N)$, $\SO(N)$, and $\Spin(N)$ YM theory, and we have explained the implications from electric--magnetic duality, large $N$ limit, as well as holographic dual descriptions.

We have considered an extended VOS model to take into account the small intercommutation probability and calculate the GW spectrum emitted from cosmic string loops. 
The GW signals can be observed through ongoing and planned GW experiments. Particularly, the SKA and LISA can observe GW signals if the confinement scale $\Lambda$ is higher than $\mathcal{O}(10^{12}) \GeV$ and $\mathcal{O}(10^{10}) \GeV$ for $N = \mathcal{O}(1)$, respectively. 
The DECIGO, BBO, CE, and ET could observe GW signals for $\Lambda \gtrsim \mathcal{O}(10^8) \GeV$. 
The recently reported NANOGrav and PPTA hints favor $\Lambda = \mathcal{O}(10^{13})\GeV$.

Some assumptions should be confirmed through numerical simulations. 
The effect of small intercommutation probability is based on a numerical simulation of Ref.~\cite{Avgoustidis:2005nv}. 
Although a common procedure used in the literature is the replacement of $\tilde{c} \to \tilde{c} P_{\rm eff}^\gamma$ with $\gamma = 1/3$, 
the extended VOS model used in this paper and the aforementioned paper fits in well with the numerical result, and it seems to be 
physically reasonable. 
These procedures cannot be distinguished by the numerical results of the aforementioned paper, and thus, a numerical calculation with a larger simulation box is necessary. 
Moreover, as far as we know, there is no simulation for the network with baryon (or monopole) vertices that connect more than three cosmic strings. 
The formation of baryon vertices at the phase transition should also be confirmed through numerical simulations if the theories with the baryon vertices, such as $\SU(N)$, are considered. We expect that those properties can be understood by field-theory numerical simulations for a model of $\bZ_N$ strings, where $\SU(N)$ symmetry is spontaneously broken to $\bZ_N$ symmetry by $N$ adjoint Higgs fields. Another field-theory model is given in \eq{eq:EMdual} based on the electric--magnetic duality.

These properties, namely the small intercommutation probability and existence of baryon vertices, 
are also realized by F-strings that form after the D-brane inflation. 
In the brane inflationary scenarios, 
bulk modes like radions and dilatons play essential roles and the extra dimensions are necessary. 
Our work provides another motivation to study the consequence of cosmic strings with the small intercommutation probability in a much simpler but phenomenologically and cosmologically interesting setup. 
Particularly, we do not need to consider extra dimensions for the UV completion.
Moreover, the intercommutation probability is determined by $N$ in our case, 
and it is observed that the string tension is naturally small and does not require fine tuning by the dimensional transmutation. 
The GW signals are within the scope of future GW observations, such as SKA and LISA. Once the GW spectrum is observed, we can determine the confinement scale $\Lambda$ as well as $N$ for the pure YM theory.

We consider the case wherein the deconfinement/confinement phase transition occurs after inflation. 
Through lattice simulations of the $\SU(N)$ pure YM theory, the confinement phase transition is a second-order phase transition for $N = 2$ and a first-order transition for $N \geq 3$~\cite{Iwasaki:1992ik,Boyd:1996bx,Lucini:2003zr,Datta:2009jn,Panero:2009tv}.\footnote{
The first order phase transition is also expected in the large $N$ limit in holographic dualities~\cite{Witten:1998zw}.
(See also Refs.~\cite{Aharony:2006da,Mandal:2011ws,Isono:2015uda} for large $N$ QCD.)
't~Hooft anomaly matching also gives some implications for the nature of phase transition
for pure $\SU(2)$~\cite{Gaiotto:2017yup}, $\SU(N)$ YM with adjoint fermions~\cite{Komargodski:2017smk,Shimizu:2017asf},
and some QCD-like theories (e.g. \cite{Tanizaki:2017qhf,Tanizaki:2017mtm,Yonekura:2019vyz}).
}
The GWs can also be produced from the collision of nucleated bubbled at the phase transition (see, e.g., Ref.~\cite{Reichert:2021cvs}). This is another source of GW signals in the pure YM theory. However, the typical frequency is too high to be detected by GW experiments when the confinement scale is higher than $\mathcal{O}(10^8)\GeV$. Thus, we can neglect it in our parameters of interest. 
The glueballs are also produced at the phase transition. 
Because the glueballs are singlet in the low-energy effective field theory, they should decay into other particles via higher-dimensional operators. 
In the presence of a light modulus, 
the glueball decay rate is estimated as $\Lambda^6/(4 \pi m_g M^4)$, where $m_g$ ($\sim \Lambda$) is the glueball mass and $M$ is a cutoff scale~\cite{Halverson:2016nfq}. 
Because we focus on the case with $\Lambda \gtrsim 10^{8} \GeV$, 
the glueballs are expected to completely decay before the BBN epoch for $M = 10^{16} \GeV$. The decay into axions is also efficient if the axion decay constant is small and/or the glueball mass is large. 
A recent work~\cite{Asadi:2022vkc} discussed that 
the glueball can be long-lived in a model of ``thermal squeezeout" and it dilutes cosmological relics, such as dark matter, by its late-time decay. 
The early matter-dominated epoch by the glueball domination may modify the GW spectrum at a high frequency even for a large dynamical scale.

Finally, we comment on the reheating after inflation. 
The phase transition need not necessarily occur after the reheating completes. 
The GWs for most frequencies of interest are emitted mainly from the string loops that are generated after the electroweak phase transition. 
The number density of string loops are determined using the scaling solution, which lose information related to the phase transition. 
Thus, our results can be applied for the cases 
wherein the cosmic strings reaches the scaling solution well before the electroweak phase transition. 
In particular, the phase transition can happen before the reheating is completed; for instance during the inflation-oscillation dominated era. 
Our scenario can be applied to the case where the maximal temperature of the Universe (which is usually much higher than the reheating temperature) and/or the Hubble parameter during inflation is higher than the confinement scale $\Lambda$. 
It is also observed that the scaling solution of cosmic strings is determined using the Hubble parameter, and it is independent of the temperature of the gauge sector (such as glueballs). 
Thus, our results do not change even if the temperature of the gauge sector is different from that of the SM sector.

\section*{Acknowledgments}
The authors would like to thank Yuya Tanizaki for intensive lectures at Tohoku University which inspired this work. 
MY thanks Alexander Vilenkin for valuable comments on metastable cosmic strings. 
MY also thanks Ken D. Olum and Jose J. Blanco-Pillado for useful discussions. 
MY was supported by MEXT Leading Initiative for Excellent Young Researchers, and by JSPS KAKENHI Grant No.\ 20H0585 and 21K13910. The work of KY is supported in part by JST FOREST Program (Grant Number JPMJFR2030, Japan), 
MEXT-JSPS Grant-in-Aid for Transformative Research Areas (A) ”Extreme Universe” (No. 21H05188),
and JSPS KAKENHI (17K14265).

\bibliography{ref}

\providecommand{\href}[2]{#2}\begingroup\raggedright\begin{thebibliography}{100}

\bibitem{Okun:1979tgr}
L.~B. Okun, \emph{{THETONS}}, {\emph{Pisma Zh. Eksp. Teor. Fiz.} {\bfseries 31}
  (1979) 156}.

\bibitem{Okun:1980mu}
L.~B. Okun, \emph{{THETA PARTICLES}},
  \href{https://doi.org/10.1016/0550-3213(80)90439-3}{\emph{Nucl. Phys. B}
  {\bfseries 173} (1980) 1}.

\bibitem{Strassler:2006im}
M.~J. Strassler and K.~M. Zurek, \emph{{Echoes of a hidden valley at hadron
  colliders}},
  \href{https://doi.org/10.1016/j.physletb.2007.06.055}{\emph{Phys. Lett. B}
  {\bfseries 651} (2007) 374}
  [\href{https://arxiv.org/abs/hep-ph/0604261}{{\ttfamily hep-ph/0604261}}].

\bibitem{Kang:2008ea}
J.~Kang and M.~A. Luty, \emph{{Macroscopic Strings and 'Quirks' at Colliders}},
  \href{https://doi.org/10.1088/1126-6708/2009/11/065}{\emph{JHEP} {\bfseries
  11} (2009) 065} [\href{https://arxiv.org/abs/0805.4642}{{\ttfamily
  0805.4642}}].

\bibitem{Morningstar:1999rf}
C.~J. Morningstar and M.~J. Peardon, \emph{{The Glueball spectrum from an
  anisotropic lattice study}},
  \href{https://doi.org/10.1103/PhysRevD.60.034509}{\emph{Phys. Rev. D}
  {\bfseries 60} (1999) 034509}
  [\href{https://arxiv.org/abs/hep-lat/9901004}{{\ttfamily hep-lat/9901004}}].

\bibitem{Lucini:2010nv}
B.~Lucini, A.~Rago and E.~Rinaldi, \emph{{Glueball masses in the large N
  limit}}, \href{https://doi.org/10.1007/JHEP08(2010)119}{\emph{JHEP}
  {\bfseries 08} (2010) 119} [\href{https://arxiv.org/abs/1007.3879}{{\ttfamily
  1007.3879}}].

\bibitem{Curtin:2022tou}
D.~Curtin, C.~Gemmell and C.~B. Verhaaren, \emph{{Simulating Glueball
  Production in $N_f = 0$ QCD}},
  \href{https://arxiv.org/abs/2202.12899}{{\ttfamily 2202.12899}}.

\bibitem{Faraggi:2000pv}
A.~E. Faraggi and M.~Pospelov, \emph{{Selfinteracting dark matter from the
  hidden heterotic string sector}},
  \href{https://doi.org/10.1016/S0927-6505(01)00121-9}{\emph{Astropart. Phys.}
  {\bfseries 16} (2002) 451}
  [\href{https://arxiv.org/abs/hep-ph/0008223}{{\ttfamily hep-ph/0008223}}].

\bibitem{Feng:2011ik}
J.~L. Feng and Y.~Shadmi, \emph{{WIMPless Dark Matter from Non-Abelian Hidden
  Sectors with Anomaly-Mediated Supersymmetry Breaking}},
  \href{https://doi.org/10.1103/PhysRevD.83.095011}{\emph{Phys. Rev. D}
  {\bfseries 83} (2011) 095011}
  [\href{https://arxiv.org/abs/1102.0282}{{\ttfamily 1102.0282}}].

\bibitem{Boddy:2014yra}
K.~K. Boddy, J.~L. Feng, M.~Kaplinghat and T.~M.~P. Tait,
  \emph{{Self-Interacting Dark Matter from a Non-Abelian Hidden Sector}},
  \href{https://doi.org/10.1103/PhysRevD.89.115017}{\emph{Phys. Rev. D}
  {\bfseries 89} (2014) 115017}
  [\href{https://arxiv.org/abs/1402.3629}{{\ttfamily 1402.3629}}].

\bibitem{Boddy:2014qxa}
K.~K. Boddy, J.~L. Feng, M.~Kaplinghat, Y.~Shadmi and T.~M.~P. Tait,
  \emph{{Strongly interacting dark matter: Self-interactions and keV lines}},
  \href{https://doi.org/10.1103/PhysRevD.90.095016}{\emph{Phys. Rev. D}
  {\bfseries 90} (2014) 095016}
  [\href{https://arxiv.org/abs/1408.6532}{{\ttfamily 1408.6532}}].

\bibitem{Soni:2016gzf}
A.~Soni and Y.~Zhang, \emph{{Hidden SU(N) Glueball Dark Matter}},
  \href{https://doi.org/10.1103/PhysRevD.93.115025}{\emph{Phys. Rev. D}
  {\bfseries 93} (2016) 115025}
  [\href{https://arxiv.org/abs/1602.00714}{{\ttfamily 1602.00714}}].

\bibitem{Kribs:2016cew}
G.~D. Kribs and E.~T. Neil, \emph{{Review of strongly-coupled composite dark
  matter models and lattice simulations}},
  \href{https://doi.org/10.1142/S0217751X16430041}{\emph{Int. J. Mod. Phys. A}
  {\bfseries 31} (2016) 1643004}
  [\href{https://arxiv.org/abs/1604.04627}{{\ttfamily 1604.04627}}].

\bibitem{Forestell:2016qhc}
L.~Forestell, D.~E. Morrissey and K.~Sigurdson, \emph{{Non-Abelian Dark Forces
  and the Relic Densities of Dark Glueballs}},
  \href{https://doi.org/10.1103/PhysRevD.95.015032}{\emph{Phys. Rev. D}
  {\bfseries 95} (2017) 015032}
  [\href{https://arxiv.org/abs/1605.08048}{{\ttfamily 1605.08048}}].

\bibitem{Soni:2017nlm}
A.~Soni, H.~Xiao and Y.~Zhang, \emph{{Cosmic selection rule for the glueball
  dark matter relic density}},
  \href{https://doi.org/10.1103/PhysRevD.96.083514}{\emph{Phys. Rev. D}
  {\bfseries 96} (2017) 083514}
  [\href{https://arxiv.org/abs/1704.02347}{{\ttfamily 1704.02347}}].

\bibitem{Forestell:2017wov}
L.~Forestell, D.~E. Morrissey and K.~Sigurdson, \emph{{Cosmological Bounds on
  Non-Abelian Dark Forces}},
  \href{https://doi.org/10.1103/PhysRevD.97.075029}{\emph{Phys. Rev. D}
  {\bfseries 97} (2018) 075029}
  [\href{https://arxiv.org/abs/1710.06447}{{\ttfamily 1710.06447}}].

\bibitem{Jo:2020ggs}
B.~Jo, H.~Kim, H.~D. Kim and C.~S. Shin, \emph{{Exploring the Universe with
  dark light scalars}},
  \href{https://doi.org/10.1103/PhysRevD.103.083528}{\emph{Phys. Rev. D}
  {\bfseries 103} (2021) 083528}
  [\href{https://arxiv.org/abs/2010.10880}{{\ttfamily 2010.10880}}].

\bibitem{Spergel:1999mh}
D.~N. Spergel and P.~J. Steinhardt, \emph{{Observational evidence for
  selfinteracting cold dark matter}},
  \href{https://doi.org/10.1103/PhysRevLett.84.3760}{\emph{Phys. Rev. Lett.}
  {\bfseries 84} (2000) 3760}
  [\href{https://arxiv.org/abs/astro-ph/9909386}{{\ttfamily
  astro-ph/9909386}}].

\bibitem{Weinberg:2013aya}
D.~H. Weinberg, J.~S. Bullock, F.~Governato, R.~Kuzio~de Naray and A.~H.~G.
  Peter, \emph{{Cold dark matter: controversies on small scales}},
  \href{https://doi.org/10.1073/pnas.1308716112}{\emph{Proc. Nat. Acad. Sci.}
  {\bfseries 112} (2015) 12249}
  [\href{https://arxiv.org/abs/1306.0913}{{\ttfamily 1306.0913}}].

\bibitem{Juknevich:2009ji}
J.~E. Juknevich, D.~Melnikov and M.~J. Strassler, \emph{{A Pure-Glue Hidden
  Valley I. States and Decays}},
  \href{https://doi.org/10.1088/1126-6708/2009/07/055}{\emph{JHEP} {\bfseries
  07} (2009) 055} [\href{https://arxiv.org/abs/0903.0883}{{\ttfamily
  0903.0883}}].

\bibitem{Juknevich:2009gg}
J.~E. Juknevich, \emph{{Pure-glue hidden valleys through the Higgs portal}},
  \href{https://doi.org/10.1007/JHEP08(2010)121}{\emph{JHEP} {\bfseries 08}
  (2010) 121} [\href{https://arxiv.org/abs/0911.5616}{{\ttfamily 0911.5616}}].

\bibitem{Halverson:2016nfq}
J.~Halverson, B.~D. Nelson and F.~Ruehle, \emph{{String Theory and the Dark
  Glueball Problem}},
  \href{https://doi.org/10.1103/PhysRevD.95.043527}{\emph{Phys. Rev. D}
  {\bfseries 95} (2017) 043527}
  [\href{https://arxiv.org/abs/1609.02151}{{\ttfamily 1609.02151}}].

\bibitem{Asadi:2022vkc}
P.~Asadi, E.~D. Kramer, E.~Kuflik, T.~R. Slatyer and J.~Smirnov,
  \emph{{Glueballs in a Thermal Squeezeout Model}},
  \href{https://arxiv.org/abs/2203.15813}{{\ttfamily 2203.15813}}.

\bibitem{Reichert:2021cvs}
M.~Reichert, F.~Sannino, Z.-W. Wang and C.~Zhang, \emph{{Dark confinement and
  chiral phase transitions: gravitational waves vs matter representations}},
  \href{https://doi.org/10.1007/JHEP01(2022)003}{\emph{JHEP} {\bfseries 01}
  (2022) 003} [\href{https://arxiv.org/abs/2109.11552}{{\ttfamily
  2109.11552}}].

\bibitem{Gaiotto:2014kfa}
D.~Gaiotto, A.~Kapustin, N.~Seiberg and B.~Willett, \emph{{Generalized Global
  Symmetries}}, \href{https://doi.org/10.1007/JHEP02(2015)172}{\emph{JHEP}
  {\bfseries 02} (2015) 172} [\href{https://arxiv.org/abs/1412.5148}{{\ttfamily
  1412.5148}}].

\bibitem{tHooft:1977nqb}
G.~'t~Hooft, \emph{{On the Phase Transition Towards Permanent Quark
  Confinement}},
  \href{https://doi.org/10.1016/0550-3213(78)90153-0}{\emph{Nucl. Phys. B}
  {\bfseries 138} (1978) 1}.

\bibitem{tHooft:1979rtg}
G.~'t~Hooft, \emph{{A Property of Electric and Magnetic Flux in Nonabelian
  Gauge Theories}},
  \href{https://doi.org/10.1016/0550-3213(79)90595-9}{\emph{Nucl. Phys. B}
  {\bfseries 153} (1979) 141}.

\bibitem{Witten:1985fp}
E.~Witten, \emph{{Cosmic Superstrings}},
  \href{https://doi.org/10.1016/0370-2693(85)90540-4}{\emph{Phys. Lett. B}
  {\bfseries 153} (1985) 243}.

\bibitem{Vachaspati:1986cc}
T.~Vachaspati and A.~Vilenkin, \emph{{Evolution of cosmic networks}},
  \href{https://doi.org/10.1103/PhysRevD.35.1131}{\emph{Phys. Rev. D}
  {\bfseries 35} (1987) 1131}.

\bibitem{Seiberg:1994rs}
N.~Seiberg and E.~Witten, \emph{{Electric - magnetic duality, monopole
  condensation, and confinement in N=2 supersymmetric Yang-Mills theory}},
  \href{https://doi.org/10.1016/0550-3213(94)90124-4}{\emph{Nucl. Phys. B}
  {\bfseries 426} (1994) 19}
  [\href{https://arxiv.org/abs/hep-th/9407087}{{\ttfamily hep-th/9407087}}].

\bibitem{tHooft:1973alw}
G.~'t~Hooft, \emph{{A Planar Diagram Theory for Strong Interactions}},
  \href{https://doi.org/10.1016/0550-3213(74)90154-0}{\emph{Nucl. Phys. B}
  {\bfseries 72} (1974) 461}.

\bibitem{Coleman:1985rnk}
S.~Coleman, \emph{{Aspects of Symmetry}: {Selected Erice Lectures}}. Cambridge
  University Press, Cambridge, U.K., 1985,
  \href{https://doi.org/10.1017/CBO9780511565045}{10.1017/CBO9780511565045}.

\bibitem{Witten:1998zw}
E.~Witten, \emph{{Anti-de Sitter space, thermal phase transition, and
  confinement in gauge theories}},
  \href{https://doi.org/10.4310/ATMP.1998.v2.n3.a3}{\emph{Adv. Theor. Math.
  Phys.} {\bfseries 2} (1998) 505}
  [\href{https://arxiv.org/abs/hep-th/9803131}{{\ttfamily hep-th/9803131}}].

\bibitem{Polchinski:2000uf}
J.~Polchinski and M.~J. Strassler, \emph{{The String dual of a confining
  four-dimensional gauge theory}},
  \href{https://arxiv.org/abs/hep-th/0003136}{{\ttfamily hep-th/0003136}}.

\bibitem{Klebanov:2000hb}
I.~R. Klebanov and M.~J. Strassler, \emph{{Supergravity and a confining gauge
  theory: Duality cascades and chi SB resolution of naked singularities}},
  \href{https://doi.org/10.1088/1126-6708/2000/08/052}{\emph{JHEP} {\bfseries
  08} (2000) 052} [\href{https://arxiv.org/abs/hep-th/0007191}{{\ttfamily
  hep-th/0007191}}].

\bibitem{Maldacena:2000yy}
J.~M. Maldacena and C.~Nunez, \emph{{Towards the large N limit of pure N=1
  superYang-Mills}},
  \href{https://doi.org/10.1103/PhysRevLett.86.588}{\emph{Phys. Rev. Lett.}
  {\bfseries 86} (2001) 588}
  [\href{https://arxiv.org/abs/hep-th/0008001}{{\ttfamily hep-th/0008001}}].

\bibitem{Vafa:2000wi}
C.~Vafa, \emph{{Superstrings and topological strings at large N}},
  \href{https://doi.org/10.1063/1.1376161}{\emph{J. Math. Phys.} {\bfseries 42}
  (2001) 2798} [\href{https://arxiv.org/abs/hep-th/0008142}{{\ttfamily
  hep-th/0008142}}].

\bibitem{Dvali:2003zj}
G.~Dvali and A.~Vilenkin, \emph{{Formation and evolution of cosmic D strings}},
  \href{https://doi.org/10.1088/1475-7516/2004/03/010}{\emph{JCAP} {\bfseries
  03} (2004) 010} [\href{https://arxiv.org/abs/hep-th/0312007}{{\ttfamily
  hep-th/0312007}}].

\bibitem{Copeland:2003bj}
E.~J. Copeland, R.~C. Myers and J.~Polchinski, \emph{{Cosmic F and D strings}},
  \href{https://doi.org/10.1088/1126-6708/2004/06/013}{\emph{JHEP} {\bfseries
  06} (2004) 013} [\href{https://arxiv.org/abs/hep-th/0312067}{{\ttfamily
  hep-th/0312067}}].

\bibitem{Vilenkin:1981bx}
A.~Vilenkin, \emph{{Gravitational radiation from cosmic strings}},
  \href{https://doi.org/10.1016/0370-2693(81)91144-8}{\emph{Phys. Lett. B}
  {\bfseries 107} (1981) 47}.

\bibitem{Vachaspati:1984gt}
T.~Vachaspati and A.~Vilenkin, \emph{{Gravitational Radiation from Cosmic
  Strings}}, \href{https://doi.org/10.1103/PhysRevD.31.3052}{\emph{Phys. Rev.
  D} {\bfseries 31} (1985) 3052}.

\bibitem{Kibble:1984hp}
T.~W.~B. Kibble, \emph{{Evolution of a system of cosmic strings}},
  \href{https://doi.org/10.1016/0550-3213(85)90596-6}{\emph{Nucl. Phys. B}
  {\bfseries 252} (1985) 227}.

\bibitem{Martins:1995tg}
C.~J. A.~P. Martins and E.~P.~S. Shellard, \emph{{String evolution with
  friction}}, \href{https://doi.org/10.1103/PhysRevD.53.R575}{\emph{Phys. Rev.
  D} {\bfseries 53} (1996) 575}
  [\href{https://arxiv.org/abs/hep-ph/9507335}{{\ttfamily hep-ph/9507335}}].

\bibitem{Martins:1996jp}
C.~J. A.~P. Martins and E.~P.~S. Shellard, \emph{{Quantitative string
  evolution}}, \href{https://doi.org/10.1103/PhysRevD.54.2535}{\emph{Phys. Rev.
  D} {\bfseries 54} (1996) 2535}
  [\href{https://arxiv.org/abs/hep-ph/9602271}{{\ttfamily hep-ph/9602271}}].

\bibitem{Martins:2000cs}
C.~J. A.~P. Martins and E.~P.~S. Shellard, \emph{{Extending the velocity
  dependent one scale string evolution model}},
  \href{https://doi.org/10.1103/PhysRevD.65.043514}{\emph{Phys. Rev. D}
  {\bfseries 65} (2002) 043514}
  [\href{https://arxiv.org/abs/hep-ph/0003298}{{\ttfamily hep-ph/0003298}}].

\bibitem{Avgoustidis:2005nv}
A.~Avgoustidis and E.~P.~S. Shellard, \emph{{Effect of reconnection probability
  on cosmic (super)string network density}},
  \href{https://doi.org/10.1103/PhysRevD.73.041301}{\emph{Phys. Rev. D}
  {\bfseries 73} (2006) 041301}
  [\href{https://arxiv.org/abs/astro-ph/0512582}{{\ttfamily
  astro-ph/0512582}}].

\bibitem{Janssen:2014dka}
G.~Janssen et~al., \emph{{Gravitational wave astronomy with the SKA}},
  \href{https://doi.org/10.22323/1.215.0037}{\emph{PoS} {\bfseries AASKA14}
  (2015) 037} [\href{https://arxiv.org/abs/1501.00127}{{\ttfamily
  1501.00127}}].

\bibitem{LISA:2017pwj}
{\scshape LISA} collaboration, P.~Amaro-Seoane et~al., \emph{{Laser
  Interferometer Space Antenna}},
  \href{https://arxiv.org/abs/1702.00786}{{\ttfamily 1702.00786}}.

\bibitem{Hidaka:2019mfm}
Y.~Hidaka, M.~Nitta and R.~Yokokura, \emph{{Emergent discrete 3-form symmetry
  and domain walls}},
  \href{https://doi.org/10.1016/j.physletb.2020.135290}{\emph{Phys. Lett. B}
  {\bfseries 803} (2020) 135290}
  [\href{https://arxiv.org/abs/1912.02782}{{\ttfamily 1912.02782}}].

\bibitem{Hidaka:2020ucc}
Y.~Hidaka, Y.~Hirono and R.~Yokokura, \emph{{Counting Nambu-Goldstone Modes of
  Higher-Form Global Symmetries}},
  \href{https://doi.org/10.1103/PhysRevLett.126.071601}{\emph{Phys. Rev. Lett.}
  {\bfseries 126} (2021) 071601}
  [\href{https://arxiv.org/abs/2007.15901}{{\ttfamily 2007.15901}}].

\bibitem{Hidaka:2020iaz}
Y.~Hidaka, M.~Nitta and R.~Yokokura, \emph{{Higher-form symmetries and 3-group
  in axion electrodynamics}},
  \href{https://doi.org/10.1016/j.physletb.2020.135672}{\emph{Phys. Lett. B}
  {\bfseries 808} (2020) 135672}
  [\href{https://arxiv.org/abs/2006.12532}{{\ttfamily 2006.12532}}].

\bibitem{Hidaka:2020izy}
Y.~Hidaka, M.~Nitta and R.~Yokokura, \emph{{Global 3-group symmetry and 't
  Hooft anomalies in axion electrodynamics}},
  \href{https://doi.org/10.1007/JHEP01(2021)173}{\emph{JHEP} {\bfseries 01}
  (2021) 173} [\href{https://arxiv.org/abs/2009.14368}{{\ttfamily
  2009.14368}}].

\bibitem{Hidaka:2021mml}
Y.~Hidaka, M.~Nitta and R.~Yokokura, \emph{{Topological axion electrodynamics
  and 4-group symmetry}},
  \href{https://doi.org/10.1016/j.physletb.2021.136762}{\emph{Phys. Lett. B}
  {\bfseries 823} (2021) 136762}
  [\href{https://arxiv.org/abs/2107.08753}{{\ttfamily 2107.08753}}].

\bibitem{Hidaka:2021kkf}
Y.~Hidaka, M.~Nitta and R.~Yokokura, \emph{{Global 4-group symmetry and 't
  Hooft anomalies in topological axion electrodynamics}},
  \href{https://arxiv.org/abs/2108.12564}{{\ttfamily 2108.12564}}.

\bibitem{Nitta:2022ahj}
M.~Nitta, \emph{{Relations among topological solitons}},
  \href{https://arxiv.org/abs/2202.03929}{{\ttfamily 2202.03929}}.

\bibitem{Cordova:2022rer}
C.~Cordova, K.~Ohmori and T.~Rudelius, \emph{{Generalized Symmetry Breaking
  Scales and Weak Gravity Conjectures}},
  \href{https://arxiv.org/abs/2202.05866}{{\ttfamily 2202.05866}}.

\bibitem{Yamamoto:2022vrh}
N.~Yamamoto and R.~Yokokura, \emph{{Unstable Nambu-Goldstone modes}},
  \href{https://arxiv.org/abs/2203.02727}{{\ttfamily 2203.02727}}.

\bibitem{Witten:1999ds}
E.~Witten, \emph{{Supersymmetric index of three-dimensional gauge theory}},
  \href{https://arxiv.org/abs/hep-th/9903005}{{\ttfamily hep-th/9903005}}.

\bibitem{Witten:2000nv}
E.~Witten, \emph{{Supersymmetric index in four-dimensional gauge theories}},
  \href{https://doi.org/10.4310/ATMP.2001.v5.n5.a1}{\emph{Adv. Theor. Math.
  Phys.} {\bfseries 5} (2002) 841}
  [\href{https://arxiv.org/abs/hep-th/0006010}{{\ttfamily hep-th/0006010}}].

\bibitem{Poppitz:2021cxe}
E.~Poppitz, \emph{{Notes on Confinement on R3 \texttimes{} S1: From
  Yang\textendash{}Mills, Super-Yang\textendash{}Mills, and QCD (adj) to
  QCD(F)}}, \href{https://doi.org/10.3390/sym14010180}{\emph{Symmetry}
  {\bfseries 14} (2022) 180}
  [\href{https://arxiv.org/abs/2111.10423}{{\ttfamily 2111.10423}}].

\bibitem{Vilenkin:1982hm}
A.~Vilenkin, \emph{{COSMOLOGICAL EVOLUTION OF MONOPOLES CONNECTED BY STRINGS}},
  \href{https://doi.org/10.1016/0550-3213(82)90037-2}{\emph{Nucl. Phys. B}
  {\bfseries 196} (1982) 240}.

\bibitem{Donagi:1995cf}
R.~Donagi and E.~Witten, \emph{{Supersymmetric Yang-Mills theory and integrable
  systems}}, \href{https://doi.org/10.1016/0550-3213(95)00609-5}{\emph{Nucl.
  Phys. B} {\bfseries 460} (1996) 299}
  [\href{https://arxiv.org/abs/hep-th/9510101}{{\ttfamily hep-th/9510101}}].

\bibitem{Jackson:2004zg}
M.~G. Jackson, N.~T. Jones and J.~Polchinski, \emph{{Collisions of cosmic F and
  D-strings}}, \href{https://doi.org/10.1088/1126-6708/2005/10/013}{\emph{JHEP}
  {\bfseries 10} (2005) 013}
  [\href{https://arxiv.org/abs/hep-th/0405229}{{\ttfamily hep-th/0405229}}].

\bibitem{Douglas:1995nw}
M.~R. Douglas and S.~H. Shenker, \emph{{Dynamics of SU(N) supersymmetric gauge
  theory}}, \href{https://doi.org/10.1016/0550-3213(95)00258-T}{\emph{Nucl.
  Phys. B} {\bfseries 447} (1995) 271}
  [\href{https://arxiv.org/abs/hep-th/9503163}{{\ttfamily hep-th/9503163}}].

\bibitem{Hanany:1997hr}
A.~Hanany, M.~J. Strassler and A.~Zaffaroni, \emph{{Confinement and strings in
  MQCD}}, \href{https://doi.org/10.1016/S0550-3213(97)00651-2}{\emph{Nucl.
  Phys. B} {\bfseries 513} (1998) 87}
  [\href{https://arxiv.org/abs/hep-th/9707244}{{\ttfamily hep-th/9707244}}].

\bibitem{Herzog:2001fq}
C.~P. Herzog and I.~R. Klebanov, \emph{{On string tensions in supersymmetric
  SU(M) gauge theory}},
  \href{https://doi.org/10.1016/S0370-2693(02)01155-3}{\emph{Phys. Lett. B}
  {\bfseries 526} (2002) 388}
  [\href{https://arxiv.org/abs/hep-th/0111078}{{\ttfamily hep-th/0111078}}].

\bibitem{Firouzjahi:2006vp}
H.~Firouzjahi, L.~Leblond and S.~H. Henry~Tye, \emph{{The (p,q) string tension
  in a warped deformed conifold}},
  \href{https://doi.org/10.1088/1126-6708/2006/05/047}{\emph{JHEP} {\bfseries
  05} (2006) 047} [\href{https://arxiv.org/abs/hep-th/0603161}{{\ttfamily
  hep-th/0603161}}].

\bibitem{Giddings:2001yu}
S.~B. Giddings, S.~Kachru and J.~Polchinski, \emph{{Hierarchies from fluxes in
  string compactifications}},
  \href{https://doi.org/10.1103/PhysRevD.66.106006}{\emph{Phys. Rev. D}
  {\bfseries 66} (2002) 106006}
  [\href{https://arxiv.org/abs/hep-th/0105097}{{\ttfamily hep-th/0105097}}].

\bibitem{Aharony:2000pp}
O.~Aharony, \emph{{A Note on the holographic interpretation of string theory
  backgrounds with varying flux}},
  \href{https://doi.org/10.1088/1126-6708/2001/03/012}{\emph{JHEP} {\bfseries
  03} (2001) 012} [\href{https://arxiv.org/abs/hep-th/0101013}{{\ttfamily
  hep-th/0101013}}].

\bibitem{Gubser:2004qj}
S.~S. Gubser, C.~P. Herzog and I.~R. Klebanov, \emph{{Symmetry breaking and
  axionic strings in the warped deformed conifold}},
  \href{https://doi.org/10.1088/1126-6708/2004/09/036}{\emph{JHEP} {\bfseries
  09} (2004) 036} [\href{https://arxiv.org/abs/hep-th/0405282}{{\ttfamily
  hep-th/0405282}}].

\bibitem{Witten:1998xy}
E.~Witten, \emph{{Baryons and branes in anti-de Sitter space}},
  \href{https://doi.org/10.1088/1126-6708/1998/07/006}{\emph{JHEP} {\bfseries
  07} (1998) 006} [\href{https://arxiv.org/abs/hep-th/9805112}{{\ttfamily
  hep-th/9805112}}].

\bibitem{Witten:1979kh}
E.~Witten, \emph{{Baryons in the 1/n Expansion}},
  \href{https://doi.org/10.1016/0550-3213(79)90232-3}{\emph{Nucl. Phys. B}
  {\bfseries 160} (1979) 57}.

\bibitem{Athenodorou:2021qvs}
A.~Athenodorou and M.~Teper, \emph{{SU(N) gauge theories in 3+1 dimensions:
  glueball spectrum, string tensions and topology}},
  \href{https://doi.org/10.1007/JHEP12(2021)082}{\emph{JHEP} {\bfseries 12}
  (2021) 082} [\href{https://arxiv.org/abs/2106.00364}{{\ttfamily
  2106.00364}}].

\bibitem{Polchinski:1988cn}
J.~Polchinski, \emph{{Collision of Macroscopic Fundamental Strings}},
  \href{https://doi.org/10.1016/0370-2693(88)90942-2}{\emph{Phys. Lett. B}
  {\bfseries 209} (1988) 252}.

\bibitem{Hanany:2005bc}
A.~Hanany and K.~Hashimoto, \emph{{Reconnection of colliding cosmic strings}},
  \href{https://doi.org/10.1088/1126-6708/2005/06/021}{\emph{JHEP} {\bfseries
  06} (2005) 021} [\href{https://arxiv.org/abs/hep-th/0501031}{{\ttfamily
  hep-th/0501031}}].

\bibitem{Martins:2005es}
C.~J. A.~P. Martins and E.~P.~S. Shellard, \emph{{Fractal properties and
  small-scale structure of cosmic string networks}},
  \href{https://doi.org/10.1103/PhysRevD.73.043515}{\emph{Phys. Rev. D}
  {\bfseries 73} (2006) 043515}
  [\href{https://arxiv.org/abs/astro-ph/0511792}{{\ttfamily
  astro-ph/0511792}}].

\bibitem{Jones:2002cv}
N.~T. Jones, H.~Stoica and S.~H.~H. Tye, \emph{{Brane interaction as the origin
  of inflation}},
  \href{https://doi.org/10.1088/1126-6708/2002/07/051}{\emph{JHEP} {\bfseries
  07} (2002) 051} [\href{https://arxiv.org/abs/hep-th/0203163}{{\ttfamily
  hep-th/0203163}}].

\bibitem{Sarangi:2002yt}
S.~Sarangi and S.~H.~H. Tye, \emph{{Cosmic string production towards the end of
  brane inflation}},
  \href{https://doi.org/10.1016/S0370-2693(02)01824-5}{\emph{Phys. Lett. B}
  {\bfseries 536} (2002) 185}
  [\href{https://arxiv.org/abs/hep-th/0204074}{{\ttfamily hep-th/0204074}}].

\bibitem{Dvali:2002fi}
G.~Dvali and A.~Vilenkin, \emph{{Solitonic D-branes and brane annihilation}},
  \href{https://doi.org/10.1103/PhysRevD.67.046002}{\emph{Phys. Rev. D}
  {\bfseries 67} (2003) 046002}
  [\href{https://arxiv.org/abs/hep-th/0209217}{{\ttfamily hep-th/0209217}}].

\bibitem{Jones:2003da}
N.~T. Jones, H.~Stoica and S.~H.~H. Tye, \emph{{The Production, spectrum and
  evolution of cosmic strings in brane inflation}},
  \href{https://doi.org/10.1016/S0370-2693(03)00592-6}{\emph{Phys. Lett. B}
  {\bfseries 563} (2003) 6}
  [\href{https://arxiv.org/abs/hep-th/0303269}{{\ttfamily hep-th/0303269}}].

\bibitem{Pogosian:2003mz}
L.~Pogosian, S.~H.~H. Tye, I.~Wasserman and M.~Wyman, \emph{{Observational
  constraints on cosmic string production during brane inflation}},
  \href{https://doi.org/10.1103/PhysRevD.68.023506}{\emph{Phys. Rev. D}
  {\bfseries 68} (2003) 023506}
  [\href{https://arxiv.org/abs/hep-th/0304188}{{\ttfamily hep-th/0304188}}].

\bibitem{Copeland:2006eh}
E.~J. Copeland, T.~W.~B. Kibble and D.~A. Steer, \emph{{Collisions of strings
  with Y junctions}},
  \href{https://doi.org/10.1103/PhysRevLett.97.021602}{\emph{Phys. Rev. Lett.}
  {\bfseries 97} (2006) 021602}
  [\href{https://arxiv.org/abs/hep-th/0601153}{{\ttfamily hep-th/0601153}}].

\bibitem{Copeland:2006if}
E.~J. Copeland, T.~W.~B. Kibble and D.~A. Steer, \emph{{Constraints on string
  networks with junctions}},
  \href{https://doi.org/10.1103/PhysRevD.75.065024}{\emph{Phys. Rev. D}
  {\bfseries 75} (2007) 065024}
  [\href{https://arxiv.org/abs/hep-th/0611243}{{\ttfamily hep-th/0611243}}].

\bibitem{Copeland:2007nv}
E.~J. Copeland, H.~Firouzjahi, T.~W.~B. Kibble and D.~A. Steer, \emph{{On the
  Collision of Cosmic Superstrings}},
  \href{https://doi.org/10.1103/PhysRevD.77.063521}{\emph{Phys. Rev. D}
  {\bfseries 77} (2008) 063521}
  [\href{https://arxiv.org/abs/0712.0808}{{\ttfamily 0712.0808}}].

\bibitem{Avgoustidis:2007aa}
A.~Avgoustidis and E.~P.~S. Shellard, \emph{{Velocity-Dependent Models for
  Non-Abelian/Entangled String Networks}},
  \href{https://doi.org/10.1103/PhysRevD.78.103510}{\emph{Phys. Rev. D}
  {\bfseries 78} (2008) 103510}
  [\href{https://arxiv.org/abs/0705.3395}{{\ttfamily 0705.3395}}].

\bibitem{Rajantie:2007hp}
A.~Rajantie, M.~Sakellariadou and H.~Stoica, \emph{{Numerical experiments with
  p F- and q D-strings: The Formation of (p,q) bound states}},
  \href{https://doi.org/10.1088/1475-7516/2007/11/021}{\emph{JCAP} {\bfseries
  11} (2007) 021} [\href{https://arxiv.org/abs/0706.3662}{{\ttfamily
  0706.3662}}].

\bibitem{Pourtsidou:2010gu}
A.~Pourtsidou, A.~Avgoustidis, E.~J. Copeland, L.~Pogosian and D.~A. Steer,
  \emph{{Scaling configurations of cosmic superstring networks and their
  cosmological implications}},
  \href{https://doi.org/10.1103/PhysRevD.83.063525}{\emph{Phys. Rev. D}
  {\bfseries 83} (2011) 063525}
  [\href{https://arxiv.org/abs/1012.5014}{{\ttfamily 1012.5014}}].

\bibitem{Sousa:2016ggw}
L.~Sousa and P.~P. Avelino, \emph{{Probing Cosmic Superstrings with
  Gravitational Waves}},
  \href{https://doi.org/10.1103/PhysRevD.94.063529}{\emph{Phys. Rev. D}
  {\bfseries 94} (2016) 063529}
  [\href{https://arxiv.org/abs/1606.05585}{{\ttfamily 1606.05585}}].

\bibitem{Ng:2008mp}
Y.~Ng, T.~W.~B. Kibble and T.~Vachaspati, \emph{{Formation of Non-Abelian
  Monopoles Connected by Strings}},
  \href{https://doi.org/10.1103/PhysRevD.78.046001}{\emph{Phys. Rev. D}
  {\bfseries 78} (2008) 046001}
  [\href{https://arxiv.org/abs/0806.0155}{{\ttfamily 0806.0155}}].

\bibitem{Spergel:1996ai}
D.~Spergel and U.-L. Pen, \emph{{Cosmology in a string dominated universe}},
  \href{https://doi.org/10.1086/311074}{\emph{Astrophys. J. Lett.} {\bfseries
  491} (1997) L67} [\href{https://arxiv.org/abs/astro-ph/9611198}{{\ttfamily
  astro-ph/9611198}}].

\bibitem{McGraw:1997nx}
P.~McGraw, \emph{{Evolution of a nonAbelian cosmic string network}},
  \href{https://doi.org/10.1103/PhysRevD.57.3317}{\emph{Phys. Rev. D}
  {\bfseries 57} (1998) 3317}
  [\href{https://arxiv.org/abs/astro-ph/9706182}{{\ttfamily
  astro-ph/9706182}}].

\bibitem{Hindmarsh:1985xc}
M.~Hindmarsh and T.~W.~B. Kibble, \emph{{BEADS ON STRINGS}},
  \href{https://doi.org/10.1103/PhysRevLett.55.2398}{\emph{Phys. Rev. Lett.}
  {\bfseries 55} (1985) 2398}.

\bibitem{Berezinsky:1997td}
V.~Berezinsky and A.~Vilenkin, \emph{{Cosmic necklaces and ultrahigh-energy
  cosmic rays}}, \href{https://doi.org/10.1103/PhysRevLett.79.5202}{\emph{Phys.
  Rev. Lett.} {\bfseries 79} (1997) 5202}
  [\href{https://arxiv.org/abs/astro-ph/9704257}{{\ttfamily
  astro-ph/9704257}}].

\bibitem{Hindmarsh:2016dha}
M.~Hindmarsh, K.~Rummukainen and D.~J. Weir, \emph{{Numerical simulations of
  necklaces in SU(2) gauge-Higgs field theory}},
  \href{https://doi.org/10.1103/PhysRevD.95.063520}{\emph{Phys. Rev. D}
  {\bfseries 95} (2017) 063520}
  [\href{https://arxiv.org/abs/1611.08456}{{\ttfamily 1611.08456}}].

\bibitem{futurework}
M.~Yamada and K.~Yonekura, \emph{{In preparation}}, .

\bibitem{Artru:1974zn}
X.~Artru, \emph{{String Model with Baryons: Topology, Classical Motion}},
  \href{https://doi.org/10.1016/0550-3213(75)90019-X}{\emph{Nucl. Phys. B}
  {\bfseries 85} (1975) 442}.

\bibitem{Copeland:2005cy}
E.~J. Copeland and P.~M. Saffin, \emph{{On the evolution of cosmic-superstring
  networks}}, \href{https://doi.org/10.1088/1126-6708/2005/11/023}{\emph{JHEP}
  {\bfseries 11} (2005) 023}
  [\href{https://arxiv.org/abs/hep-th/0505110}{{\ttfamily hep-th/0505110}}].

\bibitem{Hindmarsh:2006qn}
M.~Hindmarsh and P.~M. Saffin, \emph{{Scaling in a SU(2)$/\mathbb{Z}_{3}$ model
  of cosmic superstring networks}},
  \href{https://doi.org/10.1088/1126-6708/2006/08/066}{\emph{JHEP} {\bfseries
  08} (2006) 066} [\href{https://arxiv.org/abs/hep-th/0605014}{{\ttfamily
  hep-th/0605014}}].

\bibitem{Urrestilla:2007yw}
J.~Urrestilla and A.~Vilenkin, \emph{{Evolution of cosmic superstring networks:
  A Numerical simulation}},
  \href{https://doi.org/10.1088/1126-6708/2008/02/037}{\emph{JHEP} {\bfseries
  02} (2008) 037} [\href{https://arxiv.org/abs/0712.1146}{{\ttfamily
  0712.1146}}].

\bibitem{Spergel:1990ee}
D.~N. Spergel, N.~Turok, W.~H. Press and B.~S. Ryden, \emph{{Global texture as
  the origin of large scale structure: numerical simulations of evolution}},
  \href{https://doi.org/10.1103/PhysRevD.43.1038}{\emph{Phys. Rev. D}
  {\bfseries 43} (1991) 1038}.

\bibitem{Hiramatsu:2010yn}
T.~Hiramatsu, M.~Kawasaki and K.~Saikawa, \emph{{Evolution of String-Wall
  Networks and Axionic Domain Wall Problem}},
  \href{https://doi.org/10.1088/1475-7516/2011/08/030}{\emph{JCAP} {\bfseries
  08} (2011) 030} [\href{https://arxiv.org/abs/1012.4558}{{\ttfamily
  1012.4558}}].

\bibitem{Hiramatsu:2012sc}
T.~Hiramatsu, M.~Kawasaki, K.~Saikawa and T.~Sekiguchi, \emph{{Axion cosmology
  with long-lived domain walls}},
  \href{https://doi.org/10.1088/1475-7516/2013/01/001}{\emph{JCAP} {\bfseries
  01} (2013) 001} [\href{https://arxiv.org/abs/1207.3166}{{\ttfamily
  1207.3166}}].

\bibitem{Kawasaki:2014sqa}
M.~Kawasaki, K.~Saikawa and T.~Sekiguchi, \emph{{Axion dark matter from
  topological defects}},
  \href{https://doi.org/10.1103/PhysRevD.91.065014}{\emph{Phys. Rev. D}
  {\bfseries 91} (2015) 065014}
  [\href{https://arxiv.org/abs/1412.0789}{{\ttfamily 1412.0789}}].

\bibitem{Banks:2010zn}
T.~Banks and N.~Seiberg, \emph{{Symmetries and Strings in Field Theory and
  Gravity}}, \href{https://doi.org/10.1103/PhysRevD.83.084019}{\emph{Phys. Rev.
  D} {\bfseries 83} (2011) 084019}
  [\href{https://arxiv.org/abs/1011.5120}{{\ttfamily 1011.5120}}].

\bibitem{Montero:2017yja}
M.~Montero, A.~M. Uranga and I.~Valenzuela, \emph{{A Chern-Simons Pandemic}},
  \href{https://doi.org/10.1007/JHEP07(2017)123}{\emph{JHEP} {\bfseries 07}
  (2017) 123} [\href{https://arxiv.org/abs/1702.06147}{{\ttfamily
  1702.06147}}].

\bibitem{Harlow:2018tng}
D.~Harlow and H.~Ooguri, \emph{{Symmetries in quantum field theory and quantum
  gravity}}, \href{https://doi.org/10.1007/s00220-021-04040-y}{\emph{Commun.
  Math. Phys.} {\bfseries 383} (2021) 1669}
  [\href{https://arxiv.org/abs/1810.05338}{{\ttfamily 1810.05338}}].

\bibitem{Rudelius:2020orz}
T.~Rudelius and S.-H. Shao, \emph{{Topological Operators and Completeness of
  Spectrum in Discrete Gauge Theories}},
  \href{https://doi.org/10.1007/JHEP12(2020)172}{\emph{JHEP} {\bfseries 12}
  (2020) 172} [\href{https://arxiv.org/abs/2006.10052}{{\ttfamily
  2006.10052}}].

\bibitem{Yonekura:2020ino}
K.~Yonekura, \emph{{Topological violation of global symmetries in quantum
  gravity}}, \href{https://doi.org/10.1007/JHEP09(2021)036}{\emph{JHEP}
  {\bfseries 09} (2021) 036}
  [\href{https://arxiv.org/abs/2011.11868}{{\ttfamily 2011.11868}}].

\bibitem{Heidenreich:2020pkc}
B.~Heidenreich, J.~McNamara, M.~Montero, M.~Reece, T.~Rudelius and
  I.~Valenzuela, \emph{{Chern-Weil global symmetries and how quantum gravity
  avoids them}}, \href{https://doi.org/10.1007/JHEP11(2021)053}{\emph{JHEP}
  {\bfseries 11} (2021) 053}
  [\href{https://arxiv.org/abs/2012.00009}{{\ttfamily 2012.00009}}].

\bibitem{Heidenreich:2021xpr}
B.~Heidenreich, J.~McNamara, M.~Montero, M.~Reece, T.~Rudelius and
  I.~Valenzuela, \emph{{Non-invertible global symmetries and completeness of
  the spectrum}}, \href{https://doi.org/10.1007/JHEP09(2021)203}{\emph{JHEP}
  {\bfseries 09} (2021) 203}
  [\href{https://arxiv.org/abs/2104.07036}{{\ttfamily 2104.07036}}].

\bibitem{Buchmuller:2021mbb}
W.~Buchmuller, V.~Domcke and K.~Schmitz, \emph{{Stochastic gravitational-wave
  background from metastable cosmic strings}},
  \href{https://doi.org/10.1088/1475-7516/2021/12/006}{\emph{JCAP} {\bfseries
  12} (2021) 006} [\href{https://arxiv.org/abs/2107.04578}{{\ttfamily
  2107.04578}}].

\bibitem{Dunsky:2021tih}
D.~I. Dunsky, A.~Ghoshal, H.~Murayama, Y.~Sakakihara and G.~White,
  \emph{{Gravitational Wave Gastronomy}},
  \href{https://arxiv.org/abs/2111.08750}{{\ttfamily 2111.08750}}.

\bibitem{Lazarides:2022jgr}
G.~Lazarides, R.~Maji and Q.~Shafi, \emph{{Gravitational Waves from
  Quasi-stable Strings}},  \href{https://arxiv.org/abs/2203.11204}{{\ttfamily
  2203.11204}}.

\bibitem{Luty:2004ye}
M.~A. Luty and T.~Okui, \emph{{Conformal technicolor}},
  \href{https://doi.org/10.1088/1126-6708/2006/09/070}{\emph{JHEP} {\bfseries
  09} (2006) 070} [\href{https://arxiv.org/abs/hep-ph/0409274}{{\ttfamily
  hep-ph/0409274}}].

\bibitem{Ibe:2007wp}
M.~Ibe, Y.~Nakayama and T.~T. Yanagida, \emph{{Conformal Gauge Mediation}},
  \href{https://doi.org/10.1016/j.physletb.2007.04.002}{\emph{Phys. Lett. B}
  {\bfseries 649} (2007) 292}
  [\href{https://arxiv.org/abs/hep-ph/0703110}{{\ttfamily hep-ph/0703110}}].

\bibitem{Yanagida:2010zz}
T.~T. Yanagida and K.~Yonekura, \emph{{A Conformal Gauge Mediation and Dark
  Matter with Only One Parameter}},
  \href{https://doi.org/10.1016/j.physletb.2010.08.040}{\emph{Phys. Lett. B}
  {\bfseries 693} (2010) 281}
  [\href{https://arxiv.org/abs/1006.2271}{{\ttfamily 1006.2271}}].

\bibitem{Hamaguchi:2011kt}
K.~Hamaguchi, K.~Nakayama and N.~Yokozaki, \emph{{A Solution to the mu/Bmu
  Problem in Gauge Mediation with Hidden Gauge Symmetry}},
  \href{https://doi.org/10.1007/JHEP08(2012)006}{\emph{JHEP} {\bfseries 08}
  (2012) 006} [\href{https://arxiv.org/abs/1111.1601}{{\ttfamily 1111.1601}}].

\bibitem{Ringeval:2005kr}
C.~Ringeval, M.~Sakellariadou and F.~Bouchet, \emph{{Cosmological evolution of
  cosmic string loops}},
  \href{https://doi.org/10.1088/1475-7516/2007/02/023}{\emph{JCAP} {\bfseries
  02} (2007) 023} [\href{https://arxiv.org/abs/astro-ph/0511646}{{\ttfamily
  astro-ph/0511646}}].

\bibitem{Blanco-Pillado:2011egf}
J.~J. Blanco-Pillado, K.~D. Olum and B.~Shlaer, \emph{{Large parallel cosmic
  string simulations: New results on loop production}},
  \href{https://doi.org/10.1103/PhysRevD.83.083514}{\emph{Phys. Rev. D}
  {\bfseries 83} (2011) 083514}
  [\href{https://arxiv.org/abs/1101.5173}{{\ttfamily 1101.5173}}].

\bibitem{Blanco-Pillado:2013qja}
J.~J. Blanco-Pillado, K.~D. Olum and B.~Shlaer, \emph{{The number of cosmic
  string loops}}, \href{https://doi.org/10.1103/PhysRevD.89.023512}{\emph{Phys.
  Rev. D} {\bfseries 89} (2014) 023512}
  [\href{https://arxiv.org/abs/1309.6637}{{\ttfamily 1309.6637}}].

\bibitem{Blanco-Pillado:2017oxo}
J.~J. Blanco-Pillado and K.~D. Olum, \emph{{Stochastic gravitational wave
  background from smoothed cosmic string loops}},
  \href{https://doi.org/10.1103/PhysRevD.96.104046}{\emph{Phys. Rev. D}
  {\bfseries 96} (2017) 104046}
  [\href{https://arxiv.org/abs/1709.02693}{{\ttfamily 1709.02693}}].

\bibitem{Blanco-Pillado:2017rnf}
J.~J. Blanco-Pillado, K.~D. Olum and X.~Siemens, \emph{{New limits on cosmic
  strings from gravitational wave observation}},
  \href{https://doi.org/10.1016/j.physletb.2018.01.050}{\emph{Phys. Lett. B}
  {\bfseries 778} (2018) 392}
  [\href{https://arxiv.org/abs/1709.02434}{{\ttfamily 1709.02434}}].

\bibitem{Caldwell:1991jj}
R.~R. Caldwell and B.~Allen, \emph{{Cosmological constraints on cosmic string
  gravitational radiation}},
  \href{https://doi.org/10.1103/PhysRevD.45.3447}{\emph{Phys. Rev. D}
  {\bfseries 45} (1992) 3447}.

\bibitem{DePies:2007bm}
M.~R. DePies and C.~J. Hogan, \emph{{Stochastic Gravitational Wave Background
  from Light Cosmic Strings}},
  \href{https://doi.org/10.1103/PhysRevD.75.125006}{\emph{Phys. Rev. D}
  {\bfseries 75} (2007) 125006}
  [\href{https://arxiv.org/abs/astro-ph/0702335}{{\ttfamily
  astro-ph/0702335}}].

\bibitem{Sanidas:2012ee}
S.~A. Sanidas, R.~A. Battye and B.~W. Stappers, \emph{{Constraints on cosmic
  string tension imposed by the limit on the stochastic gravitational wave
  background from the European Pulsar Timing Array}},
  \href{https://doi.org/10.1103/PhysRevD.85.122003}{\emph{Phys. Rev. D}
  {\bfseries 85} (2012) 122003}
  [\href{https://arxiv.org/abs/1201.2419}{{\ttfamily 1201.2419}}].

\bibitem{Sousa:2013aaa}
L.~Sousa and P.~P. Avelino, \emph{{Stochastic Gravitational Wave Background
  generated by Cosmic String Networks: Velocity-Dependent One-Scale model
  versus Scale-Invariant Evolution}},
  \href{https://doi.org/10.1103/PhysRevD.88.023516}{\emph{Phys. Rev. D}
  {\bfseries 88} (2013) 023516}
  [\href{https://arxiv.org/abs/1304.2445}{{\ttfamily 1304.2445}}].

\bibitem{Martins:2003vd}
C.~J. A.~P. Martins, J.~N. Moore and E.~P.~S. Shellard, \emph{{A Unified model
  for vortex string network evolution}},
  \href{https://doi.org/10.1103/PhysRevLett.92.251601}{\emph{Phys. Rev. Lett.}
  {\bfseries 92} (2004) 251601}
  [\href{https://arxiv.org/abs/hep-ph/0310255}{{\ttfamily hep-ph/0310255}}].

\bibitem{Bennett:1989yp}
D.~P. Bennett and F.~R. Bouchet, \emph{{High resolution simulations of cosmic
  string evolution. 1. Network evolution}},
  \href{https://doi.org/10.1103/PhysRevD.41.2408}{\emph{Phys. Rev. D}
  {\bfseries 41} (1990) 2408}.

\bibitem{Allen:1990tv}
B.~Allen and E.~P.~S. Shellard, \emph{{Cosmic string evolution: a numerical
  simulation}}, \href{https://doi.org/10.1103/PhysRevLett.64.119}{\emph{Phys.
  Rev. Lett.} {\bfseries 64} (1990) 119}.

\bibitem{Sakellariadou:2004wq}
M.~Sakellariadou, \emph{{A Note on the evolution of cosmic string/superstring
  networks}}, \href{https://doi.org/10.1088/1475-7516/2005/04/003}{\emph{JCAP}
  {\bfseries 04} (2005) 003}
  [\href{https://arxiv.org/abs/hep-th/0410234}{{\ttfamily hep-th/0410234}}].

\bibitem{Sakellariadou:1990nd}
M.~Sakellariadou and A.~Vilenkin, \emph{{Cosmic-string evolution in flat
  space-time}}, \href{https://doi.org/10.1103/PhysRevD.42.349}{\emph{Phys. Rev.
  D} {\bfseries 42} (1990) 349}.

\bibitem{Auclair:2019wcv}
P.~Auclair et~al., \emph{{Probing the gravitational wave background from cosmic
  strings with LISA}},
  \href{https://doi.org/10.1088/1475-7516/2020/04/034}{\emph{JCAP} {\bfseries
  04} (2020) 034} [\href{https://arxiv.org/abs/1909.00819}{{\ttfamily
  1909.00819}}].

\bibitem{Burden:1985md}
C.~J. Burden, \emph{{Gravitational Radiation From a Particular Class of Cosmic
  Strings}}, \href{https://doi.org/10.1016/0370-2693(85)90326-0}{\emph{Phys.
  Lett. B} {\bfseries 164} (1985) 277}.

\bibitem{Garfinkle:1987yw}
D.~Garfinkle and T.~Vachaspati, \emph{{Radiation From Kinky, Cuspless Cosmic
  Loops}}, \href{https://doi.org/10.1103/PhysRevD.36.2229}{\emph{Phys. Rev. D}
  {\bfseries 36} (1987) 2229}.

\bibitem{Vilenkin:2000jqa}
A.~Vilenkin and E.~P.~S. Shellard, \emph{{Cosmic Strings and Other Topological
  Defects}}. Cambridge University Press, 7, 2000.

\bibitem{Planck:2018vyg}
{\scshape Planck} collaboration, N.~Aghanim et~al., \emph{{Planck 2018 results.
  VI. Cosmological parameters}},
  \href{https://doi.org/10.1051/0004-6361/201833910}{\emph{Astron. Astrophys.}
  {\bfseries 641} (2020) A6}
  [\href{https://arxiv.org/abs/1807.06209}{{\ttfamily 1807.06209}}].

\bibitem{Battye:1997ji}
R.~A. Battye, R.~R. Caldwell and E.~P.~S. Shellard, \emph{{Gravitational waves
  from cosmic strings}},  in \emph{{Conference on Topological Defects and
  CMB}}, pp.~11--31, 6, 1997,
  \href{https://arxiv.org/abs/astro-ph/9706013}{{\ttfamily astro-ph/9706013}}.

\bibitem{Cui:2018rwi}
Y.~Cui, M.~Lewicki, D.~E. Morrissey and J.~D. Wells, \emph{{Probing the pre-BBN
  universe with gravitational waves from cosmic strings}},
  \href{https://doi.org/10.1007/JHEP01(2019)081}{\emph{JHEP} {\bfseries 01}
  (2019) 081} [\href{https://arxiv.org/abs/1808.08968}{{\ttfamily
  1808.08968}}].

\bibitem{Shannon:2015ect}
R.~M. Shannon et~al., \emph{{Gravitational waves from binary supermassive black
  holes missing in pulsar observations}},
  \href{https://doi.org/10.1126/science.aab1910}{\emph{Science} {\bfseries 349}
  (2015) 1522} [\href{https://arxiv.org/abs/1509.07320}{{\ttfamily
  1509.07320}}].

\bibitem{KAGRA:2021kbb}
{\scshape KAGRA, Virgo, LIGO Scientific} collaboration, R.~Abbott et~al.,
  \emph{{Upper limits on the isotropic gravitational-wave background from
  Advanced LIGO and Advanced Virgo\textquoteright{}s third observing run}},
  \href{https://doi.org/10.1103/PhysRevD.104.022004}{\emph{Phys. Rev. D}
  {\bfseries 104} (2021) 022004}
  [\href{https://arxiv.org/abs/2101.12130}{{\ttfamily 2101.12130}}].

\bibitem{NANOGrav:2020bcs}
{\scshape NANOGrav} collaboration, Z.~Arzoumanian et~al., \emph{{The NANOGrav
  12.5 yr Data Set: Search for an Isotropic Stochastic Gravitational-wave
  Background}},
  \href{https://doi.org/10.3847/2041-8213/abd401}{\emph{Astrophys. J. Lett.}
  {\bfseries 905} (2020) L34}
  [\href{https://arxiv.org/abs/2009.04496}{{\ttfamily 2009.04496}}].

\bibitem{Goncharov:2021oub}
B.~Goncharov et~al., \emph{{On the Evidence for a Common-spectrum Process in
  the Search for the Nanohertz Gravitational-wave Background with the Parkes
  Pulsar Timing Array}},
  \href{https://doi.org/10.3847/2041-8213/ac17f4}{\emph{Astrophys. J. Lett.}
  {\bfseries 917} (2021) L19}
  [\href{https://arxiv.org/abs/2107.12112}{{\ttfamily 2107.12112}}].

\bibitem{Schmitz:2020syl}
K.~Schmitz, \emph{{New Sensitivity Curves for Gravitational-Wave Signals from
  Cosmological Phase Transitions}},
  \href{https://doi.org/10.1007/JHEP01(2021)097}{\emph{JHEP} {\bfseries 01}
  (2021) 097} [\href{https://arxiv.org/abs/2002.04615}{{\ttfamily
  2002.04615}}].

\bibitem{Kawamura:2011zz}
S.~Kawamura et~al., \emph{{The Japanese space gravitational wave antenna:
  DECIGO}}, \href{https://doi.org/10.1088/0264-9381/28/9/094011}{\emph{Class.
  Quant. Grav.} {\bfseries 28} (2011) 094011}.

\bibitem{Kawamura:2020pcg}
S.~Kawamura et~al., \emph{{Current status of space gravitational wave antenna
  DECIGO and B-DECIGO}},
  \href{https://doi.org/10.1093/ptep/ptab019}{\emph{PTEP} {\bfseries 2021}
  (2021) 05A105} [\href{https://arxiv.org/abs/2006.13545}{{\ttfamily
  2006.13545}}].

\bibitem{Harry:2006fi}
G.~M. Harry, P.~Fritschel, D.~A. Shaddock, W.~Folkner and E.~S. Phinney,
  \emph{{Laser interferometry for the big bang observer}},
  \href{https://doi.org/10.1088/0264-9381/23/15/008}{\emph{Class. Quant. Grav.}
  {\bfseries 23} (2006) 4887}.

\bibitem{Punturo:2010zz}
M.~Punturo et~al., \emph{{The Einstein Telescope: A third-generation
  gravitational wave observatory}},
  \href{https://doi.org/10.1088/0264-9381/27/19/194002}{\emph{Class. Quant.
  Grav.} {\bfseries 27} (2010) 194002}.

\bibitem{Maggiore:2019uih}
M.~Maggiore et~al., \emph{{Science Case for the Einstein Telescope}},
  \href{https://doi.org/10.1088/1475-7516/2020/03/050}{\emph{JCAP} {\bfseries
  03} (2020) 050} [\href{https://arxiv.org/abs/1912.02622}{{\ttfamily
  1912.02622}}].

\bibitem{Reitze:2019iox}
D.~Reitze et~al., \emph{{Cosmic Explorer: The U.S. Contribution to
  Gravitational-Wave Astronomy beyond LIGO}}, {\emph{Bull. Am. Astron. Soc.}
  {\bfseries 51} (2019) 035}
  [\href{https://arxiv.org/abs/1907.04833}{{\ttfamily 1907.04833}}].

\bibitem{Somiya:2011np}
{\scshape KAGRA} collaboration, K.~Somiya, \emph{{Detector configuration of
  KAGRA: The Japanese cryogenic gravitational-wave detector}},
  \href{https://doi.org/10.1088/0264-9381/29/12/124007}{\emph{Class. Quant.
  Grav.} {\bfseries 29} (2012) 124007}
  [\href{https://arxiv.org/abs/1111.7185}{{\ttfamily 1111.7185}}].

\bibitem{KAGRA:2020cvd}
{\scshape KAGRA} collaboration, T.~Akutsu et~al., \emph{{Overview of KAGRA :
  KAGRA science}},  \href{https://arxiv.org/abs/2008.02921}{{\ttfamily
  2008.02921}}.

\bibitem{Ellis:2020ena}
J.~Ellis and M.~Lewicki, \emph{{Cosmic String Interpretation of NANOGrav Pulsar
  Timing Data}},
  \href{https://doi.org/10.1103/PhysRevLett.126.041304}{\emph{Phys. Rev. Lett.}
  {\bfseries 126} (2021) 041304}
  [\href{https://arxiv.org/abs/2009.06555}{{\ttfamily 2009.06555}}].

\bibitem{Blasi:2020mfx}
S.~Blasi, V.~Brdar and K.~Schmitz, \emph{{Has NANOGrav found first evidence for
  cosmic strings?}},
  \href{https://doi.org/10.1103/PhysRevLett.126.041305}{\emph{Phys. Rev. Lett.}
  {\bfseries 126} (2021) 041305}
  [\href{https://arxiv.org/abs/2009.06607}{{\ttfamily 2009.06607}}].

\bibitem{Blanco-Pillado:2021ygr}
J.~J. Blanco-Pillado, K.~D. Olum and J.~M. Wachter, \emph{{Comparison of cosmic
  string and superstring models to NANOGrav 12.5-year results}},
  \href{https://doi.org/10.1103/PhysRevD.103.103512}{\emph{Phys. Rev. D}
  {\bfseries 103} (2021) 103512}
  [\href{https://arxiv.org/abs/2102.08194}{{\ttfamily 2102.08194}}].

\bibitem{Charnock:2016nzm}
T.~Charnock, A.~Avgoustidis, E.~J. Copeland and A.~Moss, \emph{{CMB constraints
  on cosmic strings and superstrings}},
  \href{https://doi.org/10.1103/PhysRevD.93.123503}{\emph{Phys. Rev. D}
  {\bfseries 93} (2016) 123503}
  [\href{https://arxiv.org/abs/1603.01275}{{\ttfamily 1603.01275}}].

\bibitem{Iwasaki:1992ik}
Y.~Iwasaki, K.~Kanaya, T.~Yoshie, T.~Hoshino, T.~Shirakawa, Y.~Oyanagi et~al.,
  \emph{{Finite temperature phase transition of SU(3) gauge theory on N(t) = 4
  and 6 lattices}}, \href{https://doi.org/10.1103/PhysRevD.46.4657}{\emph{Phys.
  Rev. D} {\bfseries 46} (1992) 4657}.

\bibitem{Boyd:1996bx}
G.~Boyd, J.~Engels, F.~Karsch, E.~Laermann, C.~Legeland, M.~Lutgemeier et~al.,
  \emph{{Thermodynamics of SU(3) lattice gauge theory}},
  \href{https://doi.org/10.1016/0550-3213(96)00170-8}{\emph{Nucl. Phys. B}
  {\bfseries 469} (1996) 419}
  [\href{https://arxiv.org/abs/hep-lat/9602007}{{\ttfamily hep-lat/9602007}}].

\bibitem{Lucini:2003zr}
B.~Lucini, M.~Teper and U.~Wenger, \emph{{The High temperature phase transition
  in SU(N) gauge theories}},
  \href{https://doi.org/10.1088/1126-6708/2004/01/061}{\emph{JHEP} {\bfseries
  01} (2004) 061} [\href{https://arxiv.org/abs/hep-lat/0307017}{{\ttfamily
  hep-lat/0307017}}].

\bibitem{Datta:2009jn}
S.~Datta and S.~Gupta, \emph{{Scaling and the continuum limit of the finite
  temperature deconfinement transition in SU$(N_c)$ pure gauge theory}},
  \href{https://doi.org/10.1103/PhysRevD.80.114504}{\emph{Phys. Rev. D}
  {\bfseries 80} (2009) 114504}
  [\href{https://arxiv.org/abs/0909.5591}{{\ttfamily 0909.5591}}].

\bibitem{Panero:2009tv}
M.~Panero, \emph{{Thermodynamics of the QCD plasma and the large-N limit}},
  \href{https://doi.org/10.1103/PhysRevLett.103.232001}{\emph{Phys. Rev. Lett.}
  {\bfseries 103} (2009) 232001}
  [\href{https://arxiv.org/abs/0907.3719}{{\ttfamily 0907.3719}}].

\bibitem{Aharony:2006da}
O.~Aharony, J.~Sonnenschein and S.~Yankielowicz, \emph{{A Holographic model of
  deconfinement and chiral symmetry restoration}},
  \href{https://doi.org/10.1016/j.aop.2006.11.002}{\emph{Annals Phys.}
  {\bfseries 322} (2007) 1420}
  [\href{https://arxiv.org/abs/hep-th/0604161}{{\ttfamily hep-th/0604161}}].

\bibitem{Mandal:2011ws}
G.~Mandal and T.~Morita, \emph{{Gregory-Laflamme as the
  confinement/deconfinement transition in holographic QCD}},
  \href{https://doi.org/10.1007/JHEP09(2011)073}{\emph{JHEP} {\bfseries 09}
  (2011) 073} [\href{https://arxiv.org/abs/1107.4048}{{\ttfamily 1107.4048}}].

\bibitem{Isono:2015uda}
H.~Isono, G.~Mandal and T.~Morita, \emph{{Thermodynamics of QCD from
  Sakai-Sugimoto Model}},
  \href{https://doi.org/10.1007/JHEP12(2015)006}{\emph{JHEP} {\bfseries 12}
  (2015) 006} [\href{https://arxiv.org/abs/1507.08949}{{\ttfamily
  1507.08949}}].

\bibitem{Gaiotto:2017yup}
D.~Gaiotto, A.~Kapustin, Z.~Komargodski and N.~Seiberg, \emph{{Theta, Time
  Reversal, and Temperature}},
  \href{https://doi.org/10.1007/JHEP05(2017)091}{\emph{JHEP} {\bfseries 05}
  (2017) 091} [\href{https://arxiv.org/abs/1703.00501}{{\ttfamily
  1703.00501}}].

\bibitem{Komargodski:2017smk}
Z.~Komargodski, T.~Sulejmanpasic and M.~\"Unsal, \emph{{Walls, anomalies, and
  deconfinement in quantum antiferromagnets}},
  \href{https://doi.org/10.1103/PhysRevB.97.054418}{\emph{Phys. Rev. B}
  {\bfseries 97} (2018) 054418}
  [\href{https://arxiv.org/abs/1706.05731}{{\ttfamily 1706.05731}}].

\bibitem{Shimizu:2017asf}
H.~Shimizu and K.~Yonekura, \emph{{Anomaly constraints on deconfinement and
  chiral phase transition}},
  \href{https://doi.org/10.1103/PhysRevD.97.105011}{\emph{Phys. Rev. D}
  {\bfseries 97} (2018) 105011}
  [\href{https://arxiv.org/abs/1706.06104}{{\ttfamily 1706.06104}}].

\bibitem{Tanizaki:2017qhf}
Y.~Tanizaki, T.~Misumi and N.~Sakai, \emph{{Circle compactification and
  \textquoteright{}t Hooft anomaly}},
  \href{https://doi.org/10.1007/JHEP12(2017)056}{\emph{JHEP} {\bfseries 12}
  (2017) 056} [\href{https://arxiv.org/abs/1710.08923}{{\ttfamily
  1710.08923}}].

\bibitem{Tanizaki:2017mtm}
Y.~Tanizaki, Y.~Kikuchi, T.~Misumi and N.~Sakai, \emph{{Anomaly matching for
  the phase diagram of massless $\mathbb{Z}_N$-QCD}},
  \href{https://doi.org/10.1103/PhysRevD.97.054012}{\emph{Phys. Rev. D}
  {\bfseries 97} (2018) 054012}
  [\href{https://arxiv.org/abs/1711.10487}{{\ttfamily 1711.10487}}].

\bibitem{Yonekura:2019vyz}
K.~Yonekura, \emph{{Anomaly matching in QCD thermal phase transition}},
  \href{https://doi.org/10.1007/JHEP05(2019)062}{\emph{JHEP} {\bfseries 05}
  (2019) 062} [\href{https://arxiv.org/abs/1901.08188}{{\ttfamily
  1901.08188}}].

\end{thebibliography}\endgroup

\end{document}